\documentclass[11pt,oneside]{kuthesis}

\usepackage{fancyhdr}
\usepackage{epsfig}
\usepackage{graphicx}
\usepackage{epsf}
\usepackage{verbatim}
\usepackage{amsmath}
\usepackage{longtable}
\usepackage{array}
\usepackage{appendix}
\usepackage{mathrsfs,amsmath}
\usepackage{ucs}
\usepackage[utf8x]{inputenc}

\begin{document}

\prelimpages
\Title{THE STRUCTURE AND DYNAMICS OF\\GENE REGULATION NETWORKS}
\Author{Murat Tuğrul}
\Year{December, 2007}
\Signature{Assist. Prof. Alkan Kabakçıoğlu}
\Signature{Assist. Prof. Deniz Yüret}
\Signature{Assoc. Prof. Attila Gürsoy}
\titlepage
\thesissignaturepage

\dedication{ \vspace*{5cm}
\begin{center}{\em\large To Peace}
\end{center}
}


\abstract{The structure and dynamics of a typical biological system are complex due
to strong and inhomogeneous interactions between its constituents. The investigation of such systems with classical mathematical tools, such as differential equations for their dynamics, is not always suitable. The graph theoretical models may serve as a rough but powerful tool in such cases.

In this thesis, I first consider the network modeling for the representation of the biological systems. Both the topological and dynamical investigation tools are developed and applied to the various model networks. In particular, the attractor features' scaling with system size and distributions are explored for model networks. Moreover, the theoretical robustness expressions are discussed and computational studies are done for confirmation.

The main biological research in this thesis is to investigate the transcriptional regulation of gene expression with synchronously and deterministically updated Boolean network models. I explore the attractor structure and the robustness of the known interaction network of the yeast, \textit{Saccharomyces Cerevisiae} and compare with the model networks. Furthermore, I discuss a recent model claiming a possible root to the topology of the yeast's gene regulation network and investigate this model dynamically.

The thesis also included another study which investigates a relation between folding kinetics with a new network representation, namely, the incompatibility network of a protein's native structure. I showed that the conventional topological aspects of these networks are not statistically correlated with the phi-values, for the limited data that is available.}

\oz{Tipik bir biyolojik sistemin yapısı ve dinamiği, ögelerinin birbirleri ile homojen olmayan ve güçlü etkileşimleri sebebiyle karmaşıktır. Dinamik incelemelerde kullanılan türevli denklemler gibi klasik diyebileceğimiz matematiksel yöntemler, bu tür karmaşık sistemlerin incelenmesinde her zaman uygun olmayabilir. Çizge kuramsal modeller ise daha yüzeysel olsa da bu tür sistemlerin incelenmesi için daha etkili bir yöntem olabilir.

Bu tezde, ilk olarak biyolojik sistemlerin sunumu için ağ modellemesi ele aldım. Topolojik ve dinamik inceleme araçları geliştirilip çeşitli model ağlara uyarlandı. Özelde, model ağlar için çekici özelliklerinin sistem büyüklüğü ile ölçeklenmesi ve dağılımları incelendi. Ayrıca, kuramsal dayanıklılık ifadeleri tartışıllıp ve hesaplamalı olarak doğrulukları sınandı.

Bu tezdeki ana biyoloji araştırması, transkripsiyonel gen ifadesinin düzenlenmesinin eşzamanlı ve deterministik güncellenen Boolyan ağ modeli ile incelenmesi olmuştur. Etkileşim ağı bilinen maya, \textit{Saccharomices Cerevisiae}'nın çekici yapısını ve dayanıklılığını inceledim ve model ağlar ile karşılaştırdım. Ayrıca, mayanın gen ifadesi ağının topolojik muhtemel temellerini irdeleyen yeni bir modeli tartışdım ve bu modeli dinamik olarak inceledim.

Bu tezde ayrıca bir başka ağ modellenmesi olan; asıl protein yapısından elde edilen bağdaşmaz (incompatibility) ağ ile protein kinetiğinin incelenmesi yer almaktadır. Elimizdeki sınırlı veri ile yapılan sınamalarda geleneksel olarak kullanılan belirli topolojik özellikler ile fi-değerleri arasında bağıntı olmadığını gösterdim.}

\acknowledgments{First of all, I would like to express my gratidute to my advisor, Assist. Prof. Alkan Kabakçıoğlu, whose understanding, patience and help added considerably to my graduate experience. I would like to thank other members of my committee Assist. Prof. Deniz Yüret and Assoc. Prof. Attila Gürsoy for critical readings of this thesis and for their valuable comments. A special thank goes to Osman Nuri Yoğurtçu for his effort in reading and critiques. I would like to thank Elif Müjen Şencan for being near me during this master period. Lastly, I would like to thank all my friends, professors, students, workers at Koç University and the people of Sarıyer.}

\tableofcontents
\listoftables
\listoffigures

\chapter*{Prologue}
\addcontentsline{toc}{chapter}{Prologue}

I remember my scientific interest related to biology at first started in my second year in physics undergraduate. I was very enthusiastic about arranging a sort of scientific article reading group with my classmates at that time. For the first of reading I was searching through the internet for a scientific article that is not very complicated so that our education let us understand the concepts. By luck I found an article\footnote{Unfortunately, I do not remember exact reference.} about the biology of human hearing and its mathematical modeling. I was impressed by this marvelous organization in the ear and made the article be our initiator reading-piece.

Unfortunately, this reading group did not gather for the second time but helped me understand deeply two important things. The first is that it is not easy and recommendable to do something "social" with the physicists. The second and more related to this context is that biology is not scary as I used to consider, on the contrary, it seems to encompass many bright inquiries about the nature.

In the following period up to the last grade in undergraduate, my interest in biology had increased gradually. I remember some of my popular scientific readings at that period: Schrödinger's book ``What is Life?'', Watson's book ``Double Helix'', Dawkins' book ``Selfish Gene'', etc.. I had taken some courses related to biology and ecology. At last grade I had already been sure to pursue in the life sciences in academy. 

Then in September 2005, I started my master degree in Computational Sciences and Engineering program at Koç University. This program has let me appreciate some necessary knowledge about computation and given a chance to do research in biology. In particular, I have investigated the protein folding problem and the transcriptional gene regulation in my thesis. And now I introduce you my studies and research throughout this thesis.

But before passing to the thesis, at this moment, let me ask myself some very basic questions and present the answers so that you can see where I am standing and where I am looking through the biology. 

\textit{- What is Biology?}

For me, the biology is a branch of science which tries to answer the questions gathered around a very philosophical question: ``What is life?''

\textit{- How did this branch of science emerge, what is its history in brief?}

I think the history of human knowledge on biology has no starting time since every species must have some information regarding their own and other organisms for surviving in evolutionary period. However, our knowledge on this history at the preliterate time is suspicious and mainly relies on the guesses. If we want to have a more concrete idea emerging from evidences or documents we should go back to Egypt at the era of 3500-1500 BC (at least for the western history). But, for the sake of the reader's wonder it might be worthy to mention that people of preliterate ages were able to classify the animals and plants, to say which plants are/are not toxic, are suitable for some basic medical purposes.

We know that starting from Egypt (3500-1500 BC) the biological knowledge had increased in developed civilizations of the time but merely due to practical needs, i.e. anatomy and agriculture. Also, it was very mixed with mystery, magic and superstition. As long as I know, the first examples of what today we call ``the scientific studies'' (more abstract questions) regarding the biology belong to ancient Greece. Later, while we see a progress in the knowledge of living things in Arab-Islam civilizations and in Europe of Renaissance I think the actual roots of today's biology belong to Mendel's genetics and Darwin's evolution\footnote{See References~\cite{Epic_History_of_Biology, Cell_Bio_KARP} for further readings.}.

When we come to the 20th century with the development in technology, we witness many improvements in biology, such as exploring the DNA as a genetic material. Today, we even decipher the genetic code of many species and possess tremendous data. With an analogy to physics, biology seems to stand at the point of physics at the beginning of 20th century.

\textit{- How about the methodology of biology?}

As a physicist by training, biology seems to be an almost pure empirical discipline to me, however, especially for the last decades there has been a tremendous flow of scientists into the life sciences from other disciplines whose methodology depends mainly upon the theoretical studies. Therefore, we can interpret this as the methodology of biology is just in change.

\textit{- What are the main difficulties of biology today?}

The biological systems, such as cells, are very complex in terms of both structure and dynamics. In other words, there are many variables that are interacting with many other variables in time and space. Today's classical approach in science is limited and it seems to me as the main difficulty in biology. Moreover, there are tremendous data which are waiting to be analyzed, however, there seems to be not a unification for collecting and interpreting these data. That has been at least very confusing and challenging for me while surveying.

\textit{- What is the future of biology?}

There is of course not a direct answer to this question but it is seen from the avalanche of scientists on biology that this century will see many developments in the knowledge of living things.

\begin{flushright}
Murat Tuğrul

Sarıyer, October 2006
\end{flushright}

\textpages

\chapter{Introduction}

The development and use of technology have accumulated a vast amount of experimental information for biology, such as DNA sequences of different species. However, our knowledge on how a cell works remains largely unexplored~\cite{Lockhart_Winzeler_Genomics,Barabasi_Oltvai_NetworkBiology}. 

An important component of functional organization in the cell is the regulation of gene expression. Many interacting gene pairs of some organisms were identified with a high accuracy~\cite{Bergman_etal_SixGRN}, especially for the yeast, \textit{Saccharomyces Cerevisiae}~\cite{Lee_Yeast}. The networks, or with a mathematical terminology: graphs~\cite{GrapTheory_WEST}, serve as a simple but powerful mathematical representation of the regulation of the gene expression, i.e. gene regulation networks (GRN). Many topological tools have been developed for the investigation of the networks~\cite{AlbertBarabasi_StatMecCompNets,Bollobas_ModernGraph,Milo_etal_Motifs,RichClubCoeffic_Colizza_etal,Newman_ScientficCollaborationNegtworks_2_2001,Barabasi_Oltvai_NetworkBiology} and they have been already applied to the yeast and other GRNs extensively~\cite{Guelzim_Yeast,Nicholas_etal_GenomicsRegNetworks,Bergman_etal_SixGRN}.

It is known that the genes of eukaryotes are not always active~\cite{Miglani_GENETICS} and their activation profiles show a very complex dynamical aspect beacuse of the strong and inhomogeneous interactions. As a consequence, the studying the dynamics of a gene regulation is not easy with the classical dynamical investigation tools like differential equations. In gene regulation literature, the deterministically and synchronously updated boolean networks have been used widely for the dynamical investigations~\cite{Kauffman_Network,Aldana_BooleanNetPLtopology2003}. The boolean model is based on a $1$/$0$ binary representation of the individuals at discrete time steps. Though such modelings are approximate~\cite{Norrell_CompareBooleanAndContinous}, it has been shown that some applications predicted the wild and mutant phenotypes correctly~\cite{Mendoza_etal_Arabidopsis,Espinosa_AThalianaFlowerDevelopment,AlbertOthemer_TopologyPredictsExpression}.

Another challenge for the dynamics of gene regulation is that the rules (functions) that govern the interactions are not known in detail, therefore; the statistical approaches with random functions gain importance. Some experimental studies established some canalazing behaviors in the regulation functions~\cite{Harris_Functions}. Later, it was argued that a subset of canalazing functions which was named as the nested canalazing function exists in yeast gene regulation~\cite{Kauffman_Nested}. A very recent study which depended on a logical formalism (AND and OR) claimed that two subclasses of the nested canalazing functions are dominant in the yeast~\cite{Nikolajewa_SpecTypeRuletable}.

Since the size of state space $2^N$, where $N$ is the number of individuals, is finite quantity with a boolean approach, a deterministically and synchronously updated dynamics fall into the state cycles which are called \textit{attractors}. The attractors are used for investigation of the dynamics and it is argued that attractors in GRN correspond to some cycles in the cells such as phenotype~\cite{Mendoza_etal_Arabidopsis,Espinosa_AThalianaFlowerDevelopment}. In particular, the number and length of attractors, the average transient length to the attractors and the basin of attractions are explored. Another notion for the dynamics is the \textit{robustness} of the system against the perturbations. There is a hypothesis related to robustness of the living systems; ``Life at the edge of the chaos'' which states that biological systems are to be robust against the perturbations but at the same time must be able to adapt the environments in order to be successful in evolution ~\cite{Aldana_BooleanNetPLtopology2003,Kauffman_Origins_of_Order,Shmulevich_Kauffman_Aldana_EukaryoticNOTCHAOTIC}.

In order to simulate gene regulation, artificially created network models have been used in literature, to my knowledge, starting from Kauffman~\cite{Kauffman_Network}. Many dynamical studies using the boolean approach have been performed with the model networks. A highly used model was called ``Kauffman Networks'' which has N nodes and exactly $2$ incoming edges (in this thesis, this model is the in-NK network with $K=2$). For these model networks, it was believed that the number and length of attractors scale with $\sqrt{N}$ with random functions, but Socolar \& Kauffman recently argued that the number of attractors scales faster than linear~\cite{Socolar_Kauffman}. Apart from the attractors, the robustness of in-NK model with random functions which can be explained analytically by Derrida~\&~Pomeau~\cite{Derrida_Pomeau} was extensively studied in literature~\cite{Aldana_BooleanNetPLtopology2003}.

In this thesis, the theoretical background is discussed first and applied to the artificially constructed model networks; of in-NK type, which has N nodes and exactly K indegree edges for each node; of in-PL type, which has a power-law indegree distribution and argued to be found in the natural systems often~\cite{Aldana_BooleanNetPLtopology2003}; of in-EXP type, which has an exponential indegree distribution as in yeast GRN~\cite{Guelzim_Yeast}. Upon investigation with conventional topological tools, the dynamical investigations take part for those model networks. In order to be able to compare the results with the literature, the investigations are performed with the structural parameter of $\langle k_{in} \rangle=2.0$ and with the boolean functions parameter $p=0.5$, which is the probability for assigning a gene to be expressed~\cite{Bhattacharjya_Liang}. The distribution of attractor features show a power-law decay~\cite{Bhattacharjya_Liang,Paul_etal_AttractorsofCF}. Also, it is observed that simple random functions produced larger average values of attractor features for these model networks than other canalazing types do. Apart from the attractor structures, the robustness was investigated. It is shown that Derrida's robustness expression for random functions~\cite{Derrida_Pomeau,Aldana_BooleanNetPLtopology2003} predicts the robustness values of the model networks within a finite-size effect.

The yeast GRN is dynamically investigated with exploring the attractor and robustness structures and compared to the model networks. The average and distribution of the yeast attractor features were compared for different types of functions. It has been seen that the special subclasses of nested canalazing functions produce high number of attractors and short length of attractors and transients in the dynamics realizations. Also, it is observed that the distributions of the number of attractors and entropy are not typical and show a different profile unobserved before. Li \textit{et al.} stated that the yeast GRN is robustly designed~\cite{Li_etal_YeastRobustness}. The robustness of the yeast GRN with various functions and $p$-values is computed. Morover, the model networks whose degree distributions are similar to the yeast are compared with the yeast dynamics and it is shown that those networks fail to mimic the attractor features while predicting the robustness structure correctly. Furthermore, a recent model, which is called in the thesis as Balcan \textit{et al.} model, is discussed and some of the the results were reproduced~\cite{Balcan_etal2007}. It is shown that while Balcan \textit{et al.} model is very successful for producing networks that are topologically similar to the yeast, it fails to mimic both attractor and robustness structures of the yeast GRN.

The thesis also included another study which investigates a relation between folding kinetics with a new network representation, namely, the incompatibility network of a protein's native structure.  Starting from the description of the protein and the protein folding problem, I discuss a novel approach to the problem proposed by my thesis advisor Assist.~Prof.~Alkan Kabakçıoğlu and tried to investigate the relation between proteins's structure and folding kinetics. I showed that the conventional topological aspects of these networks are not statistically correlated with the $\phi$-values, for the limited data that is available.

\clearpage

\paragraph{Outline:}
There are 5 chapters in this thesis. Chapter~\ref{NetworkModeling} gives the theoretical and computational backgrounds and some applications with model networks. Chapter~\ref{GeneRegulation} consists of topological and dynamical investigation of yeast gene regulation network with a comparison to exponential and Balcan\textit{et al.} model networks. Chapter~\ref{ProteinFolding} is for the investigation of a new approach to the protein folding problem. The final part of the thesis includes conclusions and appendices that contain extra informations.  

\chapter{Network Modeling}\label{NetworkModeling}

This chapter is organized as follows: Section~\ref{GraphTheory} is an brief explanation of Graph Theory which is the mathematical roots of networks. Section~\ref{NetworkTopology} is an overview of conventional tools for topological analyse of networks and gives some applications with model networks. Section~\ref{NetworkDynamics} constructs the necessary modeling and tools for dynamics, and applies to model networks.

\section{Graph Theory}\label{GraphTheory}

Considering the dynamics and structure of \textit{complex systems}, we need better mathematical tools than ordinary ones, such as differential equations, to deal with these systems \cite{AlbertBarabasi_StatMecCompNets}. For such a purpose, \textit{networks} are used in scientific representation of complex systems.

\begin{figure}[!hb] 
 \centering 
  \includegraphics[width=0.35\textwidth]{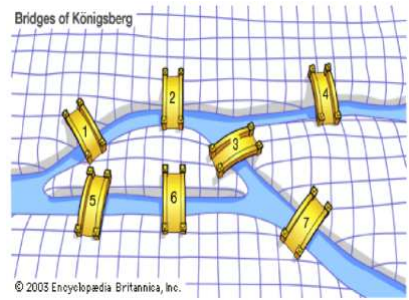}
 \caption[K\"{o}nigsberg's Bridges Problem]{The problem is whether it is possible for a citizen of K\"{o}nigsberg (today's Kaliningrad) to start from somewhere in the city and cross the bridges exactly once and return to initial place.}
 \label{KonigsbergCity} 
\end{figure}

A more mathematical term for ``Network'' is ``Graph'' and all tools of network modeling can be originated from \textit{Graph Theory} which is a popular subdiscipline of mathematics at present. It is believed that the starting point of Graph Theory goes back to 1736: Euler's negative proof to the famous \textit{K\"{o}nigsberg Bridge Problem} (Figure~\ref{KonigsbergCity}\footnote{The figure is taken from http://www.britannica.com/eb/article-9384377/Konigsberg-bridge-problem.}). Many developments in graph theory have been achieved in the previous century with an increasing interest from other sciences \cite{GrapTheory_WEST}.

Here, I emphasize only the fundamental concepts related to my thesis. The definitions of Graph Theory are given as follows (as adapted from West \cite{GrapTheory_WEST}):

\begin{quotation}
\textbf{Graph:} A graph (or network) \textbf{G} is a triple consisting of a node (or vertex) set $\textbf{V(G)}$, an edge set $\textbf{E(G)}$, and a relation that associates with each edge two vertices (not necessarily distinct) called its endpoints.

\begin{figure}[!htb]
\centering
\includegraphics[width=0.25\textwidth]{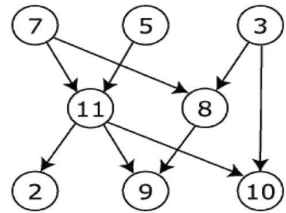}
\caption[A simple graph (network)]{A simple graph (network); numbered objects in the figure are vertices (nodes) and the arrows are directed edge.}
\label{simple_network}
\end{figure}

\textbf{Loop, Adjacent and Neighbor:} A \textbf{loop} is an edge whose endpoints are equal. When the nodes \textit{u} and \textit{v} are the endpoints of an edge, they are \textbf{adjacent} and \textbf{neighbors}, and will be shown as $u \leftrightarrow v$

\textbf{Path and Cycle:} A \textbf{path} is a simple graph whose nodes can be ordered so that two nodes are adjacent if and only if they are consecutive in the list. A \textbf{cycle} is a graph with  an equal number of nodes and edges whose nodes can be placed around a circle so that two nodes are adjacent if and only if if they appear consecutively along the circle.

\textbf{Subgraph, Connectedness, Cluster:} A \textbf{subgraph} of a graph G is a graph H such that $V(H)\subseteq V(G)$ and $E(H)\subseteq E(G)$ and the assignment of endpoints to edges in H is the same as in G. We then write $H \subseteq G$ and say that "G contains H". A graph is connected if each pair of nodes in G belongs to a path; otherwise, G is disconnected. A \textbf{cluster} is a connected subgraph in a graph and has no edge to nodes which are not in this subgraph.

\textbf{Incident, Degree, Isolated node:} If node $v$ is an endpoint of edge $e$, then $v$ and $e$ are \textbf{incident}. The \textbf{degree} of node v, $d(v)$, is the number of incident edges. An \textbf{isolated node} is a node of degree 0.

\textbf{Directed Graph (or Digraph):} A \textbf{directed graph} or \textbf{digraph} G is a triple consisting of a node set V(G), an edge set E(G), and a function assigning each edge to an ordered pair of nodes. We say that there is an edge \textbf{from} $v_i$ \textbf{to} $v_j$ and show it as $v_i \rightarrow v_j$.

\textbf{Component, Adjacency Matrix:} The \textbf{components} of a graph G are its maximal connected subgraphs. The \textbf{adjacency matrix} of G, written A(G), is the n-by-n matrix in which entry $A_{i,j}$ is the number of edges in G with $v_i \rightarrow v_j$.

\textbf{Outdegree, Indegree:} Indegree of a node $v$, $d(v)_{in}$, is the number of edges into $v$. Outdegree of $v$, $d(v)_{out}$, is the number of edges from $v$.
\end{quotation} 
\section{Network Topology}\label{NetworkTopology}

Networks are mathematical objects that represent the real complex systems. In order to classify the network structures, here I tried to examine the topological properties and quantifiers at first and later gave some examples with model networks.

\subsection{Topological Properties and Quantifiers}

Other than the number of nodes \textit{N} and of edges \textit{N(e)}, one needs some features for both comparing and classifying the topologies of networks. Here, I listed some which were used throughout this thesis. For more detailed readings one can see the review articles~\cite{Barabasi_Oltvai_NetworkBiology,AlbertBarabasi_StatMecCompNets}.

\subsubsection{a- Degree Distribution, $P(k)$}

The degree probability distribution is the probability distribution function for finding a node of network with degree $k$.

If the network is non-directed then there is only one degree notion, however, if the network is directed then one can define three different degree distributions. \textit{Total-Degree Distribution, $P(k_{tot})$}: In this case the network is pretended to be non-directed, in other words, directions of the edges are removed, and the degree distribution is explored. \textit{Out-Degree Distribution, $P(k_{out})$}: In this case, one counts only the edges \textit{outgoing} from node and calculate their distribution. \textit{In-Degree Distribution, $P(k_{in})$}: In this case, one counts only the edges \textit{incoming} to node and calculate their distribution.

\subsubsection{b- Degree-degree correlation, $k_{nn}(k)$}

Degree-degree correlation function gives us the average degree of a node which our $k$-degree node connects~\cite{Balcan_etal2007,Guelzim_Yeast}.
\begin{equation}\label{degree-degree}k_{nn}(k)=\sum_{k'}{k'p(k|k')}\end{equation}
where $p(k|k')$ is the conditional probability that the node with degree $k$ is connected to a node with degree $k'$.

\subsubsection{c- Clustering Coefficient, $C_i$}

The clustering coefficient of a node is the fraction of the existing triangles including the node in quest to maximum possible number of triangles including this node~\cite{AlbertBarabasi_StatMecCompNets}. Using the notation $\Delta_i$ for the number of triangles including $v_i$ and knowing the fact that the maximum number of triangles is $\frac{k_i(k_i-1)}{2}$, one can state the clustering coefficient of $v_i$ $C_i$ as follows:
\begin{equation}\label{cluster_coef_1}C_i=\frac{2\Delta_i}{k_i(k_i-1)}\end{equation}
One can define another quantity $C(k)$ which give us the average $C_i$ of the nodes whose degree is $k$~\cite{Balcan_etal2007}.
\begin{equation}\label{cluster_coef_2} C(k)=\langle C_i\rangle_{d(v_i)=k} \end{equation}

\subsubsection{d- Rich-Club Coefficient, $r(k)$}

One can define $N_{>k}$ as the number of nodes whose degree is greater than $k$ and $N(e_{>k})$ as the number of edges between those nodes. Then, rich-club coefficient $r(k)$~\cite{RichClubCoeffic_Colizza_etal}: 
\begin{equation}\label{rich-club} r(k)=\frac{2N(e_{>k})}{N_{>k}(N_{>k}-1)}.\end{equation}

\subsubsection{e- K-core}

$K$-core was proposed by Bollobas \cite{Bollobas_ModernGraph} as a quantity that reflects hierarchical structuring in a network. Starting from $k=0$, recursively each node whose degree is less than or equal to $k$ is labeled as the member of this $k$-core and then pruned with its edges from the network. This procedure is repeated until no node whose degree is less than $k$ remains.\footnote{For this topological feature one can use the online tool: http://xavier.informatics.indiana.edu/lanet-vi/, October 31st 2007}

\subsubsection{f- Motifs}

"Motifs" have been recently proposed by Milo \textit{et al.} \cite{Milo_etal_Motifs} to capture simple subnetwork structures in directed networks. Some motifs may appear more frequently in network at hand.\footnote{For this topological tool one can use the free-software \textbf{mfinder} from Uri Alon's webpage http://www.weizmann.ac.il/mcb/UriAlon/}

\subsubsection{g- Shortest Path, $sp$}

Another important feature in the topology of networks is \textit{geodesic} or shortest path from one node to another. There might be not a single path from $v_i$ to $v_j$ in the network and in this case the smallest length path is called shortest path, $sp_{ij}$ \cite{AlbertBarabasi_StatMecCompNets,Newman_ScientficCollaborationNegtworks_2_2001}. One can define a probability function, $P(l)$, for finding a shortest path of length $l$. Similarly, one can also define $sp_i$, average shortest path of $v_i$ to other nodes in the network.

It should be noted that if two nodes are at different clusters, then their shortest path are either taken as infinity, or ignored in the calculations as in this thesis. Networks with small shortest paths are in special attention in literature and are named as \textbf{Small World}\cite{Amaral_SmallWorld} networks \cite{AlbertBarabasi_StatMecCompNets}.

\subsubsection{h- Betweenness, $b_i$}

Betweenness $b_i$ for $v_i$ is defined as the total number of shortest paths passing through $v_i$ \cite{Newman_ScientficCollaborationNegtworks_2_2001}. Therefore, one can define a probability function $P(b)$ for finding a node with betweenness~$b$. For finding $b_i$ I have used Newman's algorithm given in Reference~\cite{Newman_ScientficCollaborationNegtworks_2_2001}. 

\subsection{Topological Investigations on Some Model Networks}

First of all, let me define what kind of model networks were used in this chapter. These network types were named according to their \textbf{indegree} distributions and this was stressed by inserting an \textbf{in-} as a prefix to the model name, such as in-XXX network. Although I did not use in this thesis, one should be aware of other sorts of model topologies, like random or gaussian.

\textbf{in-NK Network:} In this model, every node has exactly \textbf{K} incoming edges. In-NK model has been used and investigated widely in literature.

\begin{figure}[!htb]
\begin{center}
  \includegraphics[width=0.5\textwidth]{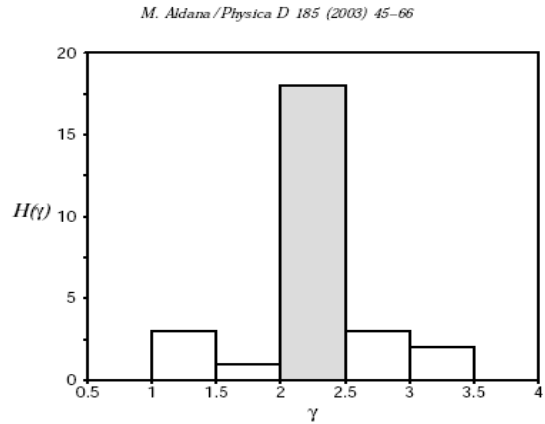}
\caption[Histogram of Power-law exponents found in nature]{Histogram shows the distribution of Power-Law exponent of 30 networks found in nature. Taken from Reference ~\cite{Aldana_BooleanNetPLtopology2003}. $\gamma$ in figure equals to $\alpha$ in my notation.}
\label{aldana_scalefreehist}
\end{center}
\end{figure}

\textbf{in-Power-Law (in-PL) Network:} A network whose degree distribution is given by power-law, i.e. $P(k)\sim k^{-\alpha}$. It has been observed that many complex systems in nature have a PL behavior in their degree distributions\footnote{In literature the term \textbf{scale-free network} is also used for power-law network}. Those PL networks found in nature come with taking the exponents ${\alpha}$ as shown in Figure~\ref{aldana_scalefreehist}~\cite{Aldana_BooleanNetPLtopology2003}.

\textbf{in-Exponential (in-EXP) Network:} A network whose degree distribution is exponential, i.e  $P(k)\sim e^{-\lambda k}$. Although to my knowledge many real-life networks are PL networks, some show an exponential behavior, in particular the gene regulation network of yeast shown in Figure~\ref{Yeast_DegreeDistr}. 

\subsubsection{a- in-NK, in-PL and in-EXP Network Topologies}

\begin{figure}[!hbt]
\begin{center}
  \includegraphics[width=0.9\textwidth]{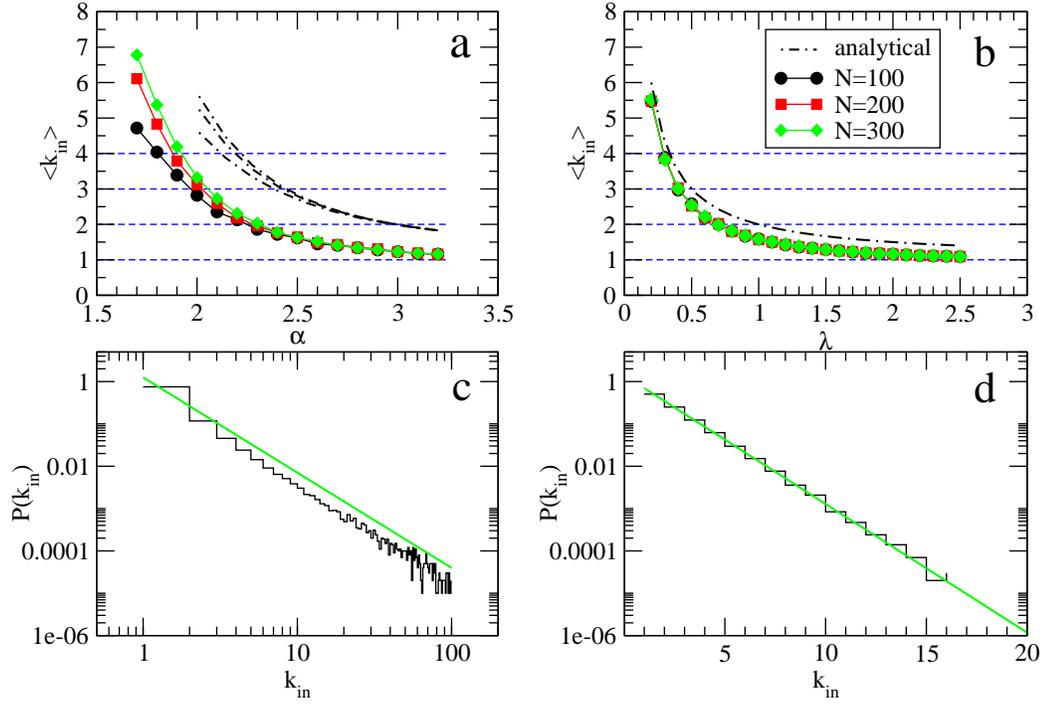}
\caption[Correspondence of in-PL and in-EXP exponents to average indegree.]{\textbf{a-)} shows the correspondence of PL exponent to $\langle k_{in}\rangle$ both for the \textit{computational} and \textit{analytical} for $N=100,200,300$ (from bottom to top). It is seen that deviation is at an important level for comp. and anal. cases. Computational results are used in this thesis. \textbf{b-)} shows the correspondence of EXP exponent to $\langle k_{in}\rangle$ both for the \textit{computational} and \textit{analytical} for $N=100,200,300$ (from bottom to top). It is seen that deviation has little effect comparing to PL case. Computational results are used in this thesis. \textbf{c-)} Indegree prob. distribution of computational and analytical cases for PL exponent $\alpha=2.25$ $N=100$; this study shows that the difference of computational and analytical cases comes from the contribution of $k_{in}=1$ in the sum. \textbf{d-)} Indegree distribution of computational and continuous analytical cases for EXP exponent $\alpha=0.7$ $N=100$; comparing the PL case, the contribution of $k_{in}=1$ to sum is small so it does not deviate from analytical case. See Appendix~\ref{ProducingNetworks} for analytical $\langle k_{in}\rangle$ expressions.}
\label{PLandEXP_gammaVSavK}
\end{center}
\end{figure}

\begin{figure}[!p]
\begin{center}
  \includegraphics[width=0.75\textwidth]{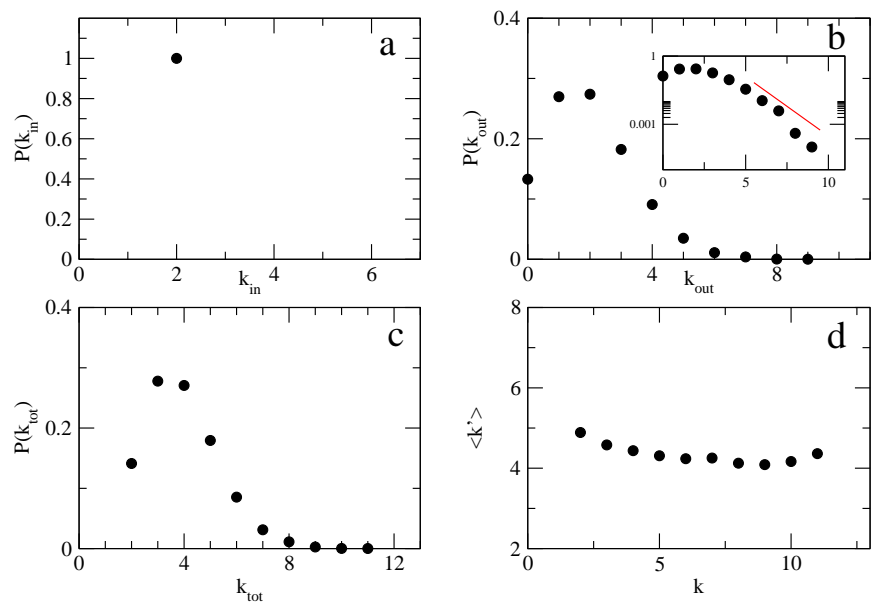}
\caption[in-NK Networks' topological investigations-1] {$100$ in-NK ($N=100$, $K=2$) networks were created. Their topological features were investigated and averaged. Figure shows the corresponding; \textbf{a-)} indegree probability distribution, \textbf{b-)} outdegree probability distribution, \textbf{c-)} totaldegree probability distribution and \textbf{d-)} degree-degree correlation.}
\label{NK2N100_DegreeDist}
\end{center}
\end{figure}

\begin{figure}[!p]
\begin{center}
  \includegraphics[width=0.75\textwidth]{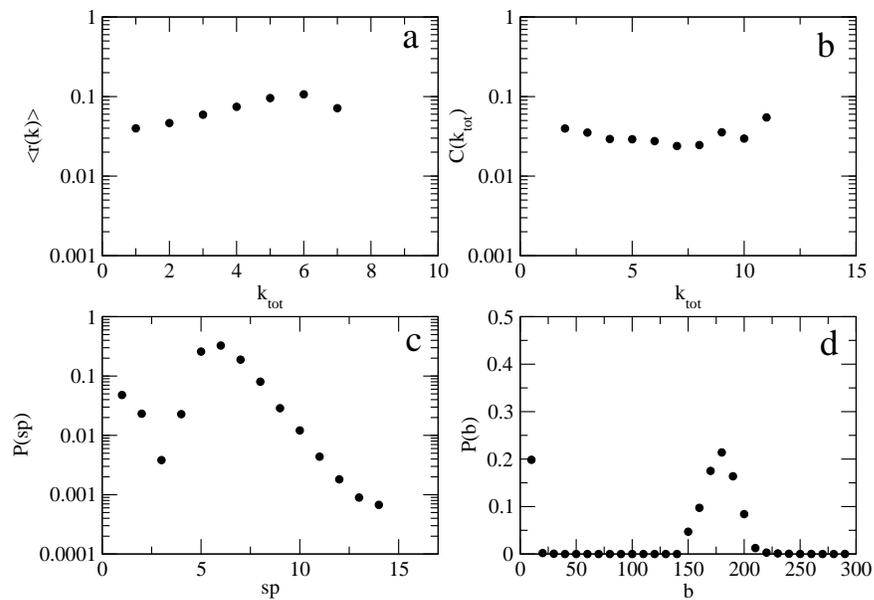}
\caption[in-NK Networks' topological investigations-2] {$100$ in-NK ($N=100$, $K=2$) networks were created. Their topological features were investigated and averaged. Figure shows the corresponding; \textbf{a-)} rich-club coefficient, \textbf{b-)} clustering coefficient distribution, \textbf{c-)} average shortest path prob. distribution and \textbf{d-)} betweenness prob. distribution.}
\label{N100K2_OtherTopologies}
\end{center}
\end{figure}

\begin{figure}[!p]
\begin{center}
  \includegraphics[width=0.72\textwidth]{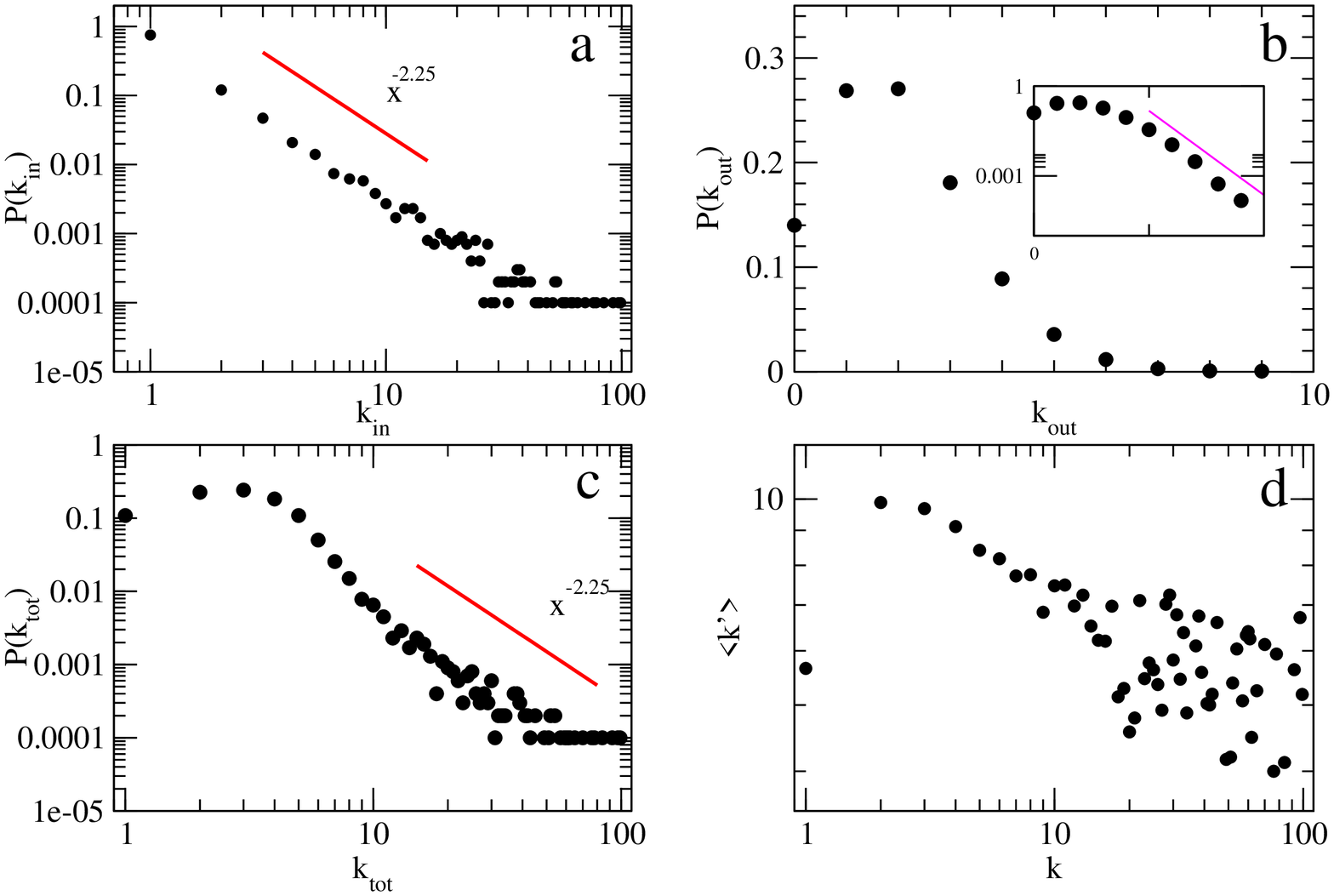}
\caption[in-PL Networks topological investigations-1] {$100$ in-PL ($N=100$, $\alpha=2.25$, $\langle k_in \rangle=2.0$) networks were created. Their topological features were investigated and averaged. Figure shows the corresponding; \textbf{a-)} indegree probability distribution, \textbf{b-)} outdegree probability distribution, \textbf{c-)} totaldegree probability distribution and \textbf{d-)} degree-degree correlation. It should be noted that $x^{-2.25}$ function was drawn in order to help the reader and is not a fitting.}
\label{inPL2.25N100_DegreeDist}
\end{center}
\end{figure}

\begin{figure}[!p]
\begin{center}
  \includegraphics[width=0.75\textwidth]{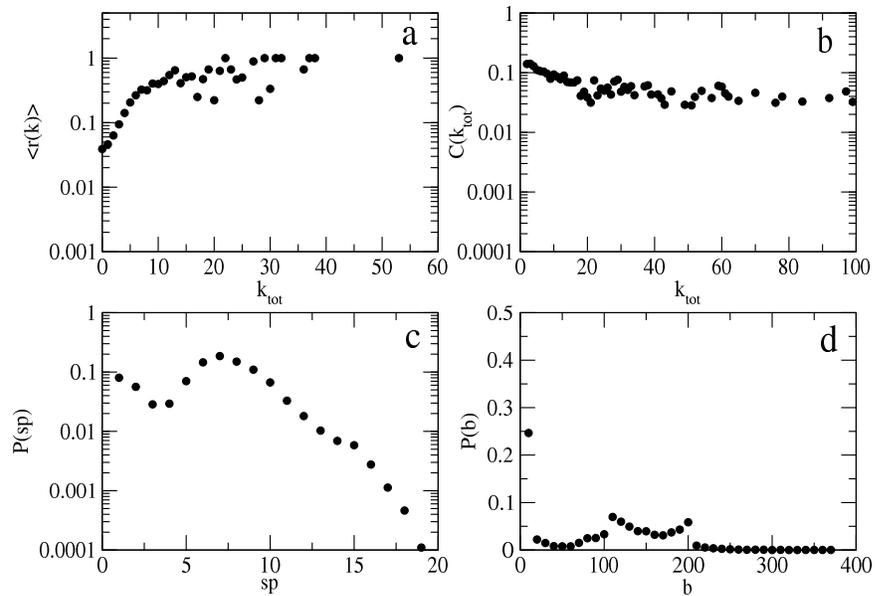}
\caption[in-PL Networks topological investigations-2] {$100$ in-PL ($N=100$, $\alpha=2.25$, $\langle k_in \rangle=2.0$) networks were created. Their topological features were investigated and averaged. Figure shows the corresponding; \textbf{b-)} clustering coefficient distribution, \textbf{c-)} average shortest path prob. distribution and \textbf{d-)} betweenness prob. distribution.}
\label{PL2.25N100_OtherTopologies}
\end{center}
\end{figure}

\begin{figure}[!p]
\begin{center}
  \includegraphics[width=0.72\textwidth]{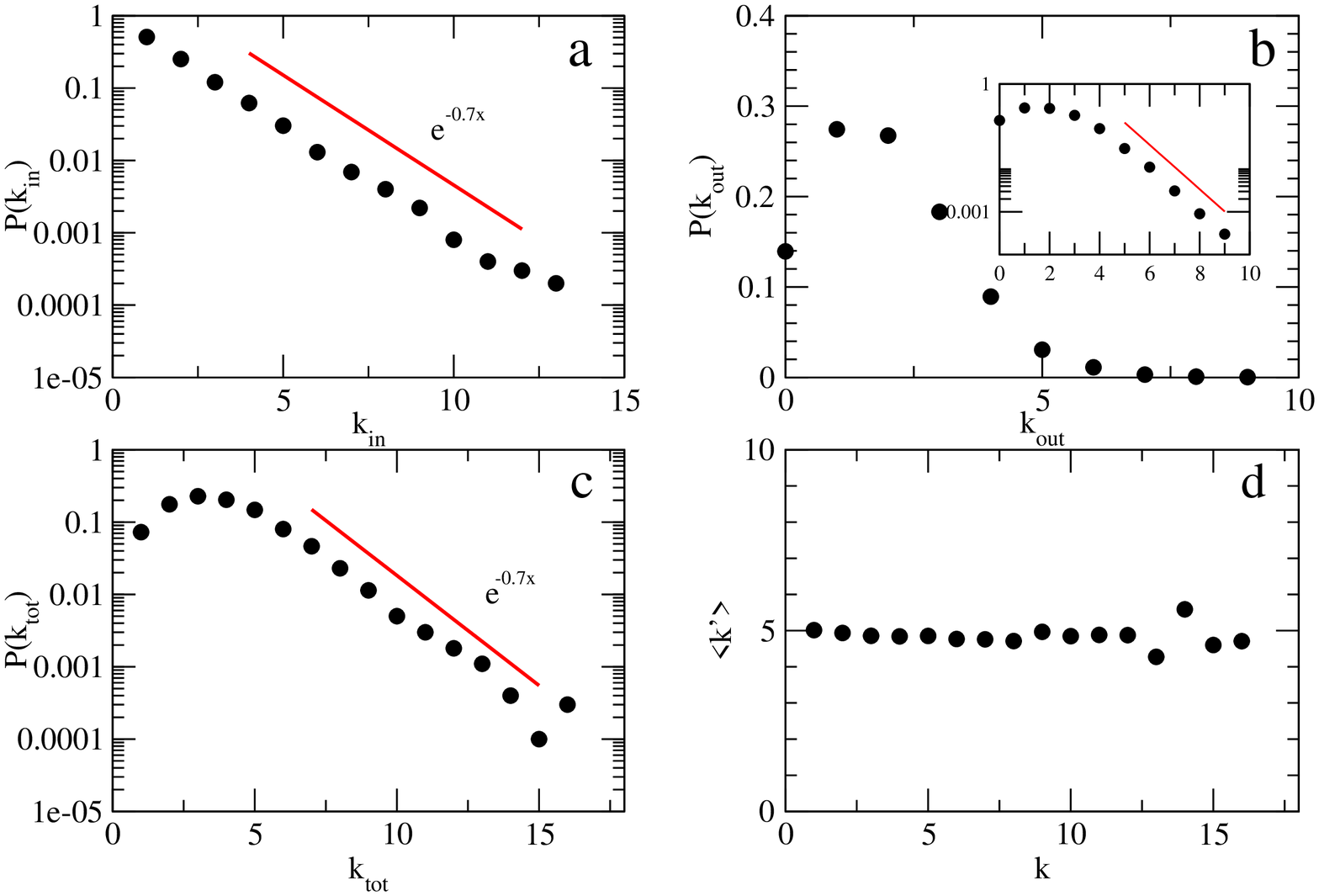}
\caption[in-EXP Networks topological investigations-1] {$100$ in-EXP ($N=100$, $\lambda=0.7$, $\langle k_{in} \rangle =2.0$) networks were created. Their topological features were investigated and averaged. Figure shows the corresponding; \textbf{a-)} indegree probability distribution, \textbf{b-)} outdegree probability distribution, \textbf{c-)} totaldegree probability distribution and \textbf{d-)} degree-degree correlation. It should be noted that $x^{-0.7}$ function was drawn in order to help the reader and is not a fitting.}
\label{inEXP0.7N100_DegreeDist}
\end{center}
\end{figure}

\begin{figure}[!p]
\begin{center}
  \includegraphics[width=0.75\textwidth]{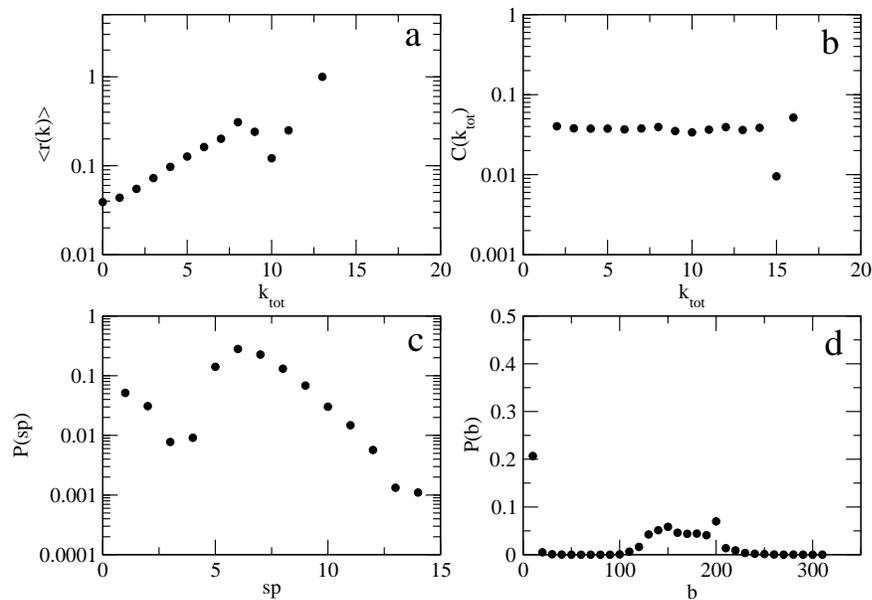}
\caption[in-EXP Networks topological investigations-2] {$100$ in-EXP ($N=100$, $\lambda=0.7$, $\langle k_{in} \rangle =2.0$) networks were created. Their topological features were investigated and averaged. Figure shows the corresponding; \textbf{a-)} rich-club coefficient, \textbf{b-)} clustering coefficient distribution, \textbf{c-)} average shortest path prob. distribution and \textbf{d-)} betweenness prob. distribution}
\label{EXP0.7N100_OtherTopologies}
\end{center}
\end{figure}

In order to be able to compare different topologies, some of the properties needed to be fixed. For the sake of comparison to the network literature, I fixed the number of nodes to $N=100$ and the average indegree to $\langle k_{in} \rangle \cong 2.00$ so that in-PL exponent $\alpha = 2.25$ and in-EXP exponent $\lambda = 0.7$ were chosen. Figure~\ref{PLandEXP_gammaVSavK} shows and discuss the correspondence of exponents to average indegree of in-EXP and in-PL networks. Next, ensembles of $100$ model networks of type in-NK, in-PL and in-EXP were created and investigated. The topological distributions of these ensembles are in Figure~\ref{NK2N100_DegreeDist}, Figure~\ref{N100K2_OtherTopologies} for in-NK networks, in Figure~\ref{inPL2.25N100_DegreeDist}, Figure~\ref{PL2.25N100_OtherTopologies} for in-PL networks, in Figure~\ref{inEXP0.7N100_DegreeDist}, Figure~\ref{EXP0.7N100_OtherTopologies} for in-EXP networks.

I found out that in general, the topological features of in-EXP networks are between in-NK and in-PL networks'.

\clearpage
\section{Network Dynamics}\label{NetworkDynamics}

For the investigation of dynamics in networks the \textbf{Boolean} model with \textit{synchronously} and \textit{deterministic update} was used. In this model, each node $v_i$ has a \textbf{node state} $\sigma_i(t)$ at a particular time \textbf{t} where $\sigma_i(t)$ is either 1 (on) or 0 (off). The \textbf{network state} $S(t)$ is the set of individual node states: $S(t) = \lbrace \sigma_1(t),\sigma_2(t),..,\sigma_N(t) \rbrace$. $\sigma_i(t+1)$ is determined by the \textbf{Boolean Function} $B_i$ of the node $v_i$. $B_i$ depends on only to indegree neighbors of the node $v_i$. It should be noted that in the case of zero indegree, $\sigma_i$ is fixed to either to $1$ or $0$ for every \textit{t}.

\begin{equation}
\sigma_i(t+1) = \left\{ \begin{array}{ll}
B_i(\sigma_{i,1}(t),\sigma_{i,2}(t),..,\sigma_{i,k_{in}}(t)), & \textrm{ $k_{in}>0$}\\
0\;or\;1\;(fixed), &\textrm{$k_{in}=0$}
\end{array} \right.
\end{equation}

where $\sigma_{i,j}$ is the node state of jth in-neighbor of $v_i$ and $k_{in}$ is the indegree of $v_i$.

During the dynamics, the node states $\lbrace \sigma_i(t)\rbrace$ are collected at time step \textbf{t} and inserted into each Boolean Function synchronously. The output of $B_i$ is assigned to the $\sigma_i(t+1)$ for time \textbf{t+1}.

\begin{table}[!htb] 
\begin{center}
\begin{tabular}{cccc||cccc}
$\sigma_{i,1}(t)$ & $\sigma_{i,2}(t)$ & ... & $\sigma_{i,k_{in}}(t)$ &   & $\sigma_i(t+1)$ & &  \\ 
\hline
0 & 0 & ... & 0 & 0 & 1 & ... & 1\\ 
0 & 0 & ... & 1 & 0 & 0 & ... & 1\\ 
. & . & ... & . & . & . & ... & 1\\
1 & 1 & ... & 0 & 0 & 0 & ... & 1\\
1 & 1 & ... & 1 & 0 & 0 & ... & 1
\end{tabular}
\caption{An example of the ruletable expression for $B_i$ where $v_i$ has $k_{in}$ indegree nodes.}
\label{Ruletable_example}
\end{center}
\end{table}

Let me explain these processes by constructing a table which includes all possible combination of incoming nodes' states and assigns an output to them for a particular node state. Such a table is called \textbf{Ruletable} as shown in Table~\ref{Ruletable_example}. If there is $k_{in}$ incoming nodes, then there are $2^{k_{in}}$ input combinations, i.e $2^{k_{in}}$ rows in the ruletable. The output of each combination is either $1$ or $0$, which makes $2^{2^k_{in}}$ different ways to construct the output column\footnote{For a better understanding of the magnitude, $2^{2^{k_{in}}} = 4, 16, 256, 65536, 4294967296$ for $K = 1, 2, 3, 4, 5$, respectively.}. Before the dynamics starts, a ruletable out of $2^{2^{k_{in}}}$ possibilities is chosen for each node, so that dynamics runs deterministically. I used the term \textbf{network realization} for one network topology with its all assigned ruletables (functions) in this thesis.

\subsection{Some Boolean Function Types}

Although in many systems we may know the interacting pairs of individuals very well, we have usually little information about which rules govern the dynamics, as in gene regulation networks \cite{Kauffman_Nested}. In other words, we do not know which combination out of $2^{2^{k_{in}}}$ to choose in the ruletable for each $v_i$. However, we are able to shape the structure of the Boolean Functions. I use $4$ types of these \textit{random} function structures found in literature.

\subsubsection{a- Simple Random Function, \textbf{RF}}

A function type whose each input combination (each row in the ruletable) is assigned to an output value of $1$ with a probability \textbf{p}.

\subsubsection{b- Canalyzing Random Function, \textbf{CF}}

A \textit{Canalyzing Random Function} has at least one canalyzing input variable, such that for at least one certain canalyzing value of that variable, the output value is fixed \cite{Kauffman_Nested,Kauffman_Origins_of_Order}.

\begin{equation}\label{CF}
B_i(\sigma_{i,1},..,\sigma_{i,j},..,\sigma_{i,k_{in}}) = \left\{ \begin{array}{ll}
 s_i & \sigma_{i,j}=s_j\\
B_i(\sigma_{i,1},..,\overline{s_j},..,\sigma_{i,k_{in}}) & \sigma_{i,j} \neq s_j
\end{array} \right.
\end{equation}
where jth in-neighbor is the canalyzing node with $s_j$ as the canalyzing value and $s_i$ as the canalazing output. As in RF only one parameter \textbf{p} is used whenever an output value is needed to be determined. It is should be mentioned that $B_i(\sigma_{i,1},..,\overline{s_j},..,\sigma_{i,k_{in}})$ in Exps.~\ref{CF} is considered to be RF.

\subsubsection{c- Nested Canalyzing Random Function, \textbf{NCF}}

Having investigated the Harris \textit{et al.}'s work on gene regulation \cite{Harris_Functions}, Kauffman \textit{et al.} have proposed a new function type known as \textit{Nested Canalyzing} or  \textit{Hierarchically Canalyzing} function which is argued to be found in the biological systems \cite{Kauffman_Nested}. In this type, there is a canalazing order in input nodes and the output is determined by first node at its canalyzing value~\cite{Kauffman_Nested} is given:
\begin{equation}\label{NCF}
B_i(\sigma_{i,1},..,\sigma_{i,j},..,\sigma_{i,k_{in}}) = \left\{ \begin{array}{ll}
s_{i,1} & \sigma_{i,1}=s_1\\
s_{i,2} & \sigma_{i,1} \neq s_1 \wedge \sigma_{i,2}=s_2\\
... & ...\\
s_{i,j} & \sigma_{i,1} \neq s_1 \wedge \sigma_{i,2} \neq s_2 \wedge ... \wedge \sigma_{i,j}=s_j\\
... & ...\\
s_{i,k_{in}} & \sigma_{i,1} \neq s_1 \wedge \sigma_{i,2} \neq s_2 \wedge ... \wedge \sigma_{i,k_{in}}=s_{k_{in}}\\
\overline{s_{i,k_{in}}} & \sigma_{i,1} \neq s_1 \wedge \sigma_{i,2} \neq s_2 \wedge ... \wedge \sigma_{i,k_{in}} \neq s_{k_{in}}
\end{array} \right.
\end{equation}
where P($s_{i,j}$=TRUE)=P($s_j$=TRUE)=$\frac{exp(-2^j\alpha)}{1+exp(-2^j\alpha)}$ j=1,2..,$k_{in}$ and j is numbered with respect to the canalyzing order.

In this thesis, the definition of NCF is modified for the sake of consistency. Firstly, a \textbf{p} parameter as in above functions were adapted. Secondly, the last statement in Exps.~\ref{NCF} in determining the output was altered. Instead of using $\overline{s_{i,k_{in}}}$, the output value was determined again by using $p$.

\subsubsection{d- Special Subclasses of Nested Canalyzing Random Function, \textbf{SNCF}}

After the proposition of Nested Canalyzing Functions (NCF) by Kauffman \textit{et al.}~\cite{Kauffman_Nested}, Nikolejewa \textit{et al.} presented ``a new minimal logical expression'' for all NCFs~\cite{Nikolajewa_SpecTypeRuletable} as follows,

\begin{equation}
\label{SNCF}
\begin{array}{ll}
\sigma_i&=B_i(\sigma_{i+1}, \sigma_{i+2}, ..., \sigma_{i+k_{in}-1}, \sigma_{i+k_{in}})\\
 &=\sigma_{i+1}^\Theta \bigodot (\sigma_{i+2}^\Theta \bigodot (...\bigodot (\sigma_{i+k_{in}-1}^\Theta \bigodot \sigma_{i+k_{in}}^\Theta)...))
\end{array}
\end{equation}
where $\bigodot$ represents either AND or OR logical function, i.e. $\bigodot\in\lbrace\wedge,\vee\rbrace$ and $\sigma^\Theta$ is for possible negation of $\sigma$, i.e. $\sigma^\Theta\in\lbrace\sigma,\overline{\sigma}\rbrace$

They classified the NCFs according to the possible chances for $\bigodot$. Upon investigation of Harris \textit{et al.} data \cite{Harris_Functions}\footnote{Nikolejewa \textit{et al.} have noted in their paper that they have taken the data from Harris by private communication} they have found that gene regulatory rules are mainly governed by two subclasses of NCF \cite{Nikolajewa_SpecTypeRuletable}:
\begin{equation}
\label{SNCF60}
\sigma_{i+1}^\Theta \wedge (\sigma_{i+2}^\Theta \wedge (...\wedge (\sigma_{i+k_{in}-1}^\Theta \wedge \sigma_{i+k_{in}}^\Theta)...))
\end{equation}
and 
\begin{equation}
\label{SNCF30}
\sigma_{i+1}^\Theta \wedge (\sigma_{i+2}^\Theta \wedge (...\wedge (\sigma_{i+k_{in}-1}^\Theta \vee \sigma_{i+k_{in}}^\Theta)...))
\end{equation}
with $66.39\%$ and $29.41\%$ probability of occurrence, respectively.
 
\begin{figure}[!ht] 
\centering 
\includegraphics[width=0.55\textwidth]{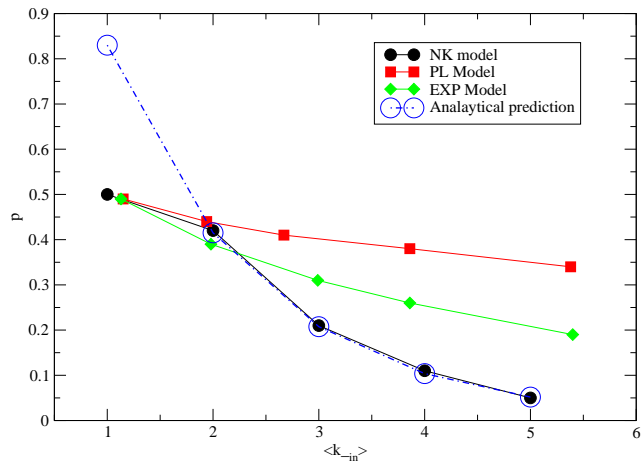}
\caption[$p$ value investigation of Special Subclasses of Nested Canalyzing Random Function.]{$p$ value investigation of Special Subclasses of Nested Canalyzing Random Function for model networks. The analytical expression for in-NK networks is $1.66\times2^{-k_{in}}$ and it is seen that it predicts correctly for $k_{in}>1$. For other types, p-values are obtained from this figure in this thesis.}  
\label{SpecType_Pinvestigation}
\end{figure}

In this type of function $p$ is not a free parameter and depends on the topology. It is easy to calculate $p$ analytically for in-NK model topologies: \textbf{p}$\approx(2/3)\times(1/2^K)+(1/3)\times(3/2^K)=1.66\times2^{-K}$ \cite{Nikolajewa_SpecTypeRuletable}. For instance, for $K=1, 2, 3, 4, 5, 6$, one finds $p=0.83, 0.41, 0.21, 0.10, 0.05, 0.03, 0.01$, respectively. Figure~\ref{SpecType_Pinvestigation} shows the validity of this formula for in-NK model where $K>1$ and presents also p values for other topologies.

\subsection{Dynamical Properties and Quantifiers}

In order to compare the dynamics of various networks, one needs quantitative measures. Here I provide quantities related to two notions: \textit{Attractor} and \textit{Robustness}.

\subsubsection{a- Attractor}

\begin{figure}[!ht] 
 \centering 
  \includegraphics[width=0.65\textwidth]{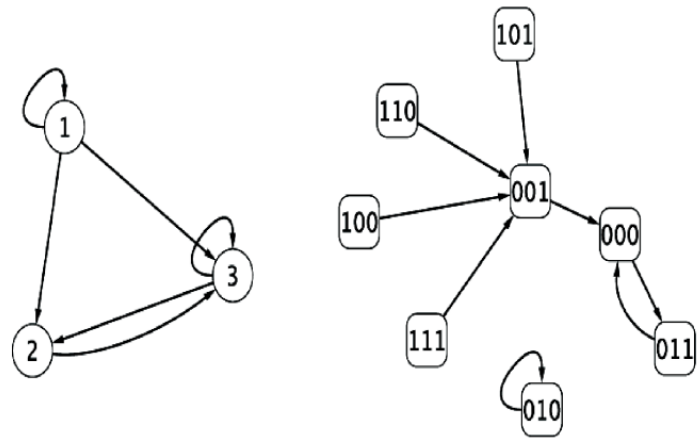}
 \caption[A 3-node network and its state space with an attractor]{A simple network of 3 nodes is shown on the left. Its realization with ruletables shown in Table~\ref{Attractor_Example_Ruletable} produces two attractor as shown on the right.} 
 \label{Attractor_Example_Scetch} 
\end{figure}

Remembering that each node state can be either 1 or 0, the size of state space is $2^N$. Once the network realization (network topology and ruletables) are fixed, the dynamics is deterministic. In other words, if one chooses a network state $S_i$ at time \textit{t} in $2^N$ states, s/he arrives at exactly \textit{one} network state at next time step \textit{t+1}. Also, since $2^N$ is a finite number, at most after traversing all the states, the dynamics starts to fall in a cycle (Figure~\ref{Attractor_Example_Scetch}). Such a cycle is called \textit{attractor} and it is an important feature of the boolean dynamics. 

\begin{table}[!htb] 
\begin{center}
\begin{tabular}{||c|c||cc|c||ccc|c||}
\hline
\hline
$\sigma_1(t)$ & $\sigma_1(t+1)$ & $\sigma_1(t)$ & $\sigma_3(t)$ & $\sigma_2(t+1)$ & $\sigma_1(t)$ & $\sigma_2(t)$ & $\sigma_3(t)$ & $\sigma_3(t+1)$\\ 
\hline
0 & 0 & 0 & 0 & 1 & 0 & 0 & 0 & 1\\ 
1 & 0 & 0 & 1 & 0 & 0 & 0 & 1 & 0\\ 
  &   & 1 & 0 & 0 & 0 & 1 & 0 & 0\\
  &   & 1 & 1 & 0 & 0 & 1 & 1 & 0\\
  &   &   &   &   & 1 & 0 & 0 & 1\\
  &   &   &   &   & 1 & 0 & 1 & 1\\
  &   &   &   &   & 1 & 1 & 0 & 1\\
  &   &   &   &   & 1 & 1 & 1 & 1\\
\hline
\hline
\end{tabular}
\caption{The ruletables of the nodes in the network shown in Figure~\ref{Attractor_Example_Scetch}.}
\label{Attractor_Example_Ruletable}
\end{center}
\end{table}

It is believed that such attractors correspond to some process cycles in the systems, such as phenotype in the cells~\cite{Kauffman_Origins_of_Order,Mendoza_etal_Arabidopsis}. Mendoza \textit{et al.} showed the correspondence between some attractors and known phenotypes in \textit{Arabidopsis thaliana} by using a similar approach. They also predicted some mutant phenotypes and confirmed them by experiments~\cite{Mendoza_etal_Arabidopsis}. Some other studies also reported similar conclusions~\cite{Espinosa_AThalianaFlowerDevelopment,AlbertOthemer_TopologyPredictsExpression}

There are some quantifiers for the attractors in boolean systems. The first one is \textbf{the number of attractors} $N_{attr}$ in the network realization. Second one is the number of network states of an attractor possesses, \textbf{length of the attractor} $L_{attr}$. Third one is the average number of network states to arrive at an attractor, \textbf{transient to attractor} $\tau_{attr}$. The last quantifier is related to the notion of \textbf{the basin of attraction}. The set of network states which go to a particular attractor is called the basin of attraction of that attractor. The size of the basin of attraction normalized by $2^N$ is $w_{attr}$. Recently, Kravitz and Schumulevich \cite{Kravitz_Shmulevich} have proposed an entropy $h$ definition for boolean dynamics:
\begin{equation}
\label{Schumulevich_Entropy}
h=-\sum_i{w_ilnw_i}
\end{equation}
and $h$ was used here to compare the basin of attractions of the network realizations.

Attractors were found in this thesis by using a heuristic algorithm (See Appendix~\ref{FindingAttractorAlgorithms} for more details about attractor finding algorithms.).

\subsubsection{b- Robustness}

For a system to be sustainable, its dynamics should not be effected drastically in every intensive or extensive changes, such as errors in individuals or environments. On the other hand, the dynamical systems like gene regulation should be open to some changes in order to survive through evolution. These arguments brings a hypothesis called ``Life at the edge of chaos'' that states the life systems should be at some where between chaotic and ordered phases~\cite{Aldana_BooleanNetPLtopology2003,Kauffman_Origins_of_Order,Shmulevich_Kauffman_Aldana_EukaryoticNOTCHAOTIC}. 

So, how \textit{robust} system is a valuable to detect for the dynamics and urges us to quantify \textit{robustness}. It is presented here as follows \cite{Aldana_BooleanNetPLtopology2003}. Consider two network states $S(t)$ and ${S}^{'}(t)$. Their \textbf{Hamming Distance} $HD(t)$ is the number of nodes that are different in their states at time \textbf{t} \cite{Shmulevich_Kauffman_Aldana_EukaryoticNOTCHAOTIC,Aldana_BooleanNetPLtopology2003}:
\begin{equation}
\label{HammingDistance}HD(t)=\sum_{i=1}^{N}{\mid\sigma_{i}(t)-{\sigma_{i}^{'}(t)}\mid}
\end{equation}

Let me define two other quantities which are the \textbf{overlapping functions} at time step \textbf{t} and \textbf{t+1}, respectively:
\begin{equation}\label{x_t}
x(t)=1-\frac{HD(t)}{N}
\end{equation}
\begin{equation}\label{Mx_t}
M(x(t))\equiv x(t+1)
\end{equation}

The robustness under small perturbations is measured at the attractor of the system. In other words, we have
\begin{equation}
\label{x_Mx}
\begin{array}{cc}
x&=\lim_{t\rightarrow\infty}x(t)
\end{array}
\end{equation}
while $x\rightarrow1^-$.

\begin{figure}[!ht] 
 \centering 
  \includegraphics[width=0.5\textwidth]{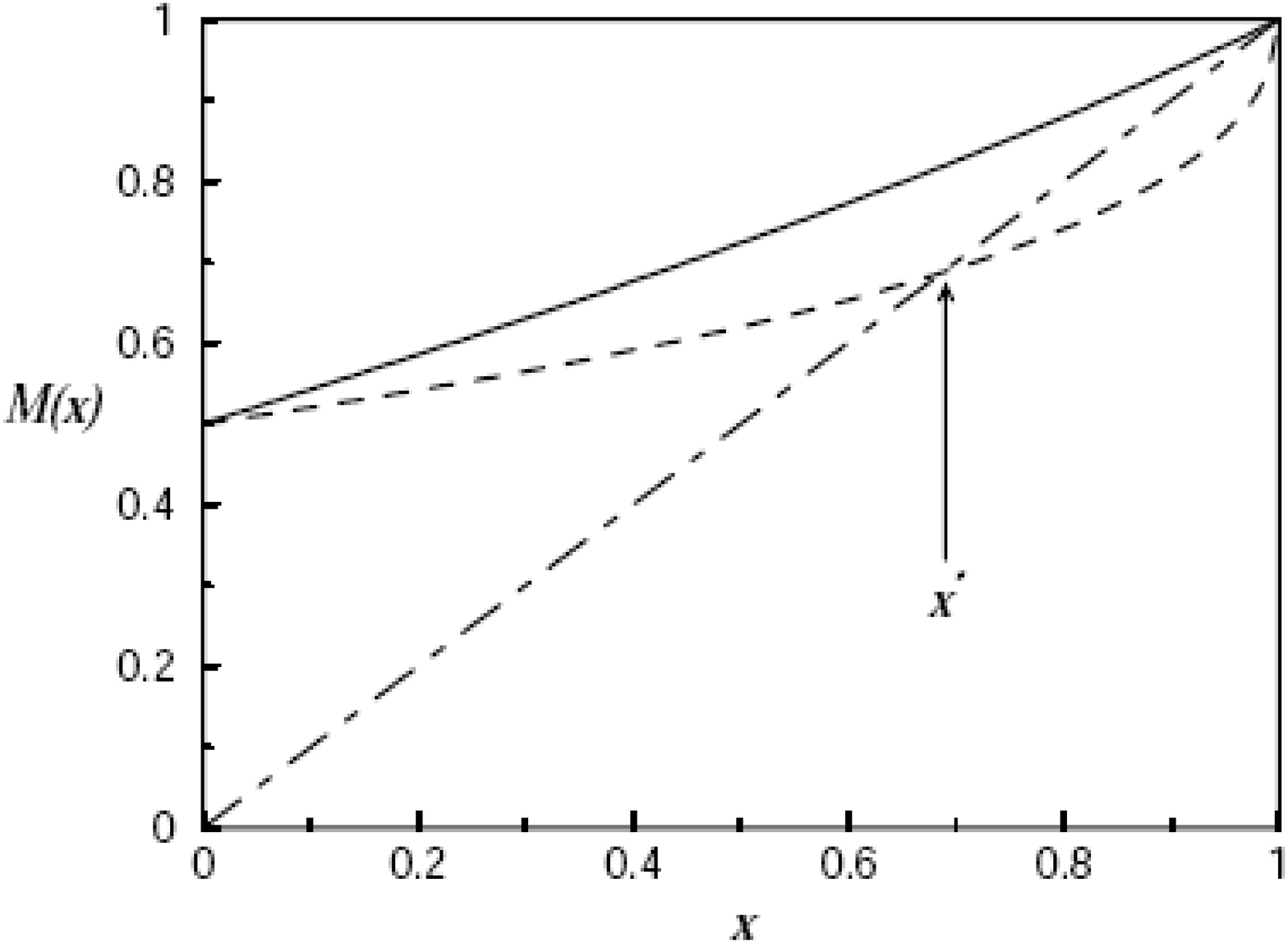}
 \caption[Robustness criteria]{Robustness criteria \cite{Aldana_BooleanNetPLtopology2003}} 
 \label{Robustness_criteria} 
\end{figure}

Referring the Figure~\ref{Robustness_criteria}, we know $M(x)\geq 0$ when $x=0$ and assume a monotic increase in $M(x)$ function. If $\lim_{x\rightarrow1^-}\frac{dM(x)}{dx}$ is gerater than $1$ then M(x) function crosses $M(x)=x$ line at some $x*<1$, which is a stable fixed point other than $x=1$. In this case, the system is forbidden to arrive at $x=1$ unless $x=1$. If $\lim_{x\rightarrow1^-}\frac{dM(x)}{dx}$ is less than $1$ then there is no stable fixed point other than $x=1$ and two network states $S$ and $S^{'}$ converge soon or later. These two cases are named as \textbf{chaotic}, \textbf{ordered} respectively and the case of $\lim_{x\rightarrow1^-}\frac{dM(x)}{dx}$ equals $1$ is named as \textbf{critical transition border/boundary} in the corresponding literature \cite{Aldana_BooleanNetPLtopology2003,Kauffman_Origins_of_Order,Shmulevich_Kauffman_Aldana_EukaryoticNOTCHAOTIC}. 

In sum, with showing the \textit{robustness} quantity with $s$,

\begin{equation}
\label{sensivity_robustness}
s=\lim_{x\rightarrow1^-}\frac{dM(x)}{dx}
\end{equation}
three important phase are summarized as follows,

\begin{equation}
\label{robustness_condtions}
\begin{array}{cccc}
s & < & 1 & Ordered \\ 
s & = & 1 & Critical\; Boundary \\
s & > & 1 & Chaotic.
\end{array}
\end{equation}
and it is concluded that if $s<1$, the system is robust against perturbations while if $s>1$, the system is very sensitive to them~\cite{Shmulevich_Kauffman_Aldana_EukaryoticNOTCHAOTIC}.

\subsection{Dynamically Relevant Subnetwork}

Some of the nodes are irrelevant to the attractor results due to topology or functions of the network realization. These nodes only serve as computational challenges and some of the nodes with their edges can be recursively removed from the network without any general change in the dynamics \cite{Socolar_Kauffman}. Such nodes were labeled as \textbf{irrelevant} and the rest of the network/nodes were called the \textbf{dynamically relevant subnetwork/nodes} in this thesis.

\begin{figure}[!hb] 
 \centering 
  \includegraphics[width=0.7\textwidth]{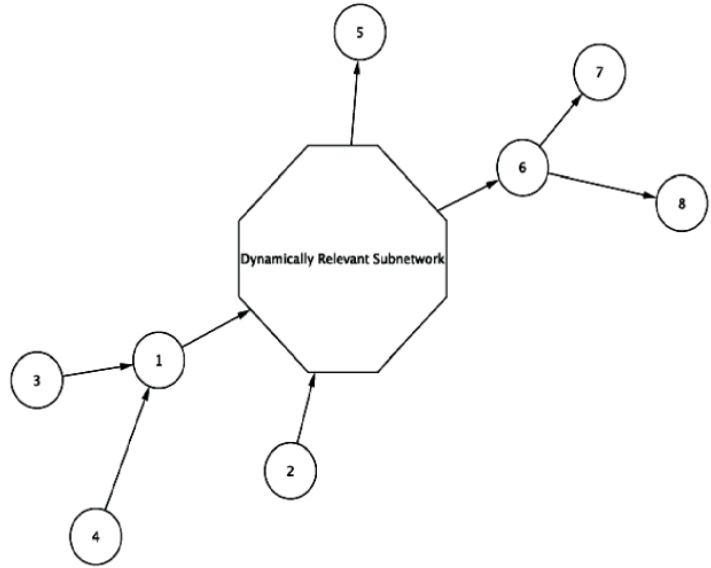}
 \caption[Dynamically Relevant Subnetwork]{The subnetwork yielded after pruning recursively the nodes with either zero outdegree or zero indegree is named \textit{dynamically relevant subnetwork} and this subnetwork was used in this thesis during the dynamics runs.}
 \label{DynamicallyRelSubnetwork}
\end{figure}

The attractor computations in this thesis were done with a minimal dynamically relevant subnetwork which is found by use of a procedure which considers only the topology of the network to obtain it . The procedure depends on the fact that a node with zero indegree stays at a fix state all time steps. Also, a node with zero outdegree does not affect any node in the system although its state may fluctuate. Removing these two types of nodes with their edges recursively results in the dynamically relevant subnetworks used in this thesis.

\begin{figure}[!htb] 
 \centering 
  \includegraphics[width=0.8\textwidth]{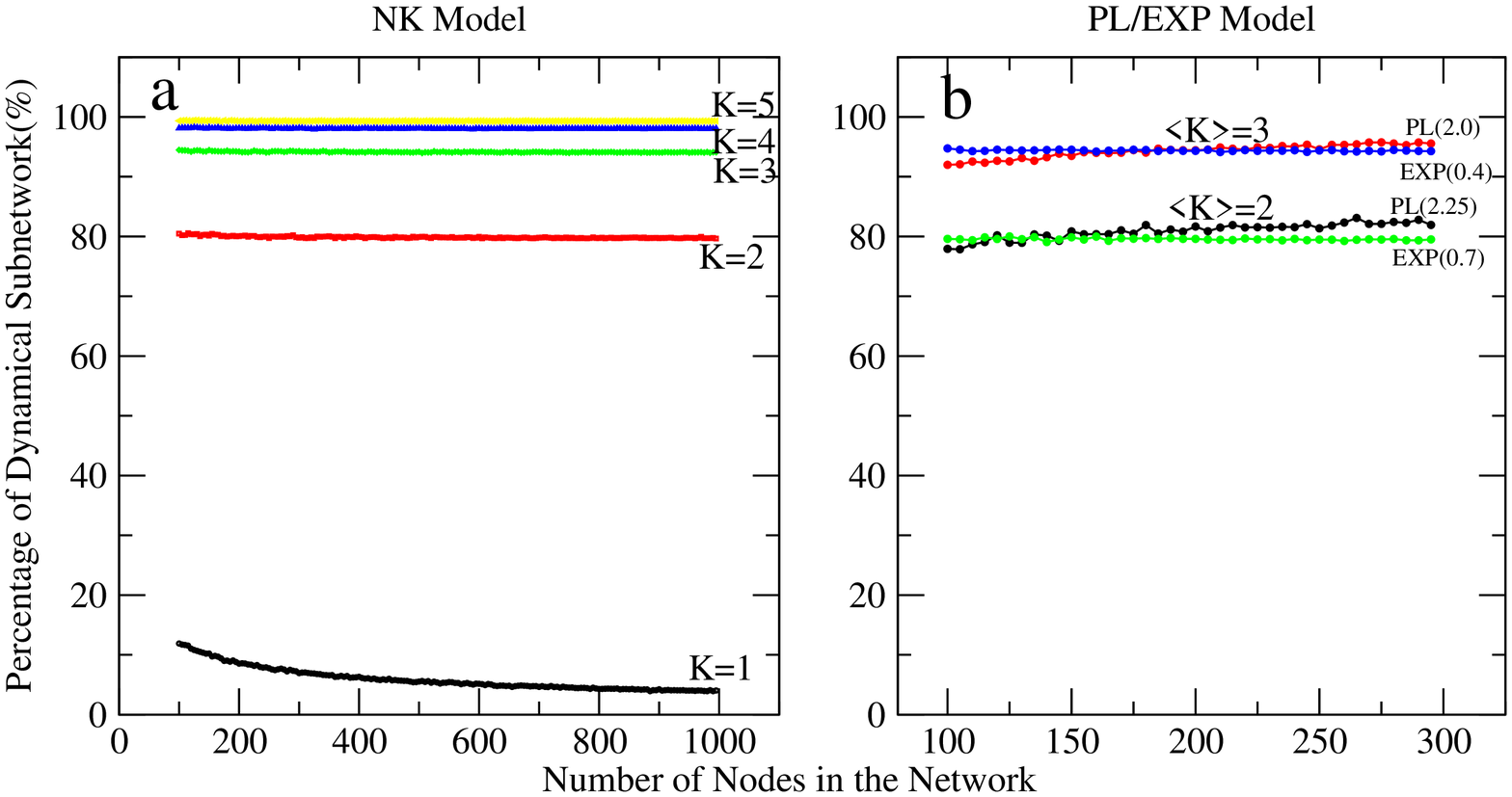}
 \caption[The fraction of dynamical relevant nodes]{The fraction of dynamically relevant nodes to network size $N$ at \textbf{a-)} in-NK, \textbf{b-)} in-PL and in-EXP networks. For in-NK, the values were found by averaging over $1000, 800, 500, 300, 200$ networks of $K = 1, 2, 3, 4, 5$, respectively. For in-PL and in-EXP, the values were found over $100, 500$ networks, respectively.} 
 \label{DynamicalSubnetwork_NKPLEXP} 
\end{figure}

Figure~\ref{DynamicalSubnetwork_NKPLEXP} presents a computational study which investigate the percentage size of the dynamically relevant nodes in the in-NK, in-PL and in-EXP networks for different $\langle k_{in} \rangle$ and $N$. Main conclusion of this study is that the percentage of relevant nodes depends only on the $\langle k_{in} \rangle$, not on the network topology.

\subsection{Dynamical Investigations on Some Model Networks}

In order to investigate the dynamics of different model networks and the effect of their topologies on the dynamics, in-NK, in-PL and in-EXP network ensembles with $\langle k_{in} \rangle \cong 2.0$ were studied (Figure~\ref{PLandEXP_gammaVSavK} shows that in-PL exponent $\alpha = 2.25$ and in-EXP exponent $\lambda = 0.7$ give $\langle k_{in} \rangle \cong 2.00$). The reason of choosing $\langle k_{in} \rangle \cong 2.0$ was its extensively use in related literature.

\subsubsection{a- Attractors of in-NK, in-PL and in-EXP Model Networks}

For distribution of the attractor features, $N$ and $p$ are fixed to $100$ and $0.5$, respectively. The reason behind using this p value was to compare the results to literature. However, since p was fixed, the SNCF whose p is not a free parameter was not used in this part. For each model type, I used $200$ networks with $10$ realization for each network in computations. Attractors were obtained after sampling $1000$ initial conditions with the limits of maximum step and maximum length of an attractor as $1000$ and $200$, respectively.

\begin{figure}[!p] 
 \centering 
  \includegraphics[width=\textwidth]{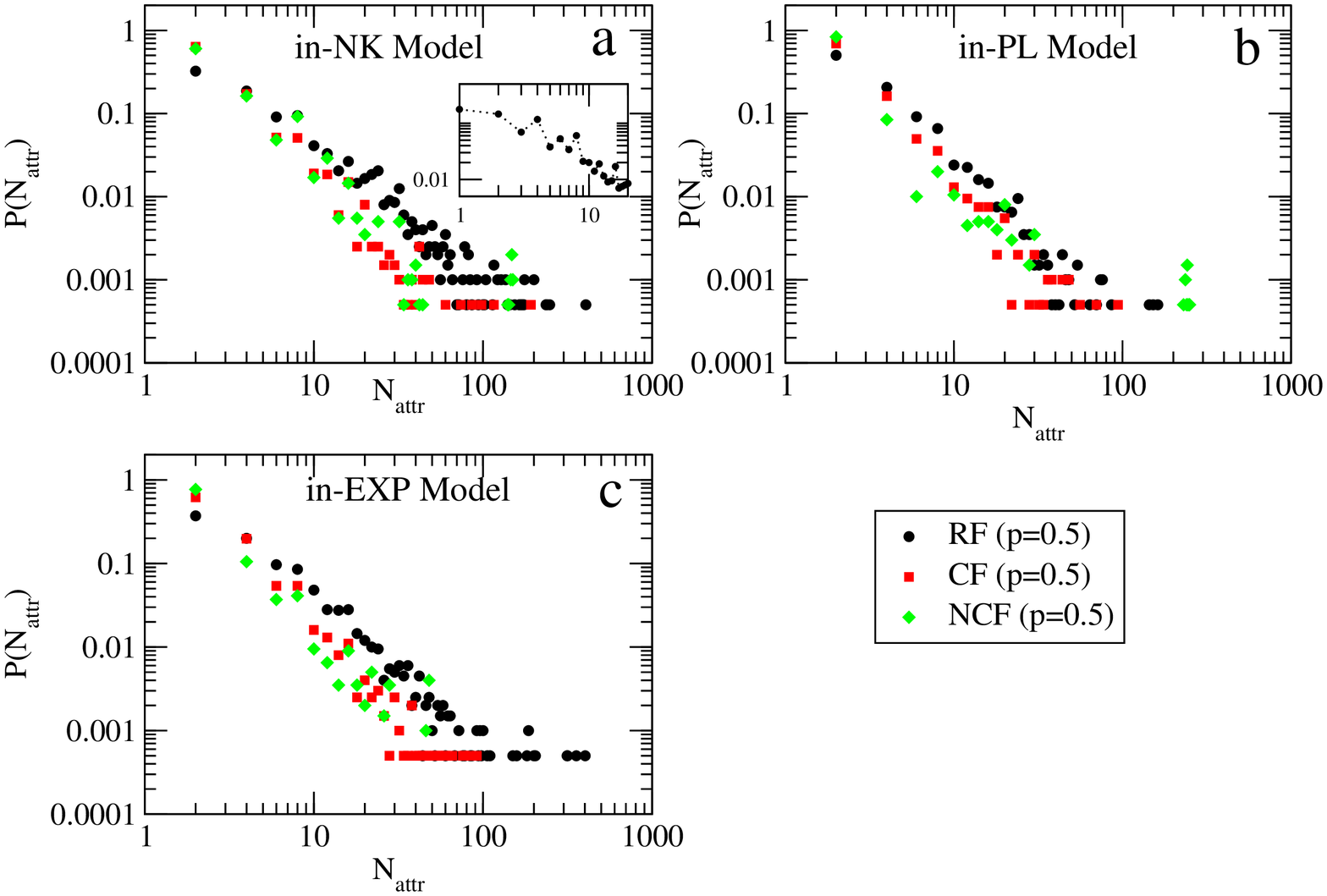}
 \caption[The Number of Attractor Distribution]{The \textbf{number of attractors} prob. distribution for model networks. \textbf{RF}: Random Function, \textbf{CF}: Canalazing Function, \textbf{NCF}: Nested Canalazing Function. For $200$ networks with $N=100$ and $10$ realizations for each network, the attractors were found by initiating from $1000$ initials conditions with limits of $1000$ maximum step size and $200$ attractor length. In order to uncover the artificial odd-even effect as shown in small frame, the data were binned in 2 units.}
 \label{NumberOfAttractors_Model} 
\end{figure}

\begin{figure}[!p] 
 \centering 
  \includegraphics[width=\textwidth]{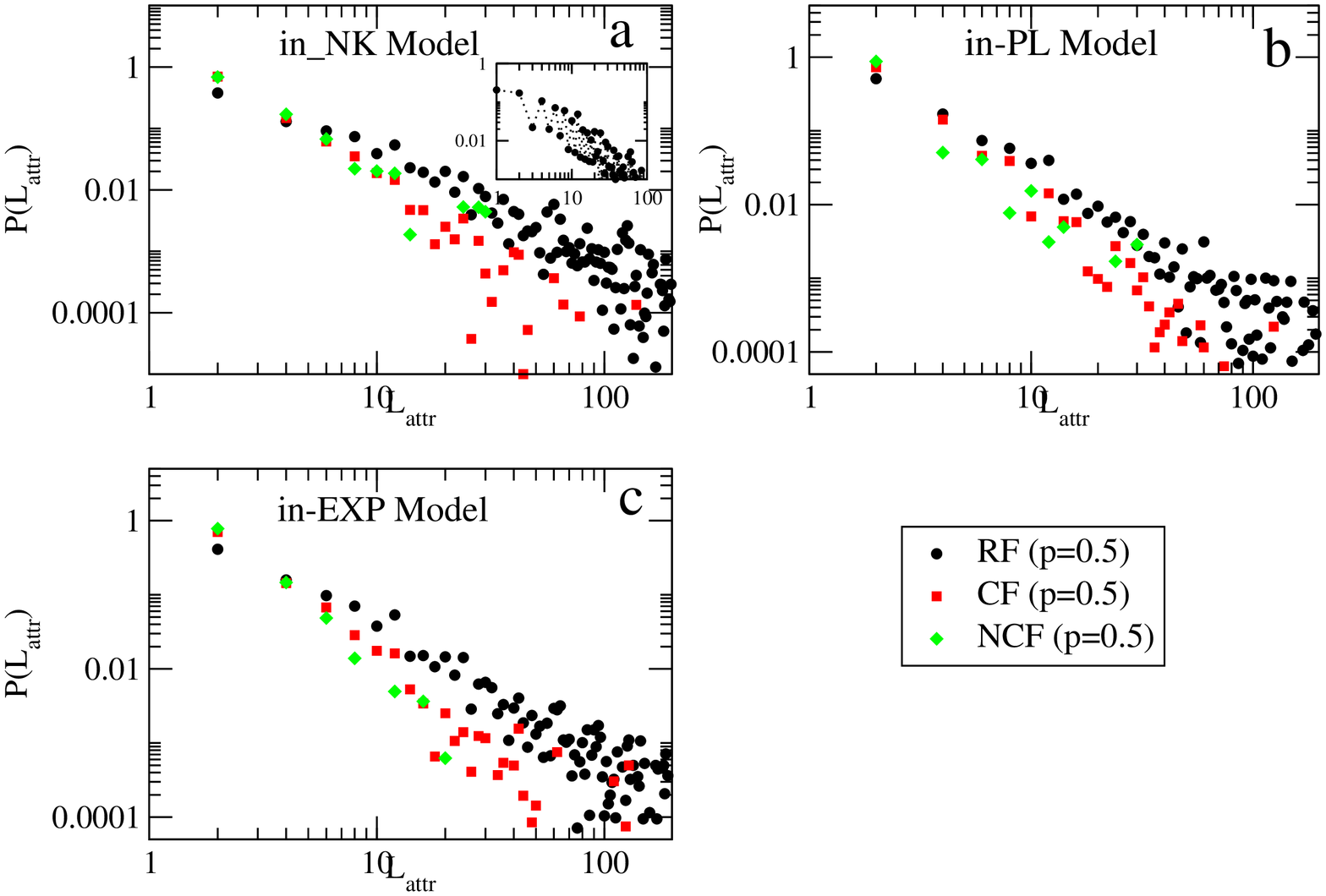}
 \caption[The Length of Attractor $\langle L_{attr} \rangle$ Distribution]{The \textbf{length of attractors} $\langle L_{attr} \rangle$ prob. distribution of model networks. \textbf{RF}: Random Function, \textbf{CF}: Canalyzing Function, \textbf{NCF}: Nested Canalyzing Function. For $200$ networks with $N=100$ and $10$ realizations for each network, the attractors were found by initiating from $1000$ initials conditions with limits of $1000$ maximum step size and $200$ attractor length. In order to uncover the artificial odd-even effect as shown in small frame, the data were binned in 2 units.}
 \label{LengthOfAttractor_Model} 
\end{figure}

\begin{figure}[!p] 
 \centering 
  \includegraphics[width=\textwidth]{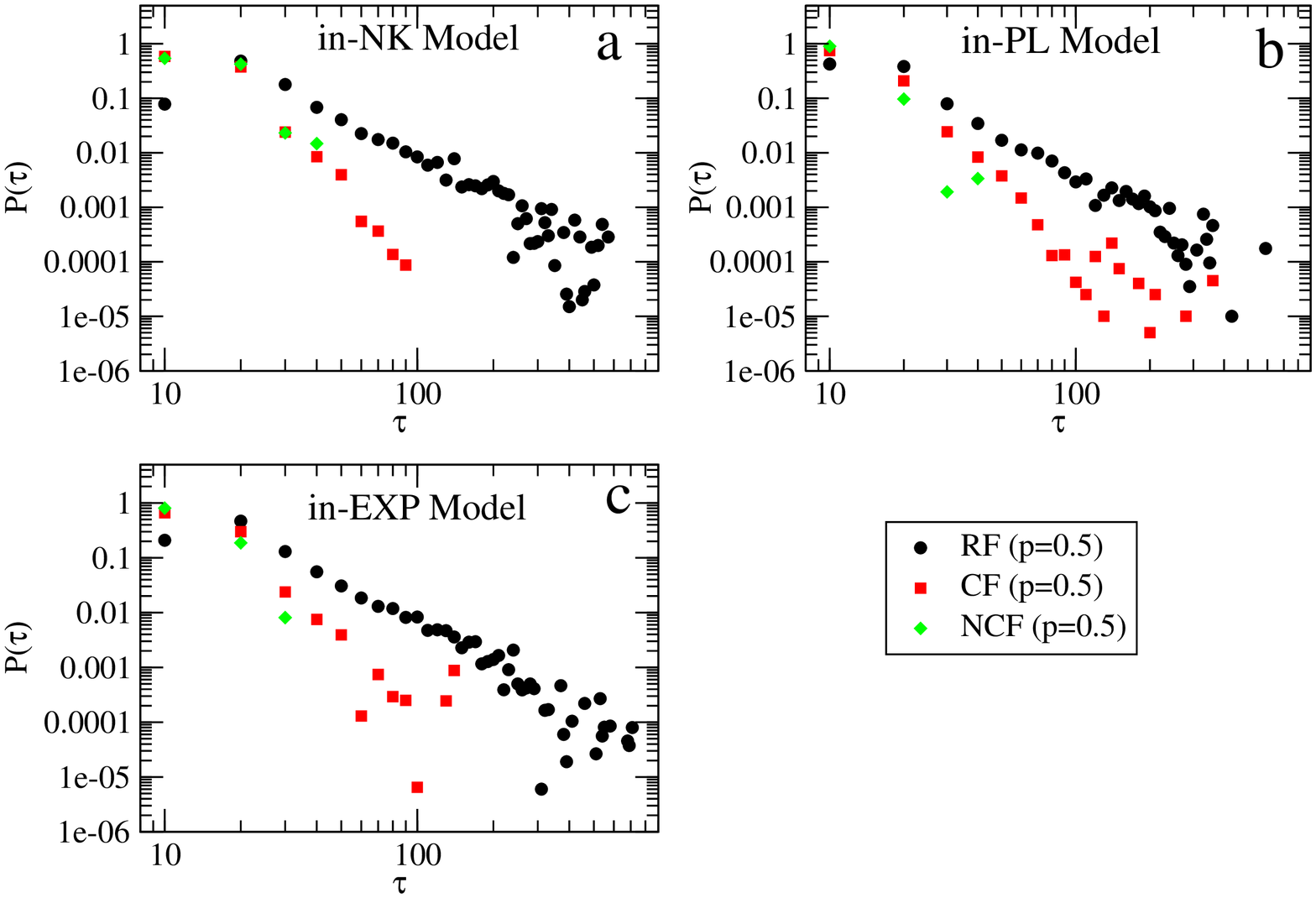}
 \caption[Transient to Attractor $\tau_{attr}$ Distribution]{\textbf{Transient to Attractors} distribution of model networks. \textbf{RF}: Random Function, \textbf{CF}: Canalyzing Function, \textbf{NCF}: Nested Canalyzing Function. For $200$ networks with $N=100$ and $10$ realizations for each network, the attractors were found by initiating from $1000$ initials conditions with limits of $1000$ maximum step size and $200$ attractor length. The data were binned in 10 units in order to have clearer distribution. }
 \label{TransOfAttractor_Model} 
\end{figure}

\begin{figure}[!p] 
 \centering 
  \includegraphics[width=\textwidth]{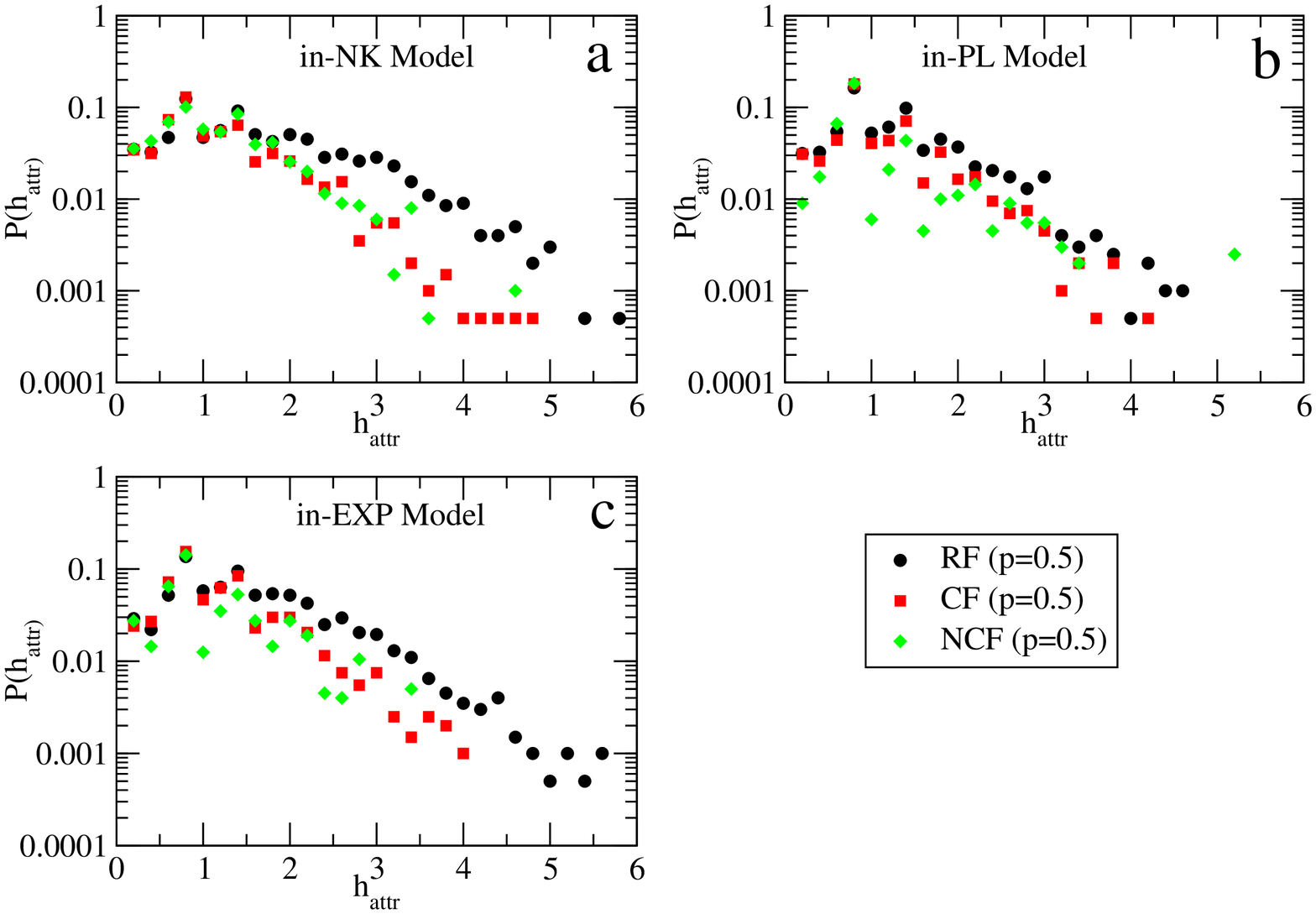}
 \caption[Entropy of Basin of Attraction $h_{attr}$ Distribution]{Entropy $h_{attr}$ distribution  of model networks. \textbf{RF}: Random Function, \textbf{CF}: Canalazing Function, \textbf{NCF}: Nested Canalazing Function. For $200$ networks with $N=100$ and $10$ realizations for each network, the attractors were found by initiating from $1000$ initials conditions with limits of $1000$ maximum step size and $200$ attractor length. The data were binned in 0.2 units in order to have clearer distribution.}
 \label{EntropyOfAttractor_Model} 
\end{figure}

The distribution of the number of attractors $N_{attr}$, the length of attractor $L_{attr}$, transient $\tau_{attr}$ and the entropy $h_{attr}$ are shown in Figure~\ref{NumberOfAttractors_Model}, Figure~\ref{LengthOfAttractor_Model}, Figure~\ref{TransOfAttractor_Model} and Figure~\ref{EntropyOfAttractor_Model}, respectively. Apart from the distributions, averages of these features are given in Table~\ref{Average_NumberOfAttractors_Model}, Table~\ref{Average_LengthOfAttractor_Model}, Table~\ref{Average_TransOfAttractor_Model} and Table~\ref{Average_EntropyOfAttractor_Model}, respectively.

I found out that the probability distribution functions for $N_{attr}$, $L_{attr}$, $\tau_{attr}$ and $h_{attr}$ in a network realization with in-NK, in-PL and in-EXP networks and RF, CF and NCF decay as a power-law function as stated for some topology and function types in References~\cite{Bhattacharjya_Liang,Paul_etal_AttractorsofCF}. Also, is was noted that both $N_{attr}$ and $L_{attr}$ shows a strange odd-even oscillations in the distributions which was also stated in Reference~\cite{Paul_etal_AttractorsofCF}. After some discussions, it was considered that  these odd-evenness due to from artificial effects, for instance, the combinations of 2-, 3-, etc. node \textit{partial} network states tends to create evenness. Furthermore, it should be noted that RF gives out considerably greater average values of those features than CF's and NCF's. While the average values were closer to each other with CF and NCF for all types of topologies, the averages with RF are higher than other for in-NK topology.

\begin{table}[!p]
\begin{center}
\begin{tabular}{|c||c|c|c|}
\hline 
$\langle N_{attr} \rangle$ &  in-NK  & in-PL  & in-EXP \\ 
\hline 
\hline 
RF & $12.00 \mp 24.06$ & $5.28 \mp 9.49$ & $9.20 \mp 21.95$ \\
\hline 
CF & $3.97 \mp 7.92 $ & $3.11 \mp 5.06$ & $3.81 \mp 6.49$ \\
\hline 
NCF & $4.59  \mp 11.18$ &$3.54 \mp 17.17$ & $2.86 \mp 4.69$ \\
\hline
\end{tabular}
\end{center}
\caption[Average number of attractors $\langle N_{attr} \rangle$ of various topologies and functions.]{The average number of attractors $\langle N_{attr} \rangle$ of model networks for random \textbf{RF}, canalyzing \textbf{CF}, nested canalyzing \textbf{NCF} functions. For $200$ networks with $N=100$ and $10$ realisations for each network, the attractors were found by initiating from $1000$ initials conditions with limits of $1000$ maximum step size and $200$ attractor length.}
\label{Average_NumberOfAttractors_Model} 
\end{table}

\begin{table}[!p]
\begin{center}
\begin{tabular}{|c||c|c|c|}
\hline 
$\langle L_{attr} \rangle$ &  in-NK  & in-PL  & in-EXP \\ 
\hline 
\hline 
RF & $11.69 \mp 22.36$ & $12.13 \mp 71.60$ & $20.15 \mp 101.82 $ \\
\hline 
CF & $3.05 \mp 4.39$ & $3.39 \mp 23.48$ & $3.77 \mp 26.58$ \\
\hline 
NCF & $2.98 \mp 3.83$ &$2.02 \mp 2.60$ & $2.12 \mp 1.92$ \\
\hline
\end{tabular}
\end{center}
\caption[Average length of attractors of model networks.]{The average length of attractors of model networks. \textbf{RF}: Random Function, \textbf{CF}: Canalazing Function, \textbf{NCF}: Nested Canalazing Function. For $200$ networks with $N=100$ and $10$ realizations for each network, the attractors were found by initiating from $1000$ initials conditions with limits of $1000$ maximum step size and $200$ attractor length.} 
\label{Average_LengthOfAttractor_Model}
\end{table}

\begin{table}[!p]
\begin{center}
\begin{tabular}{|c||c|c|c|}
\hline 
$\langle \tau_{attr} \rangle$ &  in-NK  & in-PL  & in-EXP \\ 
\hline 
\hline 
RF & $58.44 \mp 164.61$ & $23.52 \mp 74.84$ & $35.87 \mp 104.79$ \\
\hline 
CF & $10.55 \mp 5.33$ & $9.63 \mp 23.86$ & $10.64 \mp 26.70$ \\
\hline 
NCF & $10.64 \mp 4.68 $ &$7.29 \mp 3.02$ & $8.35 \mp 2.91$ \\
\hline
\end{tabular}
\end{center}
\caption[Average transient to attractors of model networks.]{Average transient to attractors of model networks. \textbf{RF}: Random Function, \textbf{CF}: Canalyzing Function, \textbf{NCF}: Nested Canalyzing Function. For $200$ networks with $N=100$ and $10$ realizations for each network, the attractors were found by initiating from $1000$ initials conditions with limits of $1000$ maximum step size and $200$ attractor length.}
\label{Average_TransOfAttractor_Model}
\end{table}

\begin{table}[!p]
\begin{center}
\begin{tabular}{|c||c|c|c|}
\hline 
$\langle h_{attr} \rangle$ &  in-NK  & in-PL  & in-EXP \\ 
\hline 
\hline 
RF & $1.30 \mp 1.10$ & $0.88 \mp 0.86$ & $1.17 \mp 1.00$ \\
\hline 
CF & $0.65 \mp 0.78 $ & $0.58 \mp 0.71$ & $0.68 \mp 0.75$ \\
\hline 
NCF & $0.71 \mp 0.81$ & $0.43 \mp 0.72 $ & $0.48 \mp 0.68$ \\
\hline
\end{tabular}
\end{center}
\caption[Average values of the entropy of attractors of model networks.]{Average values of the entropy of model networks. \textbf{RF}: Random Function, \textbf{CF}: Canalyzing Function, \textbf{NCF}: Nested Canalyzing Function. For $200$ networks with $N=100$ and $10$ realizations for each network, the attractors were found by initiating from $1000$ initials conditions with limits of $1000$ maximum step size and $200$ attractor length.} 
\label{Average_EntropyOfAttractor_Model}
\end{table}

\begin{figure}[!p] 
 \centering 
  \includegraphics[width=0.8\textwidth]{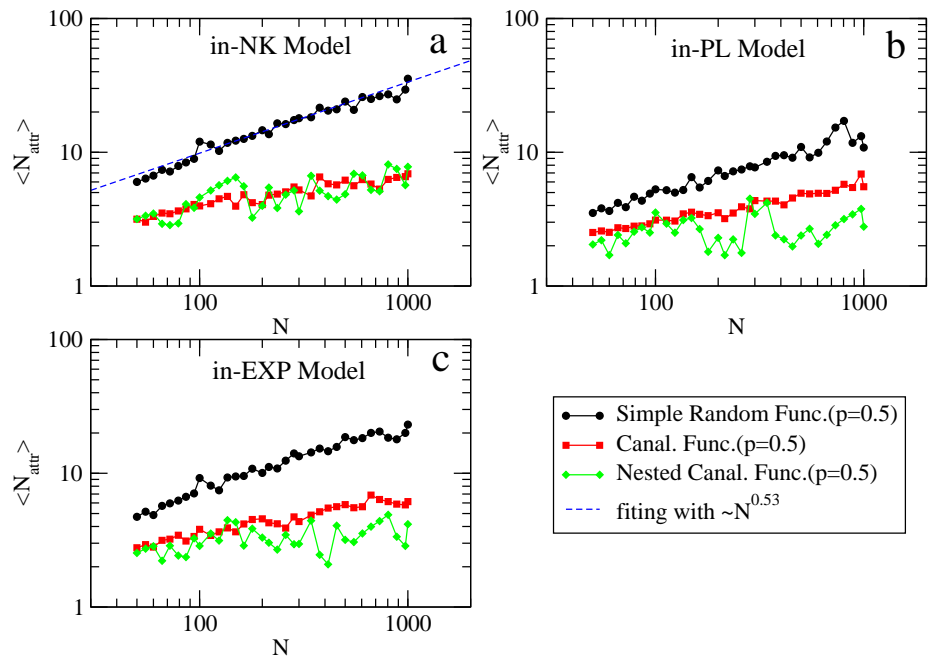}
 \caption[The scaling of the average number of attractors]{The scaling with $N$ of the average number of attractors for RF,CF,NCF with $p=0.5$.}
 \label{Scaling_NumberOfAttractors} 
\end{figure}

\begin{figure}[!p] 
 \centering 
  \includegraphics[width=0.8\textwidth]{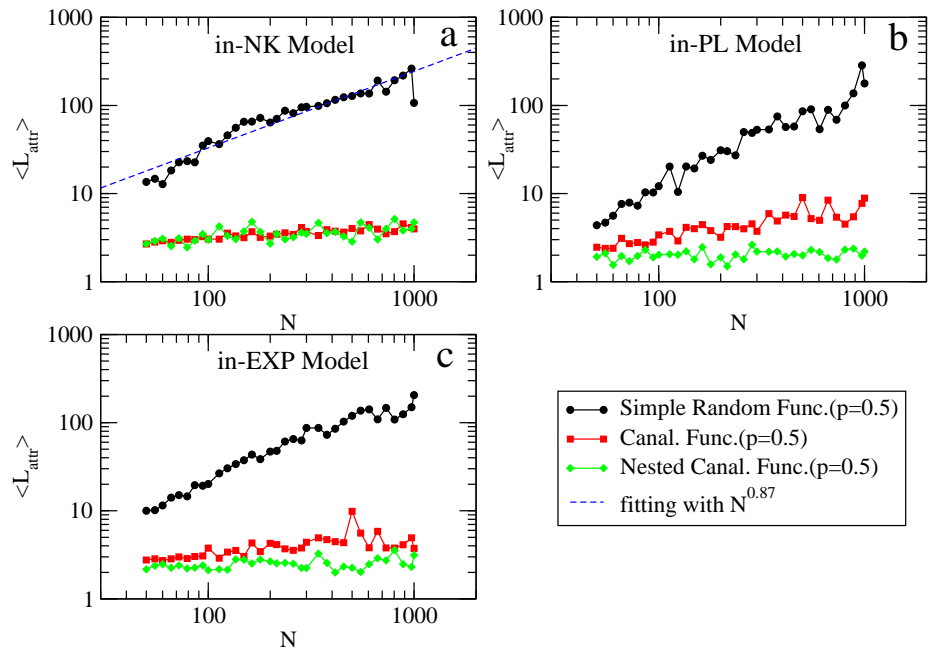}
 \caption[The scaling of the average length of attractors]{The scaling with $N$ of the average length of attractors for RF,CF,NCF with $p=0.5$.}
 \label{Scaling_LengthOfAttractors} 
\end{figure}

\begin{figure}[!p] 
 \centering 
  \includegraphics[width=0.8\textwidth]{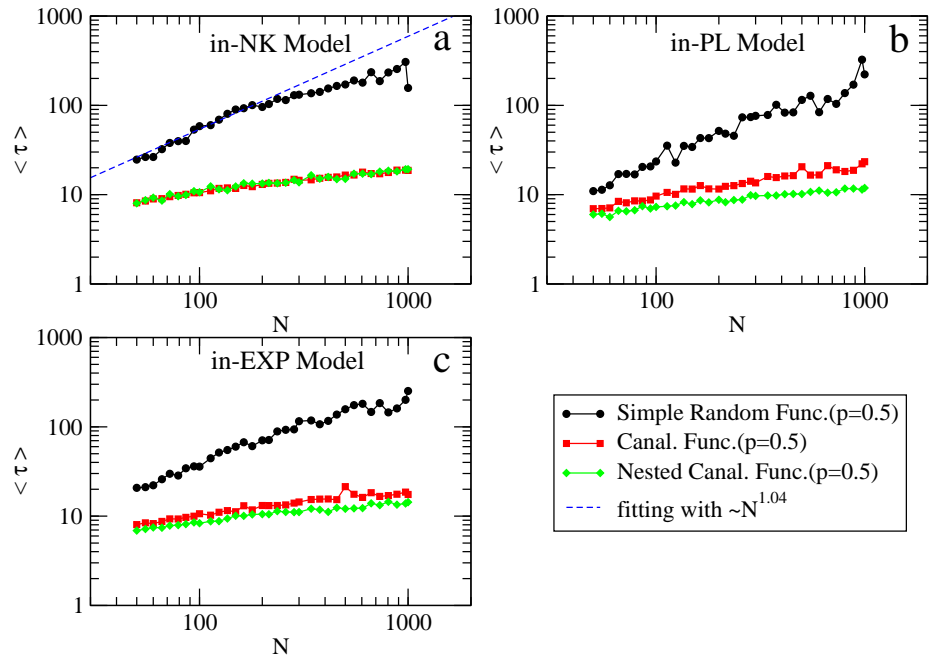}
 \caption[The scaling of transients to attractors]{The scaling with $N$ of transients to attractors for RF,CF,NCF with $p=0.5$.}
 \label{Scaling_TransToAttractors} 
\end{figure}

\begin{figure}[!p] 
 \centering 
  \includegraphics[width=0.8\textwidth]{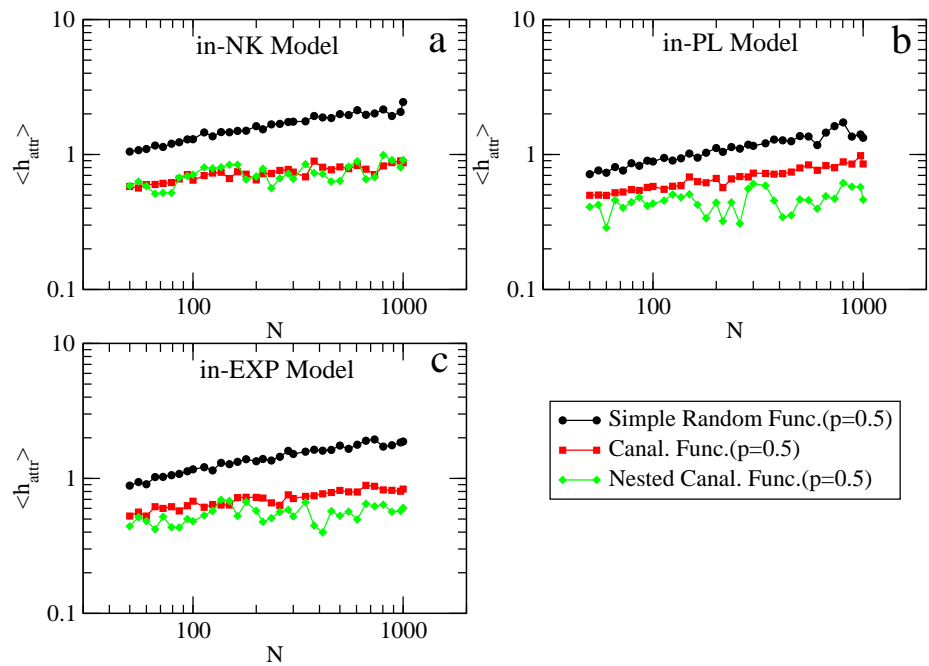}
 \caption[The scaling of the average entropy]{The scaling with $N$ of the average entropy for RF,CF,NCF with $p=0.5$.}
 \label{Scaling_Entropy} 
\end{figure}

I also checked the scaling of the average values of the quantities above with number of nodes $N$ for random, canalazing and nested canalazing functions with $p=0.5$. $N$ were chosen as 50, 55,  60, 66, 74, 82, 92, 100, 113, 124, 136, 149, 163, 179, 200, 215, 236, 259, 284, 300, 343, 377, 414, 455, 500, 550, 605, 665, 731, 804, 884, 972, 1000 in order to have a more accurate scaling behavior at small $N$s but also to check big $N$s. For $N \leq 100$, $200$ networks were used for all function types. For $N>100$, $100$ networks for CF and NCF, and $50$ networks for RF were used (RF runs slowly than others). The other parameters for dynamics were the same with $N=100$ case above. The results for $\langle N_{attr} \rangle$, $\langle L_{attr} \rangle$, $\langle \tau_{attr} \rangle$ and $\langle h_{attr} \rangle$ scalings with $N$ can be seen in Figure~\ref{Scaling_NumberOfAttractors}, Figure~\ref{Scaling_LengthOfAttractors}, Figure~\ref{Scaling_TransToAttractors} and Figure~\ref{Scaling_Entropy}, respectively.

I found out that for in-NK network with RF; $\langle N_{attr} \rangle$ scales with a fitting $N^{0.53}$, $\langle L_{attr} \rangle$ scales with a fitting $N^{0.87}$, $\langle \tau_{attr}\rangle$ scales with a fitting $N^{1.04}$. For a long time $\langle N_{attr} \rangle$ and $\langle L_{attr} \rangle$ scalings were considered as $\sqrt{N}$~\cite{Kauffman_Origins_of_Order} until Socolar \& Kauffman published Reference~\cite{Socolar_Kauffman} which states that $\langle N_{attr} \rangle$ scales with faster than linear. With this study I have shown that $\sqrt{N}$ scaling is valid for $N_{attr}$ while fails for $L_{attr}$. Also, it should be noted that for CF and NCF, scalings are very small comparing to RF which might be considered as desirable for the biological systems.

\clearpage
\subsubsection{b- Robustness of in-NK, in-PL and in-EXP Topologies}

Numerical studies have yielded that \textit{in-NK} model networks of $K=2$ with random functions have privileged dynamical aspects than others, i.e. small, less number of attractors and at the boundary between chaotic and order phase~\cite{Kauffman_Network}. Derrida \& Pomeau were the first to give an analytical argument for why there is a such critical K value \cite{Derrida_Pomeau}. Later, Aldana generalized the argument for other type of topologies and gave more pedagogic expression \cite{Aldana_BooleanNetPLtopology2003}. 

Consider $x(t)$ which was discussed in \textit{robustness} expression. One can define two sets $A(t)$ and $B(t)$ such that $A(t)$ is the set of nodes whose incoming edges come only from the nodes whose states are the same in the $S(t)$ and $S^{'}(t)$ and $B(t)$ is vice versa.

\begin{equation}
\label{TwoNodeSets}
\begin{array}{cccc}
G=\left\lbrace v_1,v_2,..v_N\right\rbrace \\ 
A(t)=\left\lbrace v_i, v_j: (v_j\rightarrow v_i) \bigwedge (\sigma_j(t)=\sigma_j'(t)) \right\rbrace \\
B(t)= G - A(t)\\
\end{array}
\end{equation}

Then we can express,
\begin{align}
M(x(t)) &=\sum_{k_{in}=1}^\infty P(k_{in})\lbrace \underbrace{\underbrace{[x(t)]^{k_{in}}}_{prob.\;of\;A(t)}}_{Contribution\; from\; A(t)}\;+\;   \underbrace{\underbrace{(1-[x(t)]^{k_{in}})}_{prob.\;of\;B(t)} \times \underbrace{(p^2+(1-p)^2)}_{prob.\;the\;same\;output}}_{Contribution\;from\; B(t)} \rbrace ,\\
	&= \sum_{k_i=1}^\infty P(k_{in}) \lbrace -[x(t)]^{k_{in}}(2p^2-2p) + (2p^2-2p+1)\rbrace.
\end{align}
If we take the derivative of the both sides with respect to $x(t)$ and use the fact that $\sum_{k_{in}=1}^\infty P(k_{in}) = 1$,
\begin{align}
\frac{dM(x(t))}{dx(t)} &= (2p^2-2p)\sum_{k_i=1}^\infty k_{in}[x(t)]^{k_{in}-1} P(k_{in}).
\end{align}
Remembering $\sum_{k_{in}=1}^\infty k_{in} P(k_{in}) = \langle k_{in} \rangle$ and Eq.~\ref{sensivity_robustness}, one finds;

\begin{align}
\lim_{x(t) \to 1^-}{\frac{dM(x(t))}{dx(t)}} &= \lim_{x(t) \to 1^-}{{\langle k_{in} \rangle}x(t)^{{\langle k_{in} \rangle}-1}(2p^2-2p)}\\
\label{DerridaFormula}&=2p(p-1){\langle k_{in} \rangle}.
\end{align}

\begin{figure}[!htb] 
 \centering
  \includegraphics[angle=-90,width=0.7\textwidth]{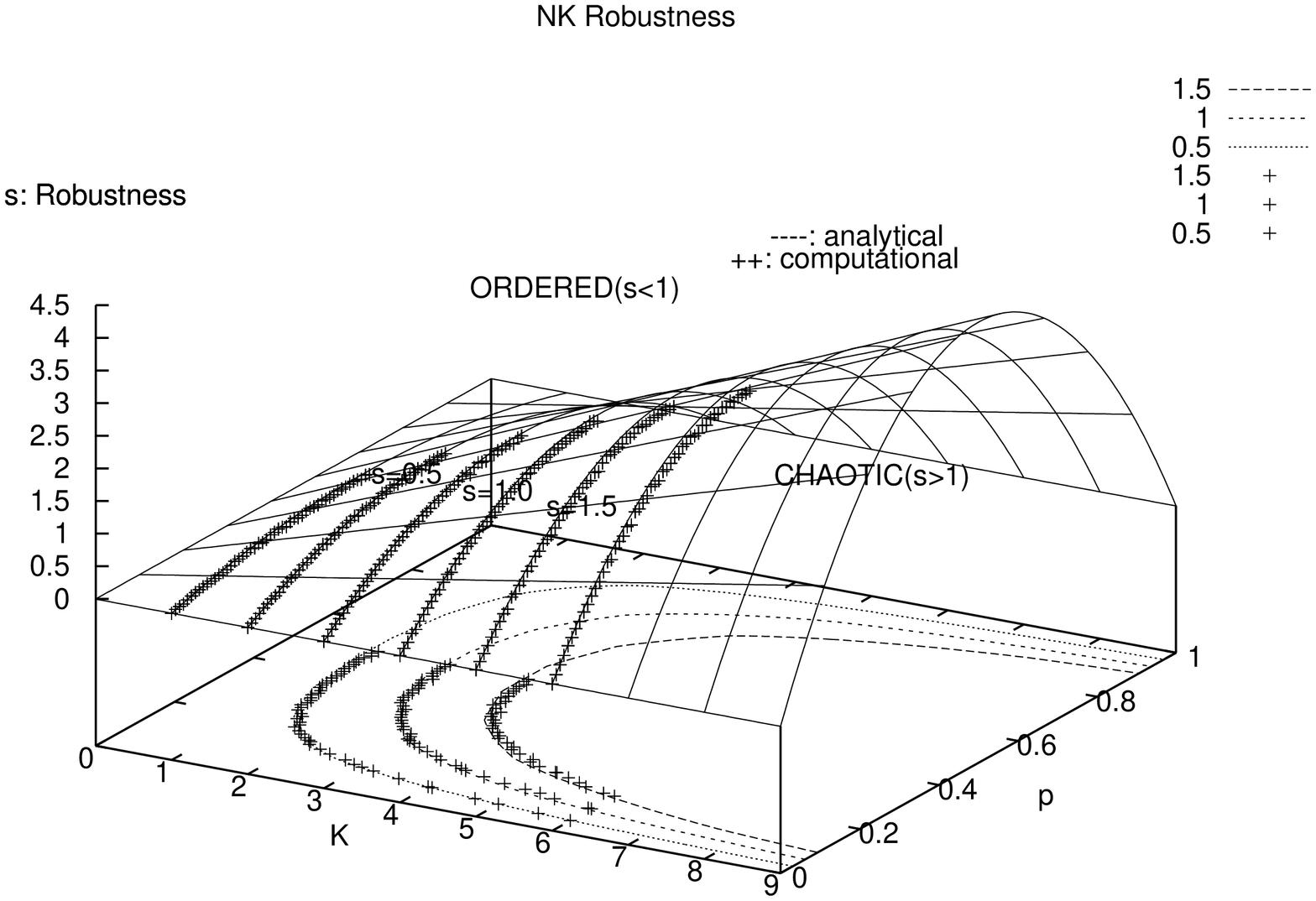}
  \hfill 
  \includegraphics[width=0.9\textwidth]{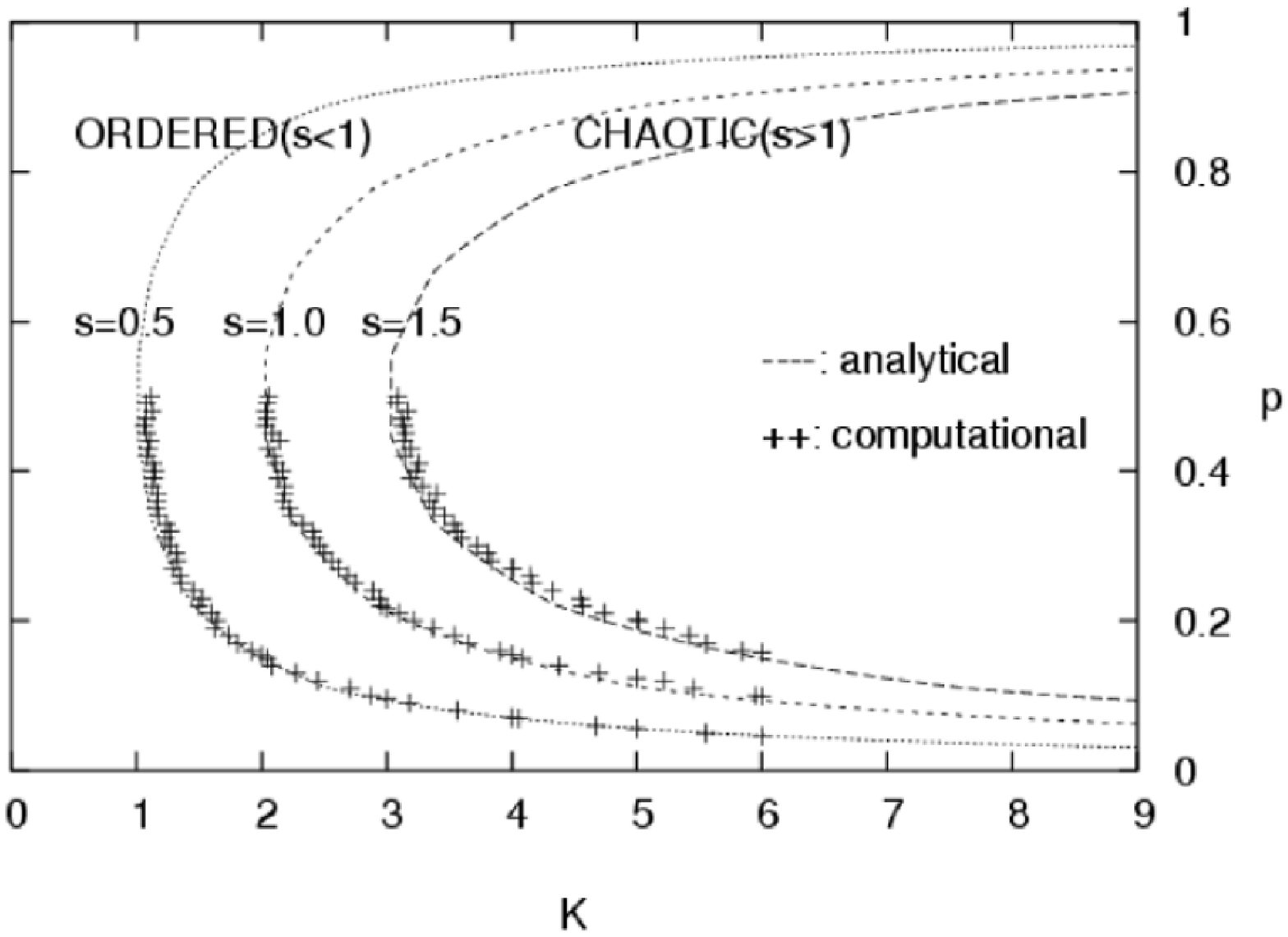}
 \caption[Comparison of analytical and computational in-NK networks robustness]{Analytical($---$) and computational($+++$) the robustness values were compared for in-NK model with $N=100$ for each $K \in \lbrace 1, 2,.., 6\rbrace$ and $p \in \lbrace 0.0, 0.01,..,0.49, 0.50\rbrace$ for simple random functions. For each $K,p$, $10$ networks and for each network $10$ realization were constructed and the robustness was calculated starting from $100$ initials conditions of each realization. Here, s=1 corresponds the critical border for transition from ordered to chaotic regimes. It seems that Derrida's results matches the s=1 border.} 
 \label{NKRobustness_AnalyticVSComputational} 
\end{figure}

\begin{figure}[!htb] 
 \centering
  \includegraphics[angle=-90,width=0.7\textwidth]{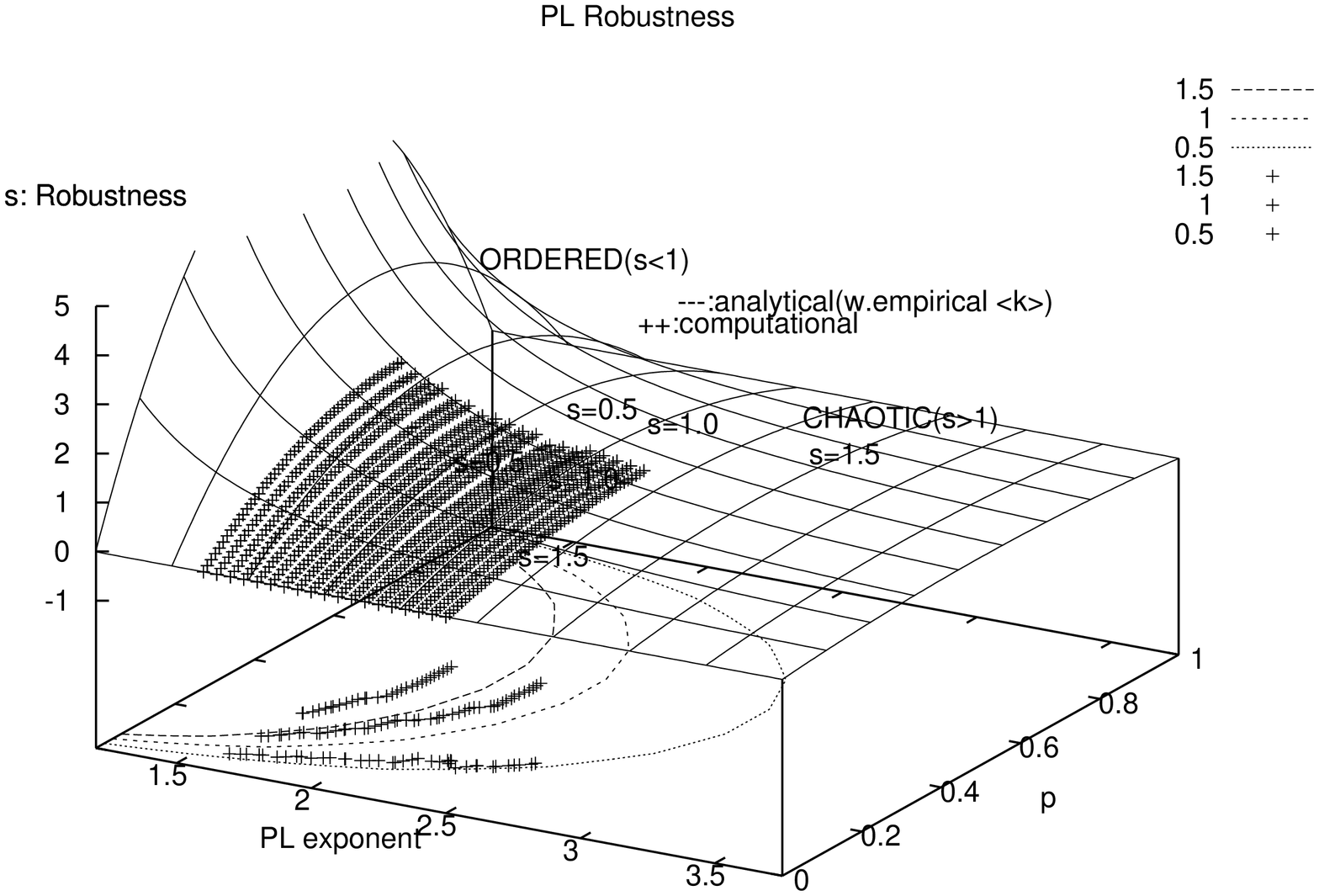}
  \hfill
  \includegraphics[width=0.9\textwidth]{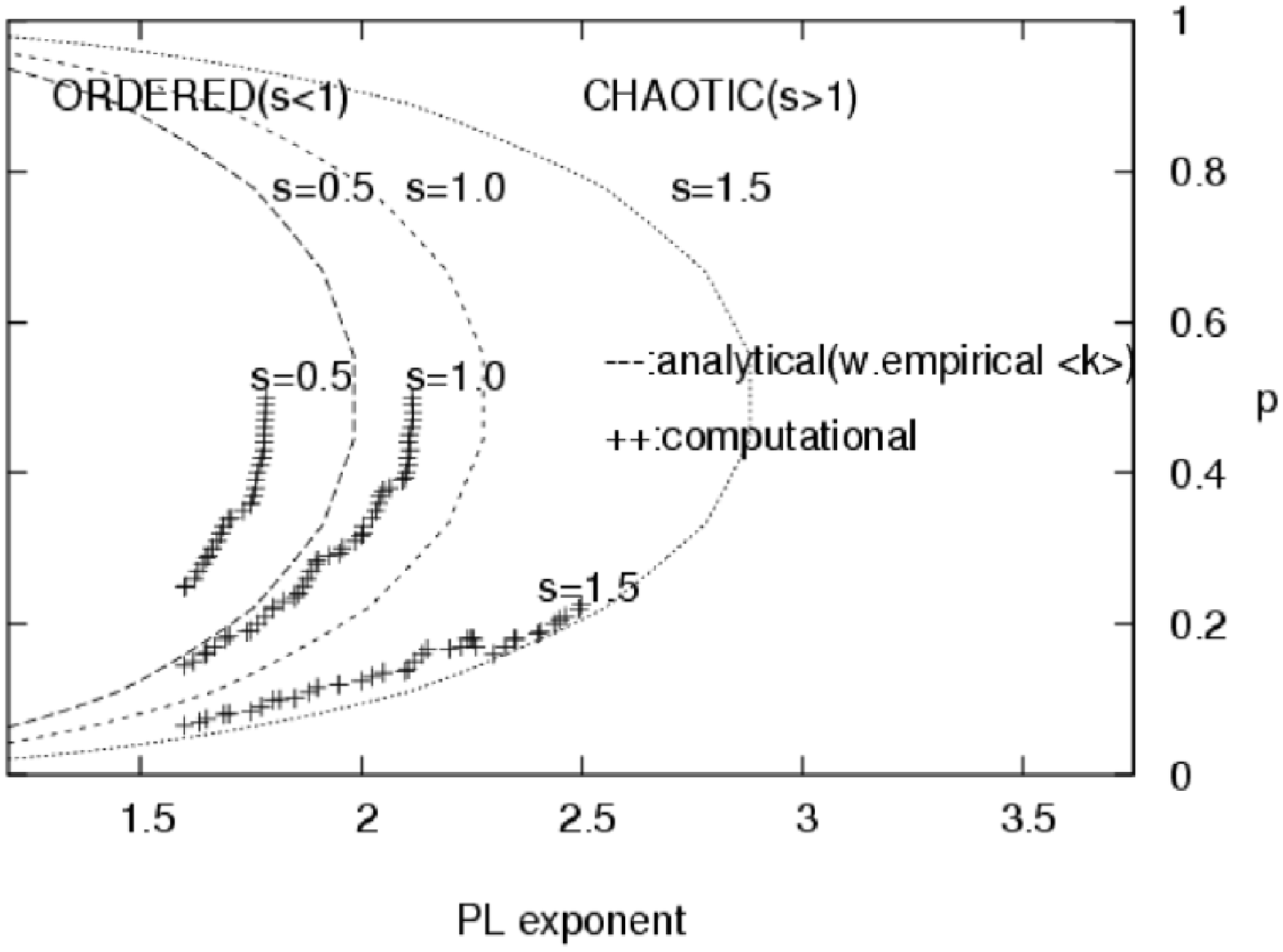}  
\caption[Comparison of analytical and computational in-PL networks robustness]{As in previous figure, the computational and analytical robustness was compared for in-PL networks with $N=100, \alpha=1.7, 1.75, .. ,2.5$ and $p \in \lbrace 0.0, 0.01,..,0.49, 0.50\rbrace$ for simple random functions. For each $\alpha,p$, $10$ networks and for each network $10$ realization were constructed and the robustness was calculated starting from $100$ initials conditions of each realization. The computational and analytical results are close to each other, it was seen with bigger $N$ values, the matching got closer which concludes a finite-size effect.} 
 \label{PLRobustness_AnalyticVSComputational} 
\end{figure}

\begin{figure}[!htb] 
 \centering
  \includegraphics[angle=-90,width=0.7\textwidth]{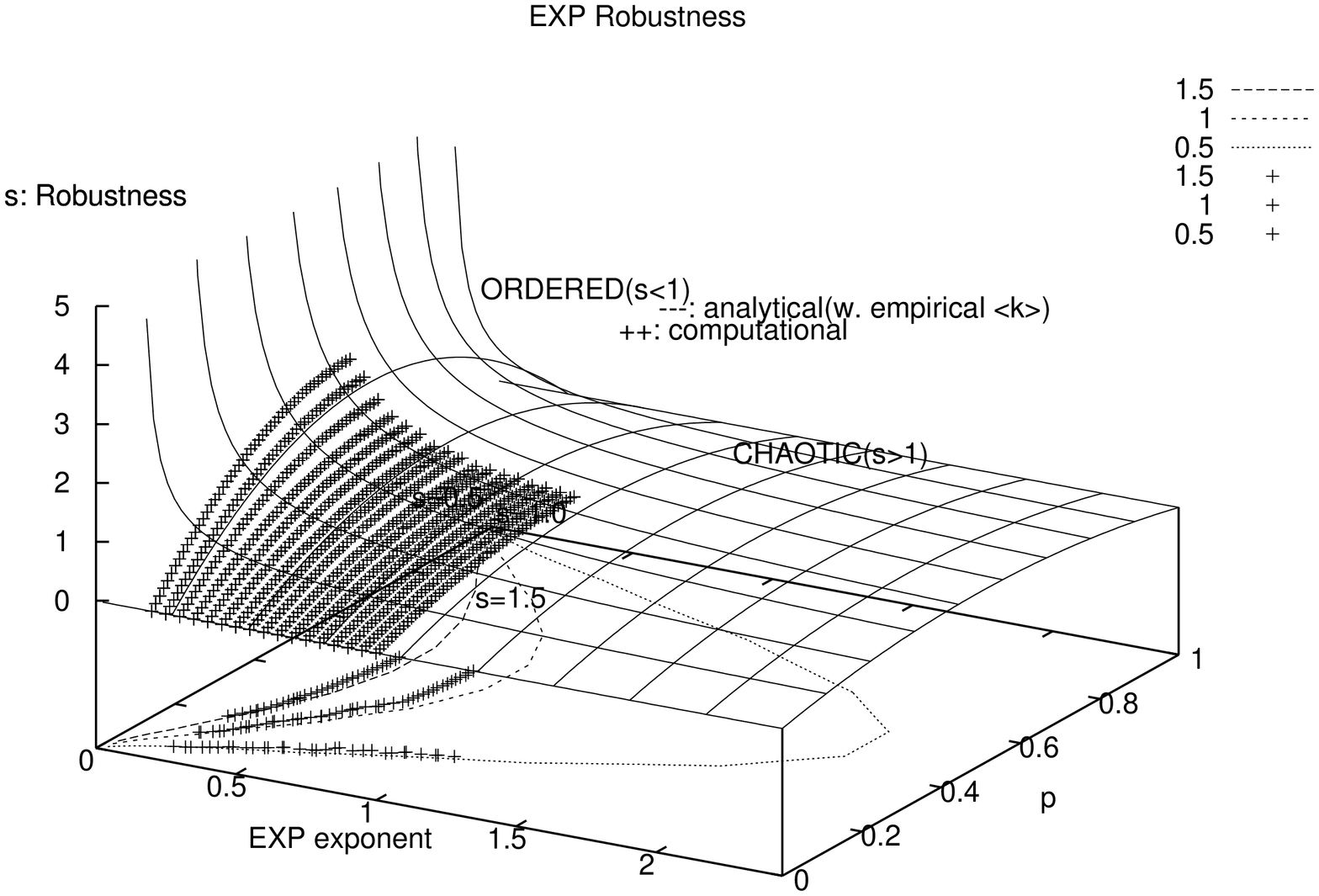}
  \hfill
  \includegraphics[width=0.9\textwidth]{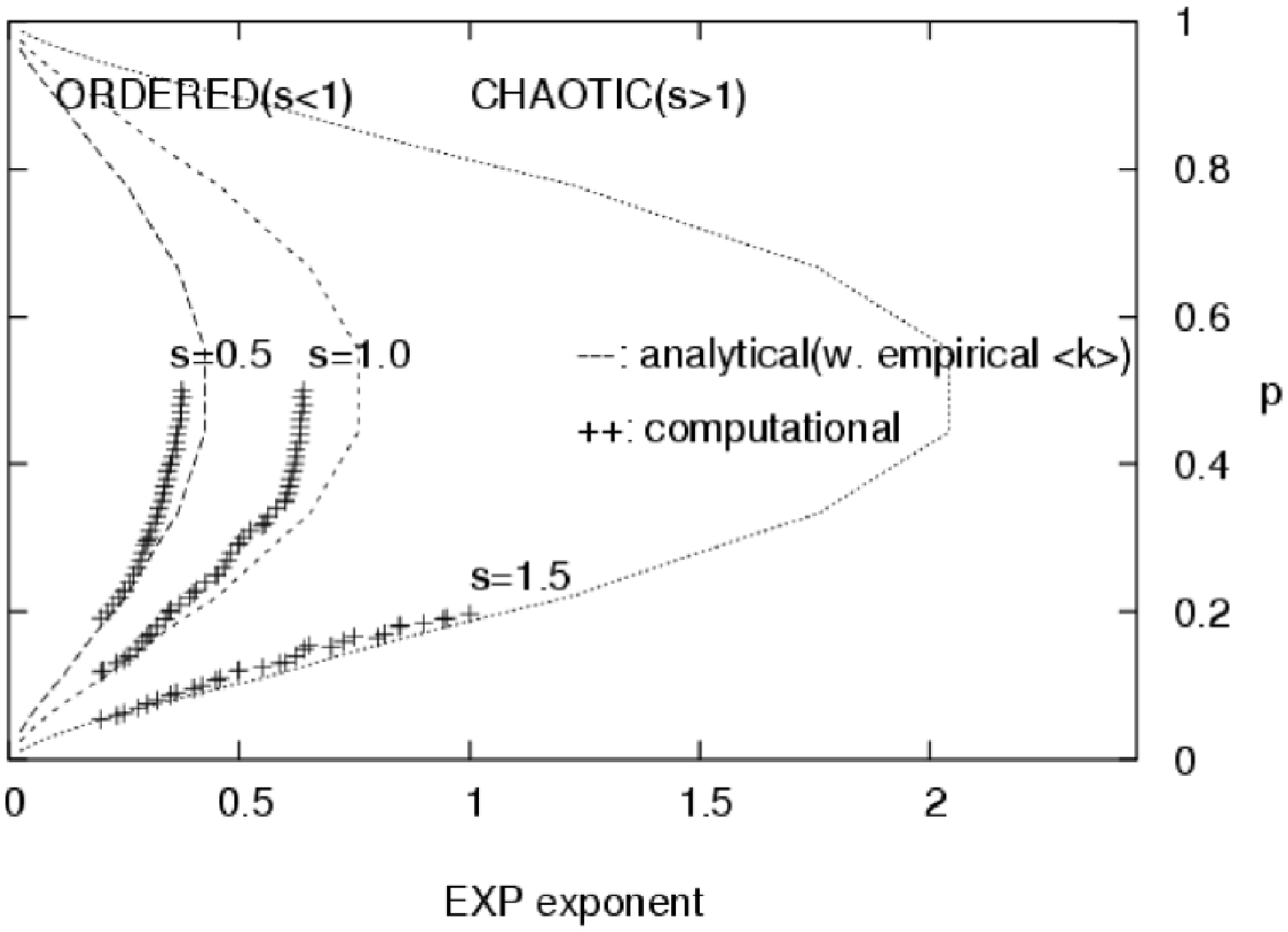}  
\caption[Comparison of analytical and computational in-EXP networks robustness]{As in previous figure, the computational and analytical robustness was compared for in-EXP networks with $N=100, \lambda=0.3,0.35,..,1.0$ and $p \in \lbrace 0.0, 0.01,..,0.49, 0.50\rbrace$ for simple random functions. For each $\lambda,p$, $10$ networks and for each network $10$ realization were constructed and the robustness was calculated starting from $100$ initials conditions of each realization. As in previous case, it was seen with bigger $N$ values, the matching got closer which concludes a finite-size effect.}  
 \label{EXPRobustness_AnalyticVSComputational} 
\end{figure}

I tried to examine the validity of Formula~\ref{DerridaFormula} for \textit{in-NK}, \textit{in-PL} and \textit{in-EXP} networks with the \textit{simple random function} (RF) by comparing the analytical and computational results. Especially, the critical chaotic-ordered boundary, $s=1$, was checked. The robustness computations were done for different sets of $p \in \lbrace 0.0, 0.01, 0.02,...,0.49,0.50\rbrace$ and corresponding parameters for ${\langle k_{in} \rangle}$: for in-NK, $K \in \lbrace 1,2,...,5,6 \rbrace$; for in-PL, $\alpha\in \lbrace 1.60,1.65,1.70,...,2.45,2.5 \rbrace$ and for in-EXP, $\lambda \in \lbrace 0.30,0.35,...,0.95,1.0 \rbrace$. For the analytical expression (Eq.\ref{DerridaFormula}), the relations between $\langle k_{in} \rangle$ with $\alpha,\lambda$ were yielded by consulting the Figure~\ref{PLandEXP_gammaVSavK}. The results can be seen in Figure~\ref{NKRobustness_AnalyticVSComputational}, in Figure~\ref{PLRobustness_AnalyticVSComputational} and in Figure~\ref{EXPRobustness_AnalyticVSComputational} for in-NK, in-PL and in-EXP networks, respectively.

I found out that robustness values, especially the critical boundary, match with the analytical values (Formula~\ref{DerridaFormula}) for in-NK Model. The analytical expression also predicted the robustness values of in-PL and in-EXP networks but not as in-NK networks' case. I saw that matching for these networks gets better while N is increased which concluded as finite-size effect. In short, Expression~\ref{DerridaFormula} is successful for predicting the robustness of the networks for simple random functions.

\begin{figure}[!htb] 
 \centering 
  \includegraphics[width=\textwidth]{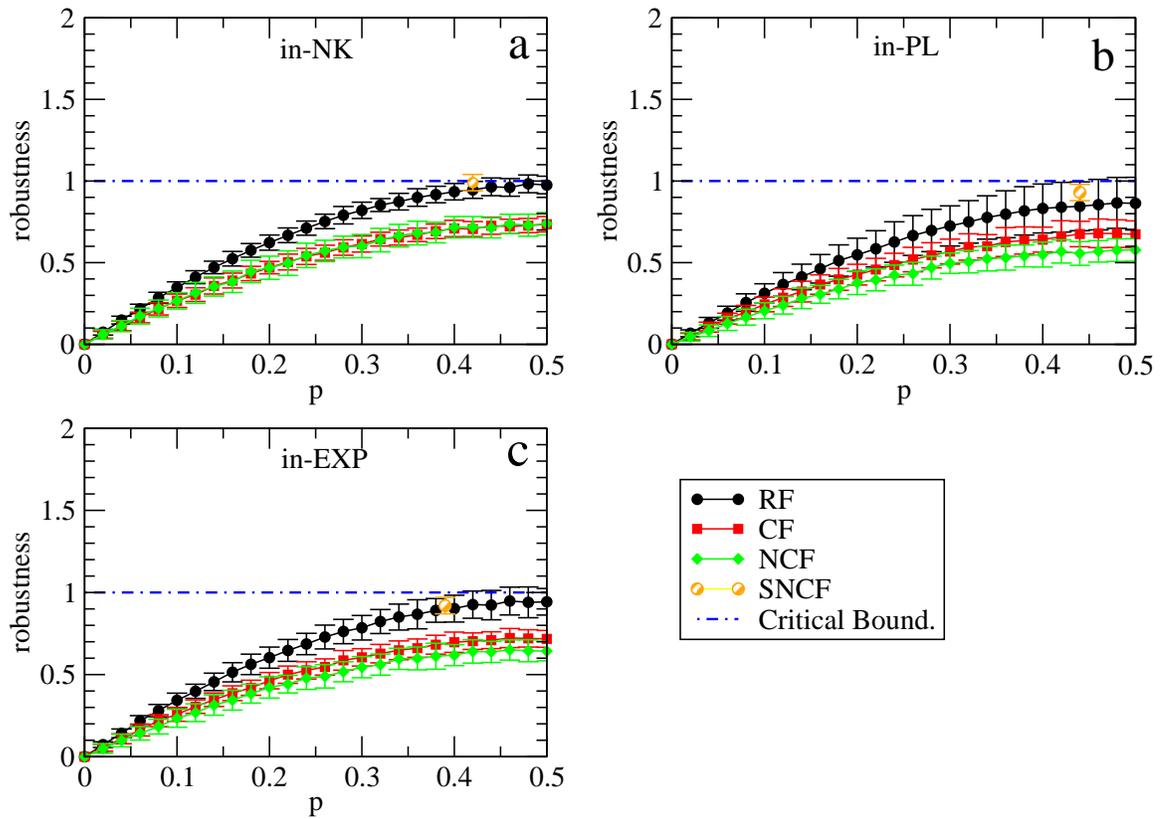}
 \caption[Robustness of model networks for 4 type functions]{Figure shows the robustness values of the \textbf{a-)}in-NK, \textbf{b-)}in-PL and \textbf{c-)}in-EXP networks of $\langle k_{in} \rangle \cong 2.0$ and $N=100$, $p \in \lbrace 0.0, 0.02,.., 0.50\rbrace$ for RF: Simple Random, CF: Canalazing, NCF: Nested Canalazing and SNCF: Special Subclasses Nested Canalazing Functions. SNCF robustness results and corresponding $p$ values for in-NK, in-PL and in-EXP are $0.99 \mp 0.05$, $0.93 \mp 0.05$, $0.92 \mp 0.05$ and $0.42$, $0.44$, $0.39$, respectively.}
 \label{Robustness_Model} 
\end{figure}

I also checked the variation of the robustness for all types of functions (RF, CF, NCF, SNCF). Again, each network topology was set to $\langle k_{in} \cong 2.0 \rangle$ and $N=100$. Each \textit{robustness} value for corresponding $\langle k_{in} \rangle$ parameter and $p$ was yielded by using the following parameters: $100$ random initials conditions for each $10$ network realizations for each $10$ networks. As it can be seen in Figure~\ref{Robustness_Model}, canalazing and nested canalazing functions resulted in more ordered robustness values than simple random and special subclasses of nested canalazing functions for all types of networks. 

\chapter{Gene Regulation}\label{GeneRegulation}

This chapter is organized as follows: Section~\ref{GRIntroduction} introduces the gene regulation concept in biology. Section~\ref{YeastGRN} gives the yeast gene regulation network (GRN) with the topological and dynamical investigations. Section~\ref{inEXPYeast} compares the yeast GRN with model networks whose indegree probability distribution is exponential. Section~\ref{BalcanModel} emphasizes a recently proposed model which produce networks that are topologically similar to yeast GRN and discusses this model dynamically.

\begin{figure}[!hbt] 
 \centering 
  \includegraphics[width=0.5\textwidth]{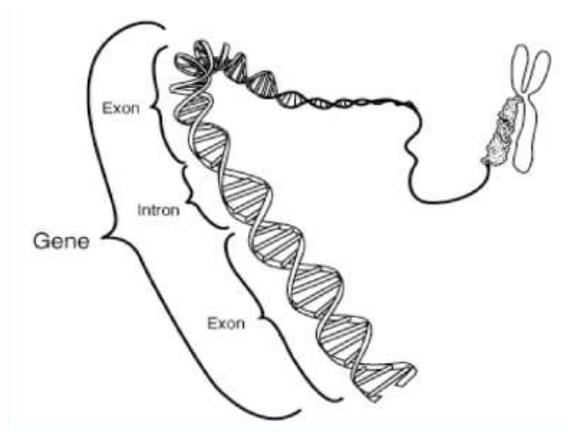}
 \caption[A gene on a chromosome]{The DNA is considered to fulfill three main function in the life systems: 1-) \textit{storage}, 2-) \textit{heritage} and 3-) \textbf{expression} of the genetic information in the cells~\cite{Cell_Bio_KARP}} 
 \label{DNAtoGENE} 
\end{figure}

\section{Introduction}\label{GRIntroduction}

%
To my knowledge, the most recent scientific definition of "gene'' (Figure~\ref{DNAtoGENE}\footnote{Taken from http://en.wikipedia.org/wiki/Gene}) was proposed by Gerstein \textit{et al.}: "A \textbf{gene} is a union of genomic sequences encoding a coherent set of potentially overlapping functional products."~\cite{Gerstein_WhatIsGene}. Genes in prokaryotes are always active and express their coded information into functional elements unless they are repressed by outside factors. However, at a particular time genes of eukaryotes (in Reference \cite{Miglani_GENETICS} it is stated as 2-15\%) are generally inactive and need to be activated, therefore, one can talk about \textit{regulation} of gene expression in eukaryotes~\cite{Miglani_GENETICS}.

There are different types of regulation of gene expression data depending of how it is detected. In this study, I used the \textbf{transcriptional} GR since the data used at this thesis was supplied by monitoring the levels of transcription materials, i.e. mRNA. For the details of experiments to yield the data, see References~\cite{Lee_Yeast,Lockhart_Winzeler_Genomics,Filkov_MicroarrayData}.

\begin{figure}[!htb] 
 \centering 
  \includegraphics[width=0.5\textwidth]{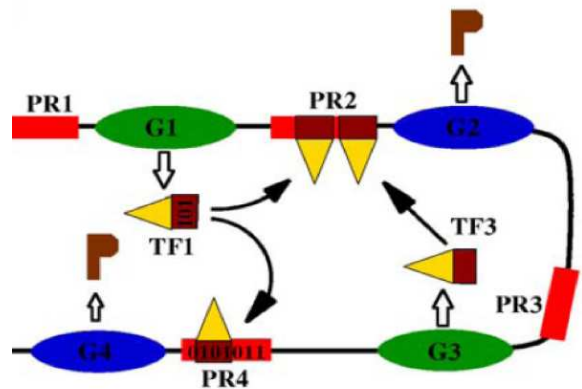}
 \caption[A brief explanation of the gene regulation process used in the thesis]{A brief explanation of the gene regulation process used in the thesis~\cite{Balcan_etal2007}.} 
 \label{GRScath} 
\end{figure}

The transcriptional regulation of gene expression, or in short, the gene regulation is modeled as follows: Each gene on the DNA possesses two main regions. The first is the \textbf{promoter region (PR)} and the second is genetic codes that are transcribed. In order to \textit{activate/inhibit} the transcription, the necessary conditions need to be hold at the PRs. Although these conditions are more complex, they are simplified as \textit{existence/absence} of \textbf{transcription factors (TFs)} that are unique proteins bind the PRs. When the conditions hold, the gene is activated/inhibited for transcription depending on the rules of this specific gene. The products of these processes could be either the functional proteins or TFs (See Figure~\ref{GRScath} for a simple scatching of the GR process~\cite{Balcan_etal2007}). More details about gene regulations can be found in References~\cite{Miglani_GENETICS,Lockhart_Winzeler_Genomics}.

My main aim in this part of the thesis is to both topologically and dynamically investigate the yeast gene regulation by using the network tools introduced in Chapter~\ref{NetworkModeling}.
 
\section{The Example at Hand: \textit{Saccharomyces Cerevisiae} (Yeast)}\label{YeastGRN}

\textit{Saccharomyces Cerevisiae} or the yeast is a unicellular eukaryotic microorganism. Its gene regulation data~\cite{Lee_Yeast} was used in this thesis since it is one of the mostly studied organism. The data was retrieved from YEASTRACT\footnote{www.yeastract.com} database~\cite{Yeastract}. When the data was taken it was including 4252 genes (146 of them are TFs) with 12541 interactions.

\subsection{Topology of the Yeast Gene Regulation Network}

\begin{figure}[!p]
\begin{center}
  \includegraphics[width=0.75\textwidth]{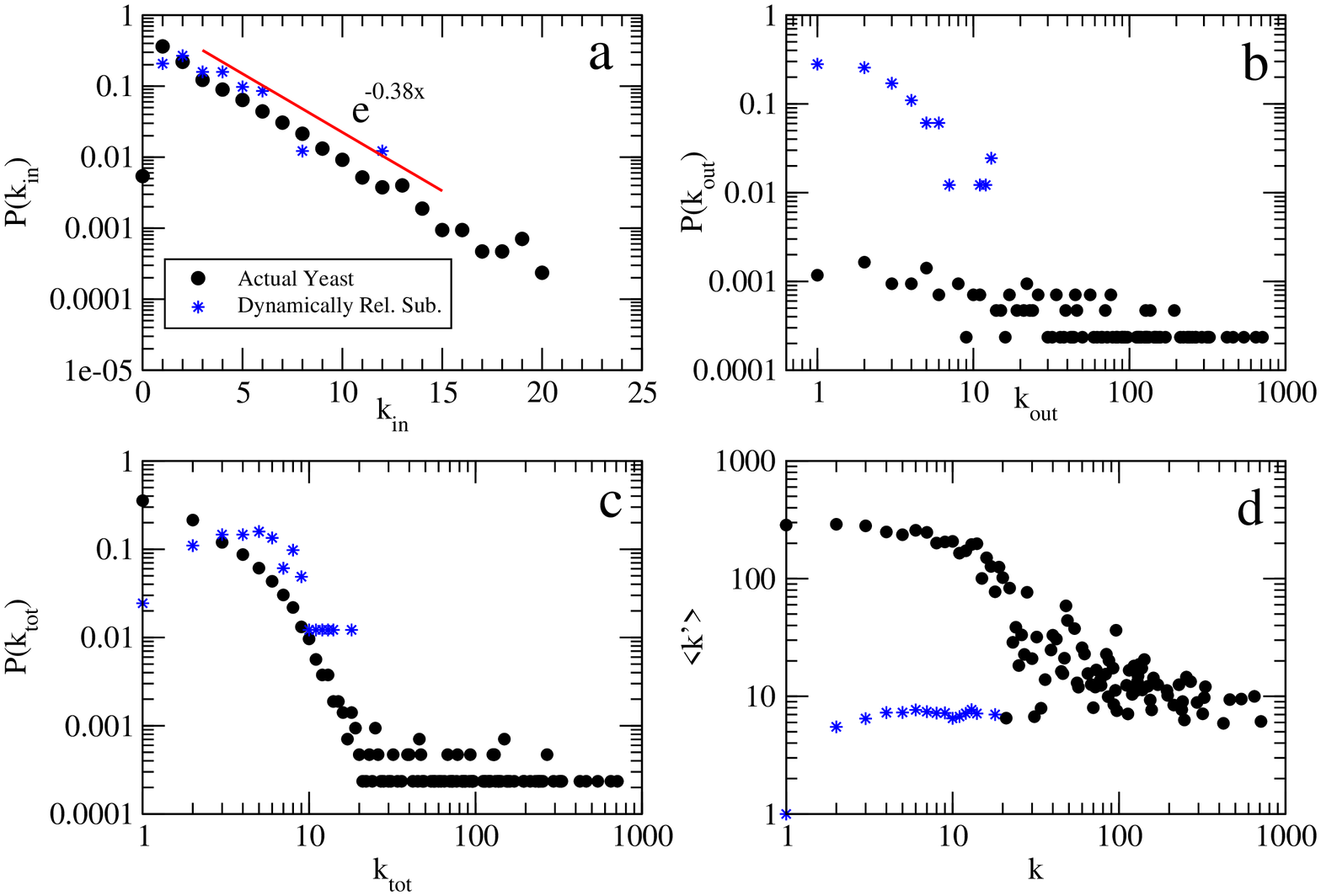}
\caption[Topological investigation of Yeast GR actual and dynamically relevant sub- network-1] {Yeast GR actual and dynamically relevant sub- network's \textbf{a-)} indegree probability distribution \textbf{b-)} outdegree probability distribution, \textbf{c-)} total degree probability distribution and \textbf{d-)} degree-degree correlation}
\label{Yeast_DegreeDistr}
\end{center}
\end{figure}

\begin{figure}[!p]
\begin{center}
  \includegraphics[width=0.75\textwidth]{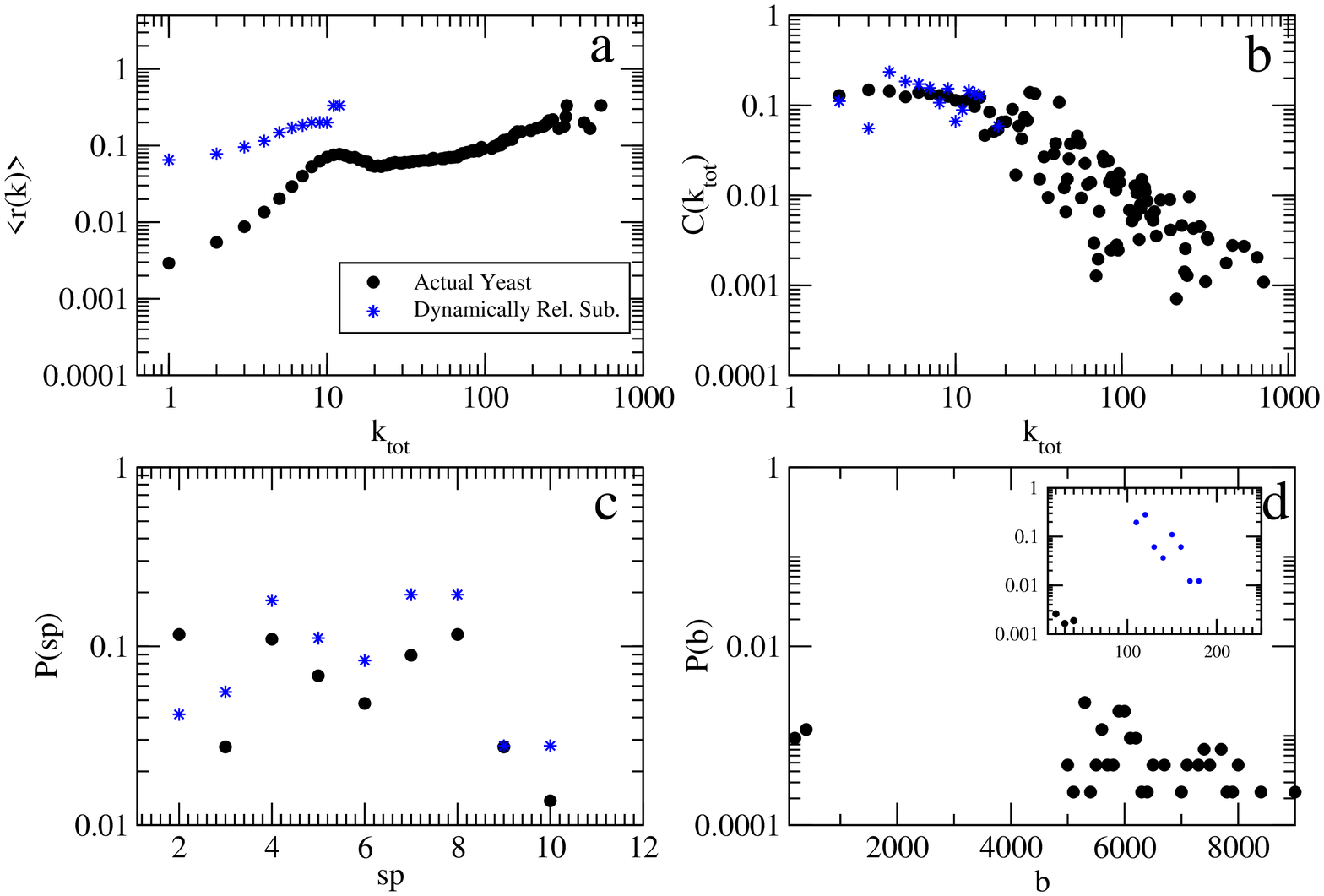}
\caption[Topological investigation of Yeast GR actual and dynamically relevant sub- network-2] {Yeast GR actual and dynamically relevant sub- network's \textbf{a-)} richclub coefficient \textbf{b-)} clustering coefficient, \textbf{c-)} shorthest path probability distribution (Binned in $1.0$ units) and \textbf{d-)} betweenness probability distribution (for actual network it is shown in big frame with binned in $100$ units, for dynamically relevant subnetwork it is shown in small frame with binned in $10$ units.).}
\label{YEAST_OtherTopologies}
\end{center}
\end{figure}

My investigations for some conventional topological features of the yeast GRN can be seen in Figure~\ref{Yeast_DegreeDistr} and Figure~\ref{YEAST_OtherTopologies}. Previously, Guelzim \textit{et al.} have topologically investigated the yeast GRN and stated an exponential decay in the indegree distribution with an exponent $\lambda=0.45$~\cite{Guelzim_Yeast}. However, my indegree distribution was fitted to an exponential decay with an exponent $\lambda=0.38 \mp 0.01$ by using GNUPLOT\footnote{http://www.gnuplot.info/} with ignoring $k_{in}=0$. The reason of this difference might be due to that my fitting was done by at first taking log of y-values and then fitting to a linear function. I also did a direct fitting resulting in $\lambda = 0.46 \mp 0.01$ exponential decay which is is almost the same as Guelzim \textit{et al.}'s. Another detailed studies for topological examination of yeast can be found in References~\cite{Balcan_etal2007,Bergman_etal_SixGRN}. 

The yeast GR network includes 4252 nodes/genes with 12541 directed edges/interactions (average degree is 2.95). 146 of those genes are TFs and there are 403 interactions between TFs (average degree is 2.76). Dynamically relevant subnetwork of yeast GRN consists of 82 TFs and 254 interactions (average degree is 3.10). 

Comparing the artificial networks, the fraction of dynamically relevant nodes to the number of nodes in yeast GRN is very low ($82/4252 \cong 2\%$). As shown in Figure~\ref{DynamicallyRelSubnetwork}, the artificial model networks with the same indegree distributions show a fraction of $~85-90\%$. Figures~\ref{Yeast_DegreeDistr}~and~\ref{YEAST_OtherTopologies} also show the topological features of dynamically relevant subnetwork of the yeast GRN and one can state the topologies are similar. This specialties of yeast GRN might be crucial in its dynamics in real case.

  \subsection{Dynamics of the Yeast Gene Regulation Network}

Although the interacting pairs are known very well, the rules governs the interactions in the yeast gene regulation are not identified in detail yet. For this reason, I used the 4 types of random functions introduced in Chapter~\ref{NetworkModeling}; simple random function (RF), canalazing random function (CF), nested canalazing random function (NCF) and special subclasses of nested canalazing random function (SNCF) with a statistical approach for investigations.

\begin{figure}[!p] 
 \centering 
  \includegraphics[width=0.95\textwidth]{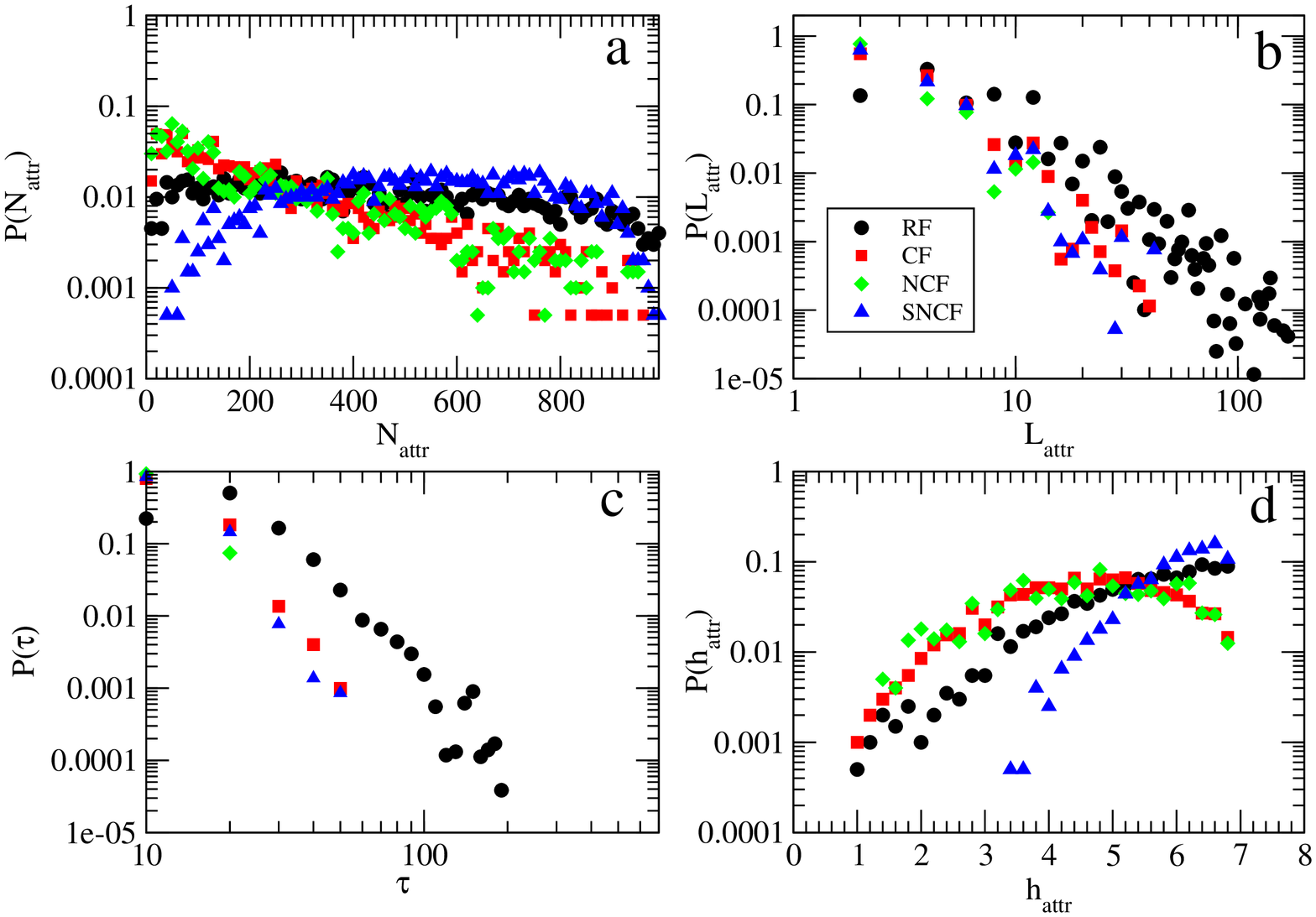}
 \caption[Attractor investigation of Yeast]{Distribution of attractor features of Yeast for Random Function (RF), Canalyzing Function (CF), Nested Canalyzing Function (NCF) and  Special Subclasses of Nested Canalyzing Function (SNCF); \textbf{a-)} number of attractors $N_{attr}$ probability distribution (Binned in $10$ units) \textbf{b-)} length of attractor $L_{attr}$ probability distribution (Binned in $2$ units), \textbf{c-)} transient to attractor $\tau_{attr}$ probability distribution (Binned in $10$ units), \textbf{d-)} entropy $h_{attr}$  probability distribution (Binned in $0.2$ units). Attractors were found with starting from $1000$ initial conditions of each $2000$ network realizations with maximum steps and maximum $L_{attr}$ limits as $1000$ and $200$, respectively. For RF, CF and NCF, $p$ was fixed to $0.27$ which is the $p$ of SNCF for yeast.} 
 \label{Yeast_Attractor} 
\end{figure}

In order to investigate and compare the attractors of Yeast GRN for all function types, one needs to fix the $p$ value. SNCF for dynamically relevant subnetwork of actual Yeast GRN gives out an effective p-value $p=0.27\mp0.05$, therefore, I used this p value as the parameter for other type of functions, too. $2000$ network realization were done for investigation. For each realization, attractors were explored by starting from $1000$ random initial conditions. In the dynamics, the maximum $1000$ steps and the maximum attractor length $200$ limits were set. The distributions and averages of  the number of attractors $N_{attr}$, the length of attractor $L_{attr}$, transient $\tau_{attr}$ and the entropy $h_{attr}$ can be seen in Figure~\ref{Yeast_Attractor} and Table~\ref{Yeast_Attractor_Averages}.

\begin{table}[!p]
\begin{center}
\begin{tabular}{|c||c|c|c|c|}
\hline 
 & $\langle N_{attr} \rangle$ &  $\langle L_{attr} \rangle$ & $\langle \tau_{attr} \rangle$ & $\langle h_{attr} \rangle$ \\ 
\hline 
\hline 
RF & $430.78 \mp 263.78$ & $8.82 \mp 13.45$ & $18.33 \mp 15.86$ & $5.40 \mp 1.10$\\
\hline 
CF & $222.77 \mp 201.04$ & $3.77 \mp 3.12$ & $8.48 \mp 3.99$ & $4.48 \mp 1.20$\\
\hline 
NCF & $221.15 \mp 209.14$ &$2.84 \mp 1.93$ & $6.68 \mp 2.36$ & $4.45 \mp 1.27$\\
\hline
SNCF & $538.72 \mp 212.18$ &$3.40 \mp 2.77$ & $7.84 \mp 3.50$ & $5.96 \mp 0.61$\\
\hline
\end{tabular}
\end{center}
\caption[Average values of attractor features of Yeast GRN.]{Average values of the number of attractors $N_{attr}$, the length of attractor $L_{attr}$, transient $\tau_{attr}$ and the entropy $h_{attr}$ of Yeast GRN for \textbf{RF}: Random Function, \textbf{CF}: Canalyzing Function, \textbf{NCF}: Nested Canalyzing Function, \textbf{SNCF}: Special Subclasses of Nested Canalyzing Function. For the details of the study, see the caption of Figure~\ref{EXPModelYEAST_Attractor}.}
\label{Yeast_Attractor_Averages}
\end{table} 

First of all, it should be noted that $N_{attr}$ and $h_{attr}$ distribution are not decreasing for all functions, especially SNCF  type shows a not ordinary profile. This type of the distribution was not observed while studying with model networks in Chapter~\ref{NetworkModeling} and should be discussed further. Secondly, the averages with SNCF type shows a significant difference with other types. As it is seen in Table~\ref{Yeast_Attractor_Averages}, it has a big $\langle N_{attr} \rangle$ while having small $\langle L_{attr} \rangle$ and $\langle \tau_{attr} \rangle$ which might be desirable in biological systems~\cite{Kauffman_Origins_of_Order}. As a conclusion to the attractor results, SNCF type seems to be very appropriate for the dynamics of the yeast gene regulation and should be continued to be investigated.

\begin{figure}[!hbt] 
 \centering 
  \includegraphics[width=\textwidth]{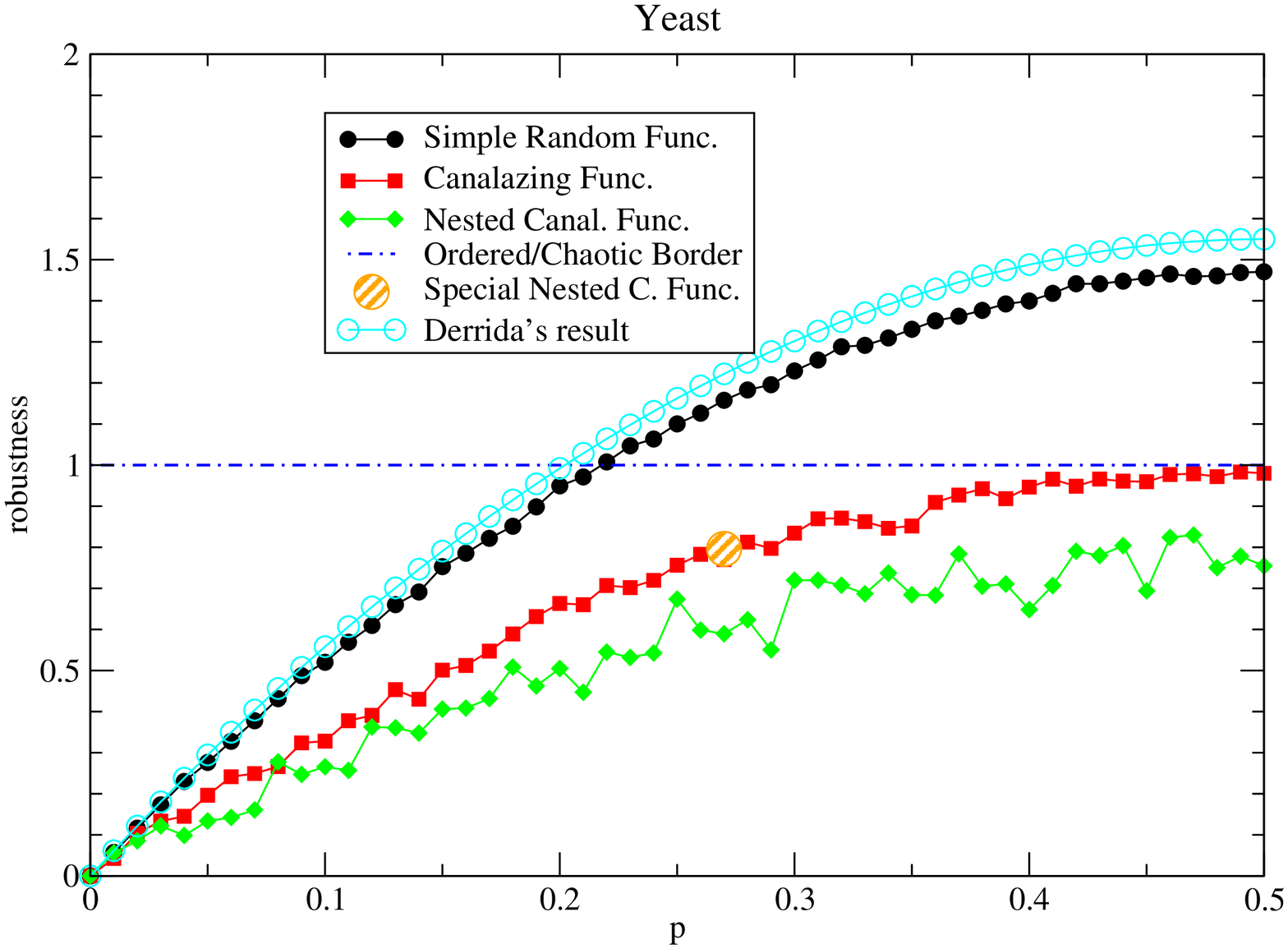}
 \caption[Yeast Robustness]{Robustness of yeast GRN for all types of functions. For each $p$ value, robustness was computed with averaging over $1000$ random initial conditions of each $10$ realization. Also, the Derrida's Exp.\ref{DerridaFormula}, $s=2p(1-p)\langle k_{in} \rangle$ was drawn with $\langle k_{in} \rangle$ of the dyn. rel. subnetwork, $3.1$.} 
 \label{Yeast_Robustness} 
\end{figure}

I also obtained the \textit{robustness} for each p $\in \lbrace 0.00, 0.01,..,0.50\rbrace$ of RF,CF and NCF, and for $p=0.27$ of SNCF. $10$ dynamics realization and for each realization $1000$ random initial conditions were created and as explained in chapter 2, dynamics were runned for $10\times N$ steps ($N=82$ in this case) in order to be able to achieve an attractor. After these steps, \textit{robustness} was measured numerically and averaged for all values. The results can be seen in Figure~\ref{Yeast_Robustness}.

It is shown that for $p=0.27$ (attractor statistics was done at this p-value) RF functions at the chaotic side near critical boundary while others are at the ordered side. Also, it seems that the Derrida's Exp.\ref{DerridaFormula} predicts RF results quite well. The results also shows that although SNCF and CF gives the same robustness value for the attractor investigation paramater $p=0.27$, they could produce different attractor structures. In other words, there might be no direct relation between attractor and robustness structure.

\clearpage
\section{in-EXP Model Networks for Yeast GR}\label{inEXPYeast}

To my knowledge Kauffman is the first who used the model networks to
elucidate the gene regulation~\cite{Kauffman_Network}. He used in-NK networks\footnote{$K=2$ case is also known as Kauffman networks in gene regulation literature} for modeling the gene regulations. However, in order to compare the dynamics of the yeast, I used $100$ in-EXP networks with this exponent since indegree of yeast GRN was an exponential decay with exponent $0.38$. I set $N=94$ for the sake of achieving a similar number of nodes of the dynamically relevant subnetwork of yeast, $N=82$ (See Figure~\ref{DynamicallyRelSubnetwork} in Chapter~\ref{NetworkModeling} for fractions of dynamically relevant nodes to system size for model networks.).

\subsection{Topological Investigation}

\begin{figure}[!p] 
 \centering 
  \includegraphics[width=0.8\textwidth]{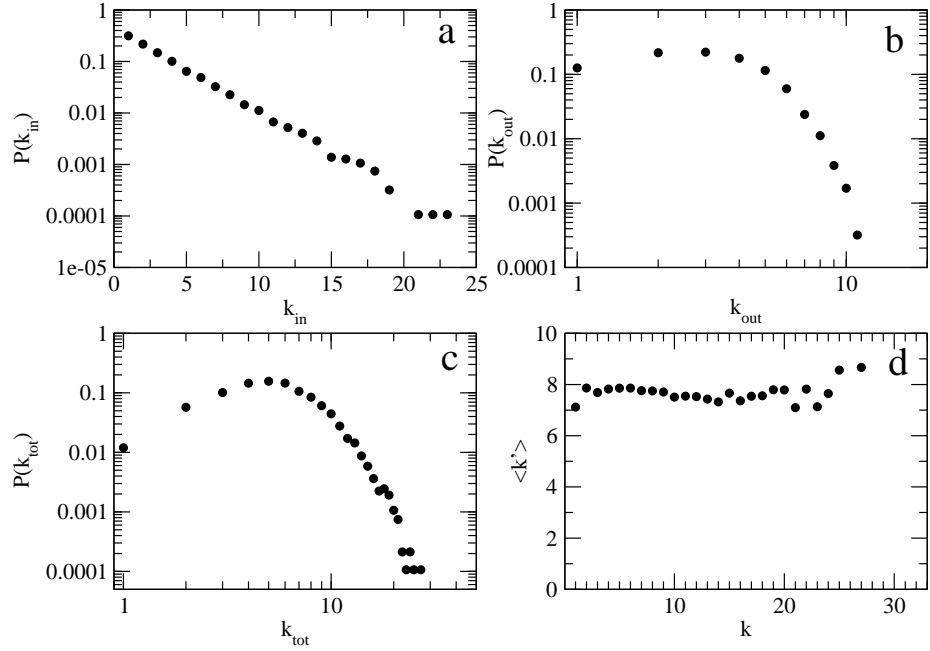}
 \caption[in-EXP Model Yeast Network's topologies-1]{in-EXP Model Yeast Network's \textbf{a-)} indegree probability distribution \textbf{b-)} outdegree probability distribution, \textbf{c-)} total degree probability distribution and \textbf{d-)} degree-degree correlation} 
 \label{EXPYEAST_DegreDist} 
\end{figure}

\begin{figure}[!p] 
 \centering 
  \includegraphics[width=0.8\textwidth]{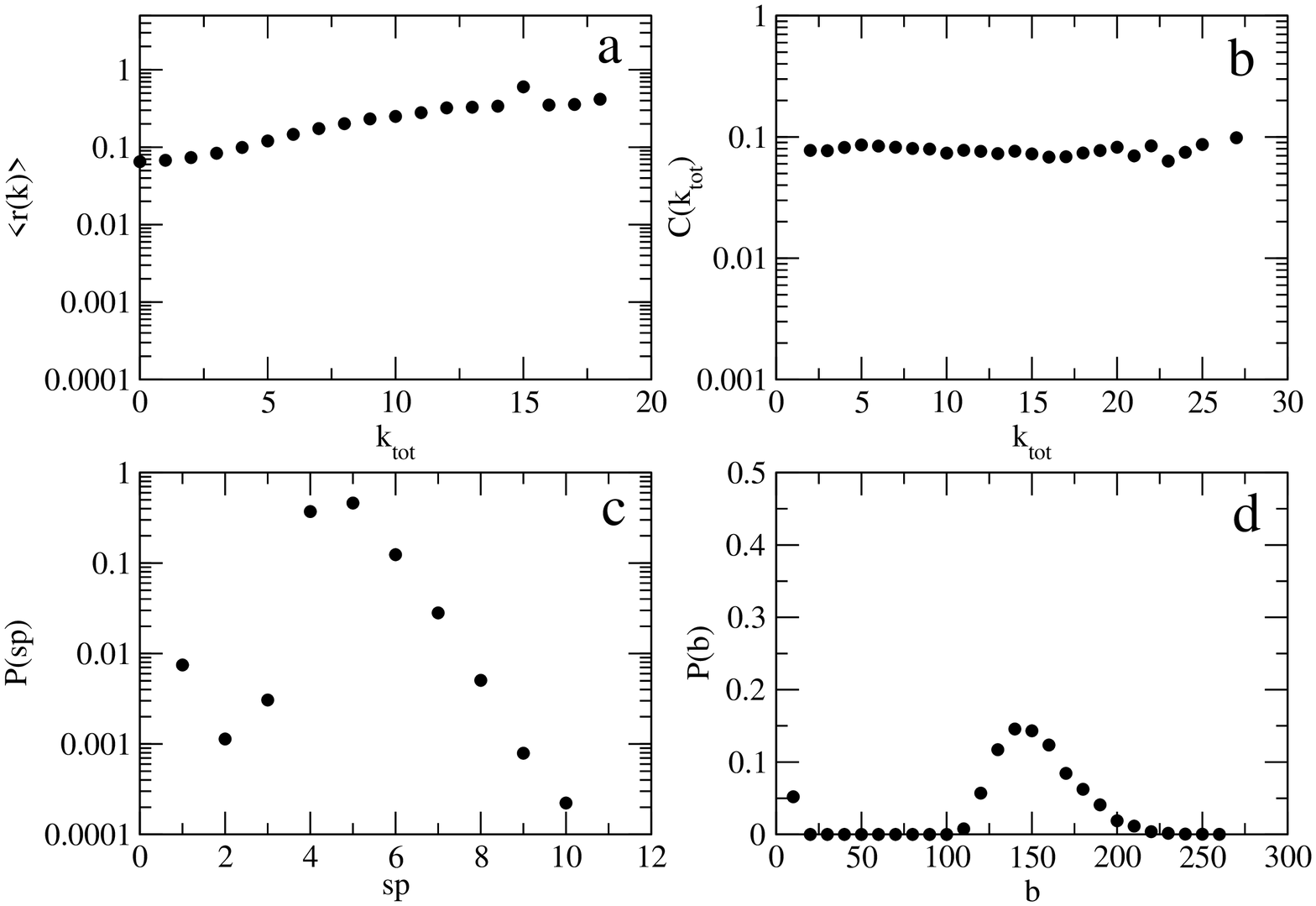}
 \caption[in-EXP Model Yeast Network's topologies-2]{in-EXP Model Yeast Network's \textbf{a-)} richclub coefficient \textbf{b-)} clustering coefficient, \textbf{c-)} shorthest path probability distribution (Binned in $1.0$ units) and \textbf{d-)} betwenness probability distribution (Binned in $10$ units)} 
 \label{EXPYEAST_OtherTopologies} 
\end{figure}

Before passing to the dynamics, let me compare the topological features of this model networks  with yeast dynamically relevant subnetwork'. Figure~\ref{EXPYEAST_DegreDist}~and~\ref{EXPYEAST_OtherTopologies} show the investigations. It can be seen that out- and total-degree distributions and degree-degree correlations resemble that of yeast dynamically relevant subnetwork while others are different.

\subsection{Dynamical Investigation}

\begin{figure}[!p] 
 \centering 
  \includegraphics[width=\textwidth]{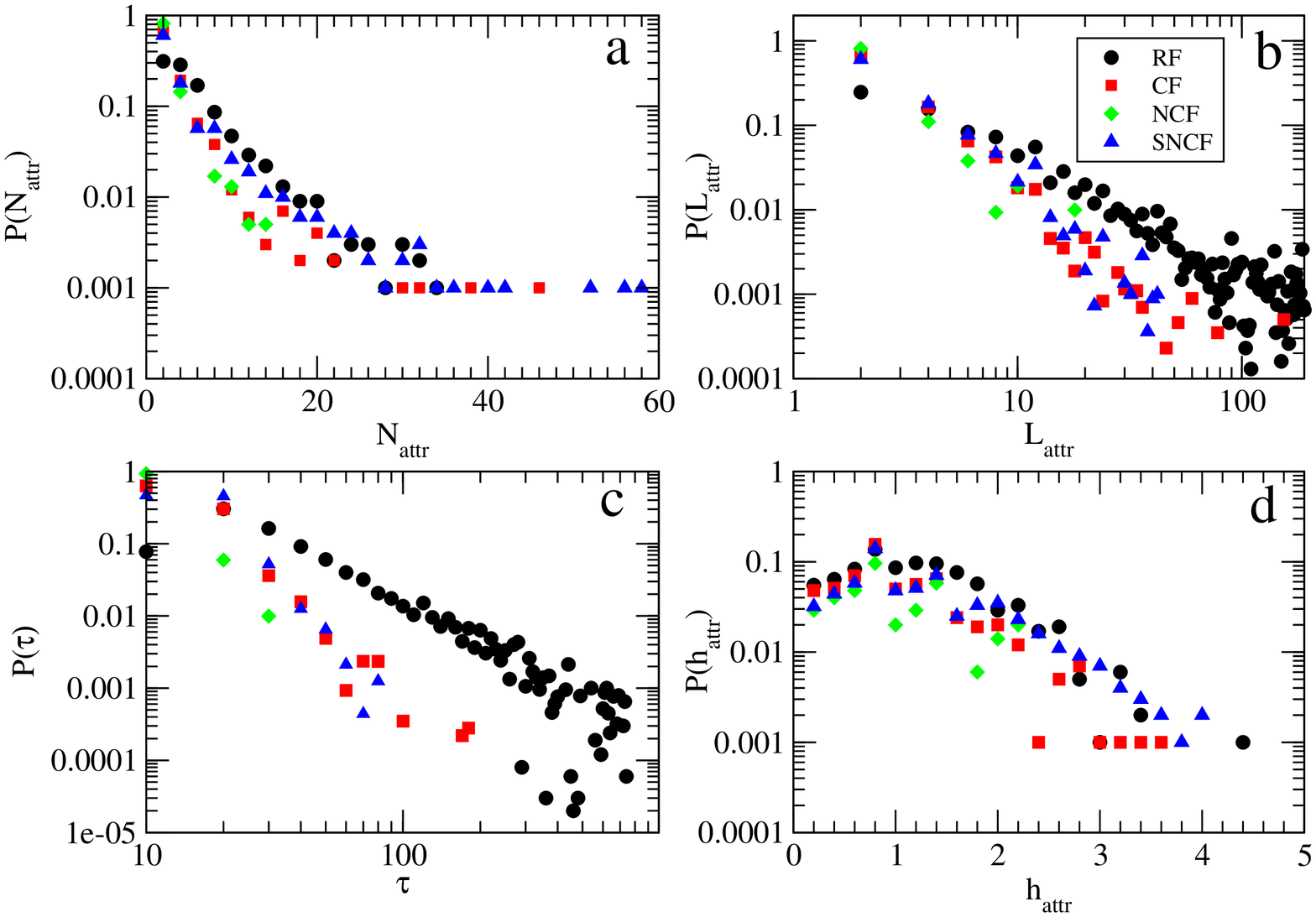}
 \caption[Attractor investigation of in-EXP Model Yeast Networks.]{Attractor investigation of in-EXP Model Yeast Networks for Random Function (RF), Canalyzing Function (CF), Nested Canalyzing Function (NCF) and  Special Subclasses of Nested Canalyzing Function (SNCF); \textbf{a-)} Number of Attractors Distribution (binned in 2 units), \textbf{b-)} Length of Attractors Distribution (binned in 2 units) \textbf{c-)} Transient time distribution (binned in 10 units), \textbf{d-)} Entropy distribution (binned in 0.1 units). Attractor were found by starting from $100$ initial conditions for each realizations. $10$ realizations were done for each of $100$ networks. The limits for maximum step size and length of attractors were $1000$ and $200$, respectively. $p$ of the functions were $0.27$} 
 \label{EXPModelYEAST_Attractor} 
\end{figure}

Similar to Yeast's case, the dynamics was investigated for $p=0.27$. It should be noted that the p-value for SNCF is found to be $p=0.27$ which is the same as Yeast's. In dynamics, $100$ networks, for each network $10$ network realization and for each network realization $100$ initial conditions were created. Again maximum $1000$ steps and $200$ attractor length were set. The distributions and averages can be found in Figure \ref{EXPModelYEAST_Attractor} and Table \ref{EXPModelYeast_Attractor_Averages}.

\begin{table}[!p]
\begin{center}
\begin{tabular}{|c||c|c|c|c|}
\hline 
 & $\langle N_{attr} \rangle$ &  $\langle L_{attr} \rangle$ & $\langle \tau_{attr} \rangle$ & $\langle h_{attr} \rangle$ \\ 
\hline 
\hline 
RF & $5.15 \mp 5.17$ & $68.29 \mp 216.64$ & $101.73 \mp 221.50$ & $0.93 \mp 0.70$\\
\hline 
CF & $2.87 \mp 3.50$ & $4.31 \mp 32.04$ & $11.83 \mp 32.34$ & $0.53 \mp 0.63$\\
\hline 
NCF & $1.90 \mp 1.88$ &$2.07 \mp 2.37$ & $7.06 \mp 3.06 $ & $0.31 \mp 0.52$\\
\hline
SNCF & $3.94 \mp 5.65$ &$3.70 \mp 4.56$ & $11.99 \mp 6.55$ & $0.69 \mp 0.79$\\
\hline
\end{tabular}
\end{center}
\caption[Average values of attractor investigation of in-EXP Model Yeast GRN.]{Average values of attractor features of in-EXP Model Yeast GRN. \textbf{RF}: Random Function, \textbf{CF}: Canalyzing Function, \textbf{NCF}: Nested Canalyzing Function, \textbf{SNCF}: Special Subclasses of Nested Canalyzing Function. For the details of the study, see the caption of Figure~\ref{EXPModelYEAST_Attractor}.}
\label{EXPModelYeast_Attractor_Averages} 
\end{table}

\begin{figure}[!ht] 
 \centering 
  \includegraphics[width=\textwidth]{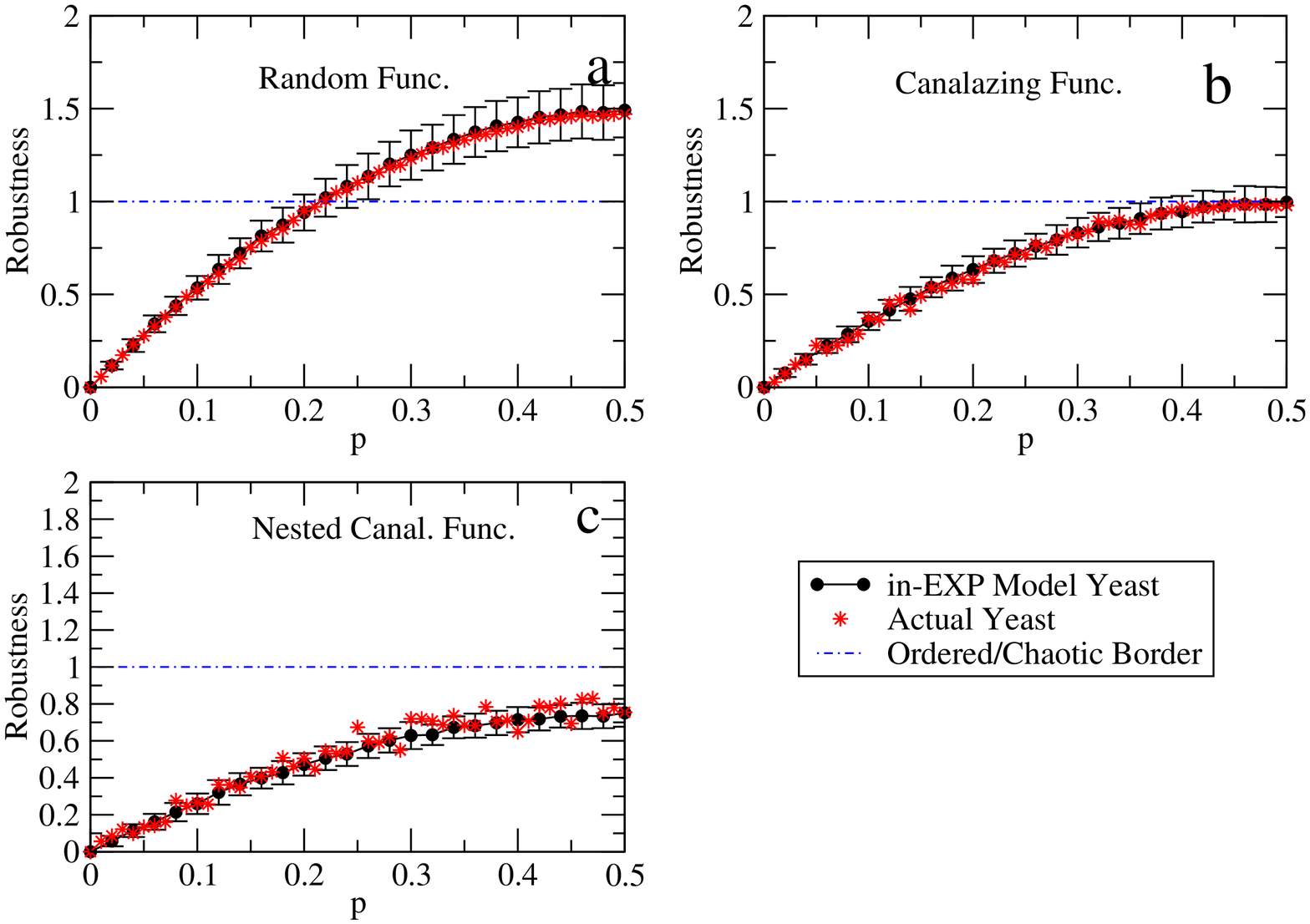}
 \caption[Robustness Comparison of Actual Yeast and in-EXP Model Networks.]{Robustness Comparison of Actual Yeast and in-EXP Model Networks. \textbf{a-)} Simple Random Func. \textbf{b-)} Canalyzing Func. \textbf{c-)} Nested Canal. Func.. The robustness was computed by averaging over $100$ networks, $10$ dynamics realization for each network and $1000$ random initial conditions for each realization.} 
 \label{EXPModelYEAST_Robustness} 
\end{figure}

\textit{Robustness} was also studied for each p $\in \lbrace 0.0, 0.02,..,0.50\rbrace$ for RF, CF and NCF, and $p=0.27$ for SNCF. The robustness was computed by averaging over $100$ networks, $10$ dynamics realization for each network and $1000$ random initial conditions for each realization. The results are in Figure~\ref{EXPModelYEAST_Robustness}.


The yeast and in-EXP model networks produce different attractors features while show similar robustness profiles. The main conclusion of that part is that the model networks which are constructed by knowing only indegree distributions and the network size fail for attractor features predictions while win for robustness. 

\clearpage
\section{A Model: Root of the Yeast Gene Regulation Network Topology}\label{BalcanModel}

Starting from some previous studies \cite{Balcan_Erzan2004,Mungan_etal2005}, Balcan \textit{et al.} have arrived at a novel model produces the complex
networks whose \textit{topological} properties resemble that of the yeast gene regulation
network~\cite{Balcan_etal2007}.

 \subsection{Description of Model}

The model is initiated with a starting \textit{fixed number} of \textbf{genes}. Next, each gene is assigned to be \textbf{Transcription Factor(TF) coding gene} with a probability \textbf{p}. The numbers are optimized as 6000 genes at the beginning and $p=200/6000$ for determining TFs by consulting available actual Yeast data \cite{Yeastract}.

The model then assigns two types of random binary sequences, i.e. 110100... The former is to all genes being called \textbf{Prometer Sequence, PS} and the latter is to only TF coding ones being called \textbf{Regulatory Sequence, RS}. Thus, TF coding genes should have two
labels where the others have only one.

\begin{figure}[!hbt] 
 \centering 
  \includegraphics[width=0.55\textwidth]{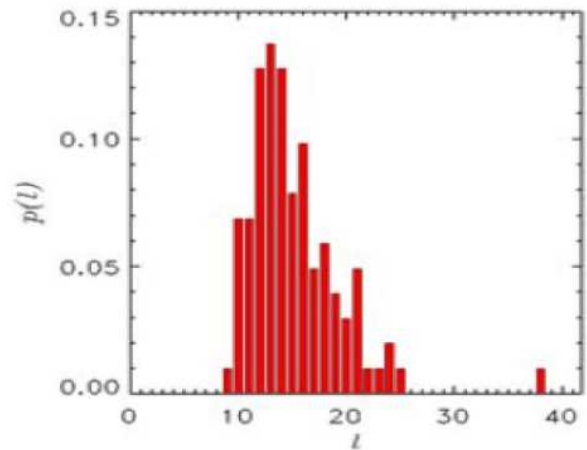}
 \caption[Regulatory sequence distribution of yeast]{Regulatory sequence distribution of yeast~\cite{Balcan_etal2007}} 
 \label{RS_Harbisson_distr} 
\end{figure}

The most important part in assigning these random binary sequences is determining the lengths of the sequence attached to genes. The RS lengths are taken from a distribution yielded from a study of Harbison~\cite{Harbison_RSdistr} as shown in Figure~\ref{RS_Harbisson_distr}. However, there is no available distribution for the PR lengths; therefore, authors have assumed that the distribution obeys the power-law with $P(l) \sim l^{-1.1}$. This assumption was based on the investigation of the intergenic portions of the DNA and the result of $P(l) \sim l^{-1-\mu}$ where $\mu$ is applicable in $0.0<\mu<2.0$~\cite{Almirantis_PRdistr}. Morover, again referring the Harbison study~\cite{Harbison_RSdistr} which argues that the most of the likelyhood for determining a TF binding site occurs in a 250 bps window on the DNA, they bounded $l$ with the expression $l_{max}-l{min}+1=250$ where $l_{min}$ is taken to be pick-value of RS distribution.

After assigning the sequences, a directed edge is placed from TF coding gene (RS) to gene (PS), if and only if a RS is fully inside a PS. These processes can be runned several times and an ensemble can be created.

 \subsection{Topological Investigation (Reproduction of Some Results)}

\begin{figure}[!ht] 
 \centering 
  \includegraphics[width=\textwidth]{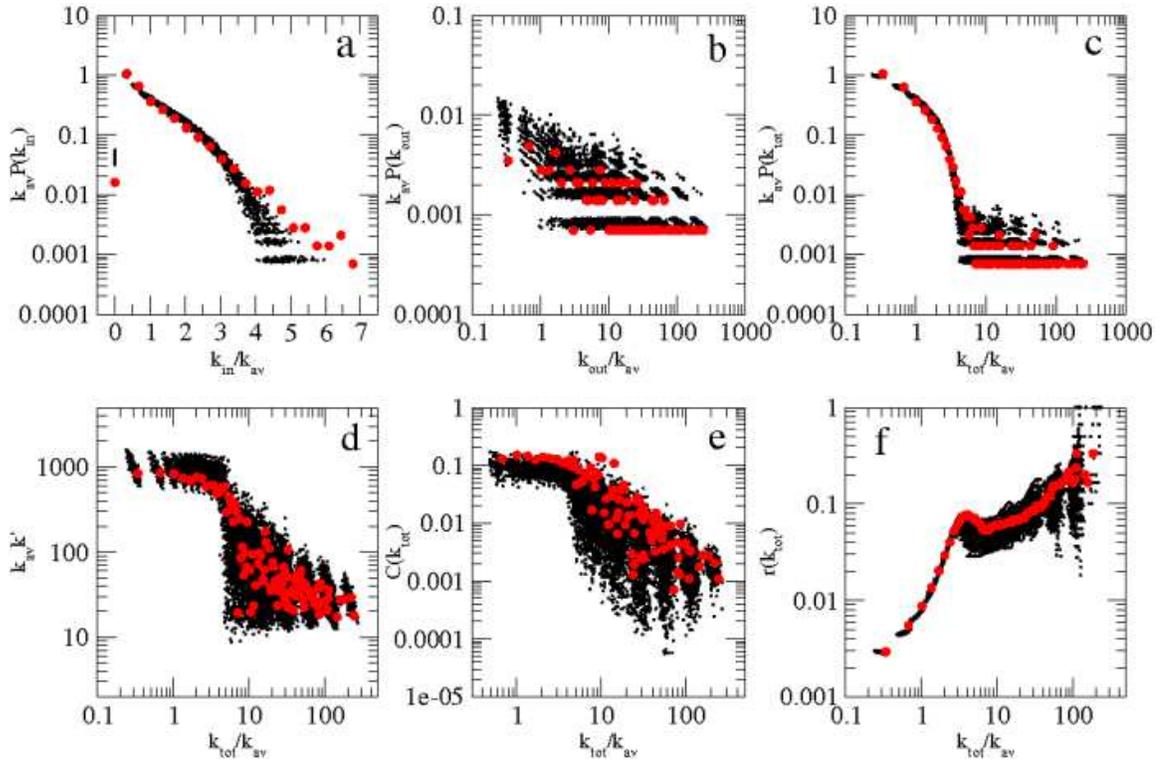}
  \caption[Some reproductions of Balcan \textit{et al.} models]{Some reproductions of Balcan \textit{et al.} models' results; a-) indegree prob. dist., b-)outdegree prob. dist., c-)total-degree prob. dist., d-)degree-degree correlation, e-)clustering coefficient, f-)richclub coefficent. x- and y-values of the data of corresponding network were multiplied/divided by the average degree.} 
 \label{BalcanetalArticleResults} 
\end{figure}

Balcan \textit{et al.} represented in their study that their model creates networks whose topological properties resembles to the actual yeast GRN. I reproduced some of the investigations found in the article~\cite{Balcan_etal2007} as shown in Figure~\ref{BalcanetalArticleResults} in order to be able to elucidate the model. Apart from these topology features, they also stated a similarity in \textit{K-core} structure. In general, the model seems to be very successful for producing the similar topologies. Only topological dissimilarity they found is a detail difference in \textit{motifs}.

I also found a detail mistake in the article which can empower the results. The average number of TFs of the model networks were stated as $202 \mp 14$ in the article, however, according to my computations it was $167 \mp 14$ which is closer to actual number of TFs in the yeast: $146$. 

 \subsection{Dynamical Investigation of Model}
My aim at this part of the thesis is to investigate the Balcan \textit{et al.} model networks in order to detect whether they are also similar with respect to dynamics as they do w.r.t. topology.

\begin{figure}[!p] 
 \centering 
  \includegraphics[width=\textwidth]{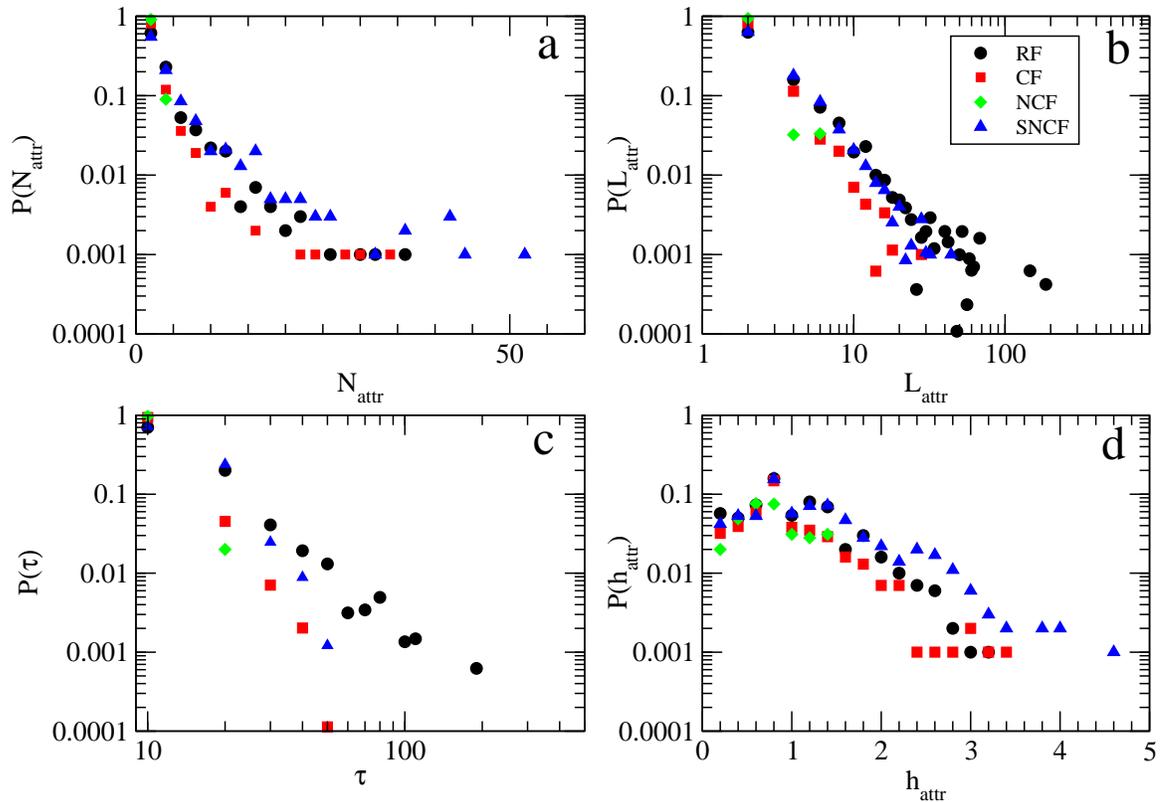}
 \caption[Attractor Investigation of Model]{Probability distributions for attractor features of Balcan \textit{et al.} Model Networks. \textbf{RF}: Random Function, \textbf{CF}: Canalyzing Function, \textbf{NCF}: Nested Canalyzing Function, \textbf{SNCF}: Special Subclasses of Nested Canalyzing Function. $100$ model Balcan \textit{et al.} networks and for each network, $10$ realizations were created. Attractors were found by starting from $100$ initials conditions for each realization where $p=0.27$ was set for the functions.} 
 \label{Balcan_Attractor} 
\end{figure}

\begin{figure}[!htp]
\centering
  \includegraphics[width=\textwidth]{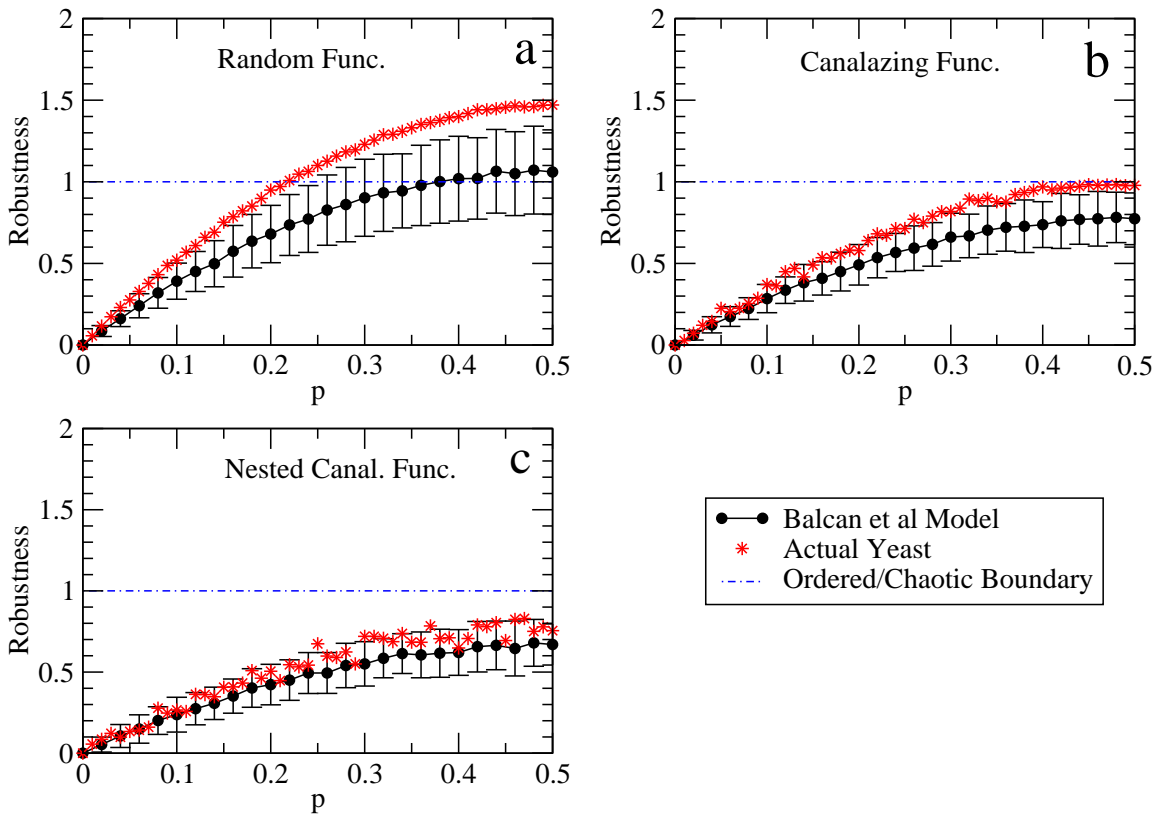}
\caption[Robustness of Model Yeast GRN]{Robustness investigation of Model Yeast GRN with comparison to Actual Yeast; \textbf{a-)}, \textbf{b-)}, \textbf{c-)}. Also it should be noted that for Special subclasses of Nested Canalyzing Functions, robustness measure is $0.83\mp0.08$ where for Yeast it is $0.78$. }
\label{Balcan_Robustness}
\end{figure}

For the dynamical investigation $100$ Balcan \textit{et al.} model networks were used. As previous cases, investigations were done with using dynamically relevant subnetworks which are found to have $36 \mp 15$ nodes. Comparing to the actual yeast case ($N=82$), the dynamically relevant nodes are very less in number. Next, both the attractors and robustness were calculated by starting from $100$ initial conditions for each of $10$ network realization for each network. For finding attractor, the limits for the maximum length of attractor and maximum step size were set to $200$ and $1000$, respectively. The $p$ value for RF,CF and NCF were set to $p=0.27$ while p of SNCF was found to be $0.30$. The results for attractor features are shown in Table~\ref{BalcanModelYeast_Attractor_Averages} and Figure~\ref{Balcan_Attractor}. The results for robustness is in Figure~\ref{Balcan_Robustness}.

\begin{table}[!p]
\begin{center}
\begin{tabular}{|c||c|c|c|c|}
\hline 
 & $\langle N_{attr} \rangle$ &  $\langle L_{attr} \rangle$ & $\langle \tau_{attr} \rangle$ & $\langle h_{attr} \rangle$ \\ 
\hline 
\hline
RF & $3.14 \mp 3.51$ & $4.28 \mp 8.48$ & $10.31 \mp 12.28$ & $0.57 \mp 0.62$\\
\hline 
CF & $2.09 \mp 2.54$ & $2.04 \mp 2.16$ & $5.47 \mp 3.28$ & $0.35 \mp 0.52$\\
\hline 
NCF & $1.44 \mp 0.77$ &$1.47 \mp 1.01$ & $4.68 \mp 2.03$ & $0.21 \mp 0.36$\\
\hline
SNCF & $4.36 \mp 7.95$ & $3.30 \mp 3.85$ & $9.02 \mp 5.07$ & $0.74 \mp 0.78$\\
\hline
\end{tabular}
\end{center}
\caption[Average values of attractor investigation of Balcan \textit{et al.} Model Networks.]{Average values of attractor features of Balcan \textit{et al.} Model Networks. \textbf{RF}: Random Function, \textbf{CF}: Canalyzing Function, \textbf{NCF}: Nested Canalyzing Function, \textbf{SNCF}: Special Subclases of Nested Canalyzing Function. For details, see the captions of Figure~\ref{Balcan_Attractor}.}
\label{BalcanModelYeast_Attractor_Averages}
\end{table}

As a result, Balcan \textit{et al.} model networks failed to mimic the yeast. The results for the average and distributions resemble in-EXP yeast model except that the attractor averages for random functions are considerably bigger. Also, the robustness values for Balcan \textit{et al.} model networks seems to be successful for NCF and SNCF whereas do not give appropriate results for RF and CF. I consider that weak part of the model is that it does not produce a network having as much dense dynamical core as the yeast. This could be related to dissimilarities they stated for the \textit{motifs}. Balcan \textit{et al.} model may be enhanced with considering the dynamcally relevant subnetwork procedure used in this thesis.

\chapter{Protein Folding}\label{ProteinFolding}

This chapter is organized as follows: Section~\ref{PFIntroduction} summarizes the protein folding problem in biology. Section~\ref{NewApproachToPF} introduces a new approach to the problem by networks. Section~\ref{2CI2Section} gives the investigations on protein \textit{Serine Proteinase Inhibitor}. Section~\ref{OtherProteins} emphasizes the results for other proteins.

\section{Introduction}\label{PFIntroduction}

Living organisms consist of five types of organic compounds: carbohydrates, lipids, nucleic acids, vitamins and proteins. Among these organic compounds, the protein has a privileged place due to its functionality, in the cells~\cite{Life_Ricki}. Because of this importance, proteins have been studied highly for decades.

\begin{figure}[!htb] 
 \centering 
  \includegraphics[width=0.4\textwidth]{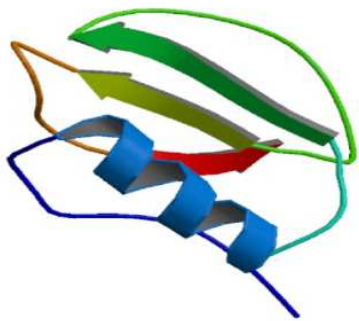}
 \caption{The structure of protein \textit{Serine Proteinase Inhibitor} (PDB ID: 2CI2)} 
 \label{2CI2_ribbon} 
\end{figure}

A protein is a chain of aminoacids which are bound to the neighbors in the chain by covalent interactions. The natural amino acids are 20 types and the number of amino acids in a protein can vary from 20 to 3000. The amino acid sequence is determined by responsible gene on the DNA. The DNA sends the necessary information to ribosomes via mRNA and they use mRNA to produce the proteins. After this process in ribosomes, this linear chain immediately folds and becomes more compact in 3D shape named as the \textbf{native structure}. This structure is crucial for the functionality of proteins in the cell. The sequence of the amino acids in a protein is unique for this protein and is called the \textbf{primary structure} of the protein~\cite{Life_Ricki,Cell_Bio_KARP}. Both the primary and native structures of proteins are available in a free-database: Protein Data Bank (PDB)\footnote{http://www.pdb.org}~\cite{PDB}.

\subsection{The Folding Problem}

\begin{figure}[!htb] 
 \centering 
  \includegraphics[width=0.6\textwidth, height=0.15\textheight]{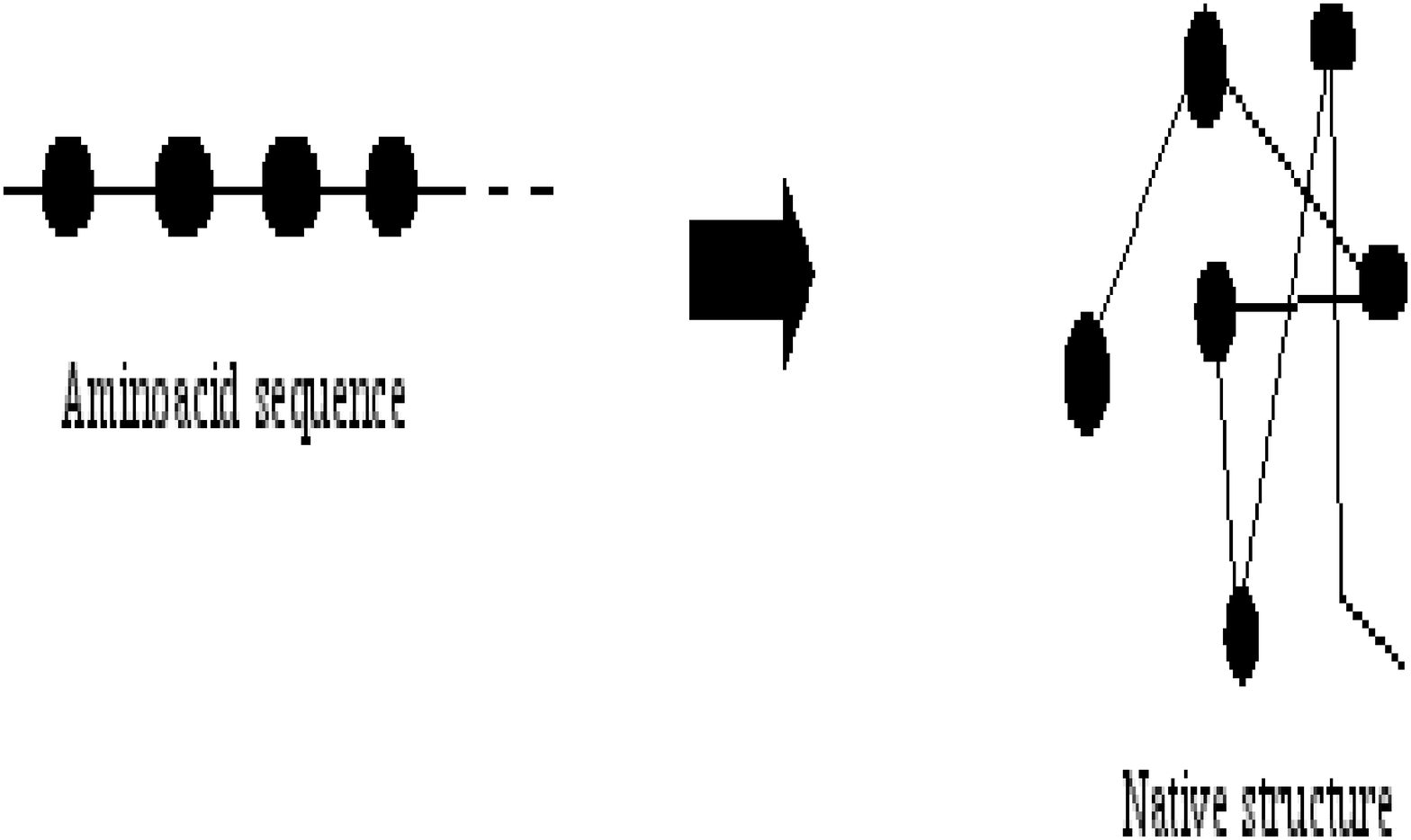}
 \caption{The Protein Folding Problem} 
 \label{ProteinFoldingScatch} 
\end{figure}

Since the native structure of a protein mainly affects its functionality in the cell, elucidating the mechanism from primary structure to native structure has been one of the leading tasks in protein studies \cite{Fersht_ProteinFoldingREVIEW}. Shortly, \textbf{the protein folding problem} is about finding out the “native” structure of protein from known and unique primary sequence (Figure \ref{ProteinFoldingScatch}). A more detailed review regarding to the protein folding can be found in Reference \cite{Fersht_ProteinFoldingREVIEW}.

\subsection{The Quantifier for Protein Folding}

\textbf{$\phi$-value, $\phi_{i}$:} To my knowledge, $\phi$-value is the only experimental analysis of two-state folding proteins~\cite{Fersht_ProteinFoldingREVIEW}. Two-state folding means that the folding occurs at first from unfolding state (U) to transition state (TS) and next, from transition state to folded state (F). In this analysis a particular aminoacid $i$ is mutated to another aminoacid type which is generally \textit{Alanine}~\cite{Fersht_ProteinFoldingREVIEW}. Later, $\phi_i$ is determined by using the experimental Gibbs-Free energies of mutant and wild-type protein according to Eq.~\ref{phi_value}~\cite{Finke_PHIvalues2CI2}:
\begin{equation}
 \label{phi_value}
\phi_i=\frac{\Delta G_{TS-U}^{wild-type} - \Delta G_{TS-U}^{mutant}}{\Delta G_{N-U}^{wild-type} - \Delta G_{N-U}^{mutant}}.
\end{equation}

\section{A New Approach to the Protein Folding}\label{NewApproachToPF}

Before getting into new approach, let me define the network of a protein, G(P). Each aminoacid of the protein is a node of G(P) and an edge is assigned to each pair of nodes in G(P) if and only if the real distance between aminoacids in native structure of the protein is less and equal to a certain threshold distance, $r_{thr}$. One should be aware that G(P) is also named as contact network (map) in literature~\cite{Protein_Folding_Contact_Map,Demirel_IdentificationOfKineticallyHotResidues}

\begin{figure}[!htp] 
 \centering 
 \includegraphics[width=0.45\textwidth]{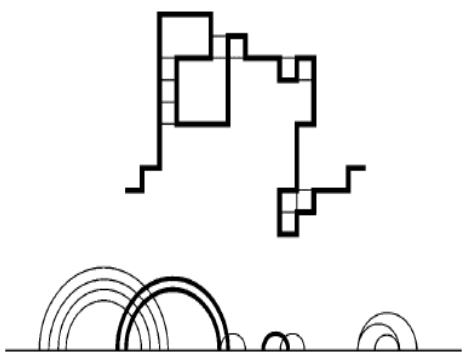}
 \caption{Definition of incompatibility networks} 
 \label{Definiton_incompatible_network} 
\end{figure}

Consulting the literature \cite{Demirel_IdentificationOfKineticallyHotResidues, Protein_Folding_Contact_Map} and after some trials, I fixed the threshold distance as $r_{thr}=6.5$ \AA{}. Also, I prohibited the edges between the aminoacid $i$ and $j$ if $\vert i-j \vert \leq 3$.

New approach starts with a definition of another network from G(P) which is named as \textbf{incompatibility} network: IG(P)~\cite{Kabakcioglu_PseudoHomoploymer}. For definition of IG(P),  let me refer to Figure \ref{Definiton_incompatible_network}. In this figure, a protein is extended like a linear chain and starting from $v_i$ and finishing at $v_k$ a half circle is drawn if an edge $e_{ik}$ exists between $v_i$ and $v_k$ in G(P). In incompatibility network IG(P); $e_{ik}$ in G(P) is defined as a node $v_{ik}$ and an edge between $v_{ik}$ and $v_{jl}$ is attached if their half circles in the figure are crossed~\cite{Kabakcioglu_PseudoHomoploymer}.

\begin{figure}[!htb] 
 \centering 
  \includegraphics[width=0.6\textwidth]{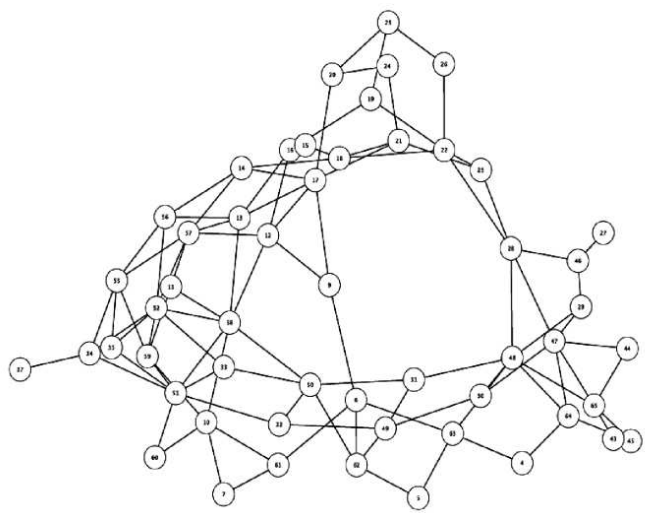}
 \caption{Network of protein 2CI2, i.e. G(CI2)} 
 \label{CI2_network} 
\end{figure}

Such incompatibility networks have been studied as a tool for understanding mRNA \cite{Tinoco_HowRNAfolds} functionality \cite{Kabakcioglu_PseudoHomoploymer} such as determining important regions of mRNA, etc.~\cite{Du_etal_mRNA,Wills_etal_mRNA,McPheeters_etal_mRNA,Farabaugh_mRNA}. Here my aim was to apply this concept to proteins.

\section{Example at hand: \textit{Serine Proteinase Inhibitor CI-2} (PDB ID:2CI2)}\label{2CI2Section}

\textit{Serine Proteinase Inhibitor} (PDB ID: 2CI2)~\cite{CI2} is a simple two-state folding protein with 65 aminoacids. It has been highly studied in literature and this is the main reason to be chosen in this thesis. A ribbon structure of protein 2CI2 is shown in Figure~\ref{2CI2_ribbon}.

$\phi$-value analysis of 2CI2 can be found in Reference~\cite{Finke_PHIvalues2CI2}. However, I used an extended $\phi$-value data of 2CI2 (Figure~\ref{2CI2_G_OtherTopologies}-b ) and some other proteins (in Appendix~\ref{OtherPhiValues}) found by a private communication with Prof.~Michele Vendruscolo. 

\begin{figure}[!htb] 
 \centering 
  \includegraphics[width=0.5\textwidth]{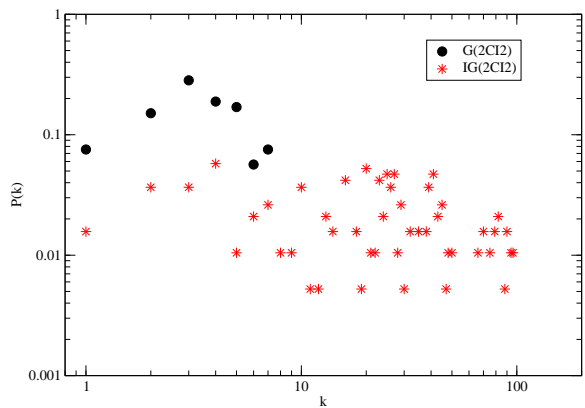}
 \caption{Degree probability distribution of normal (G) and incompatible (IG) network of protein 2CI2.} 
 \label{2CI2_Degree_dist} 
\end{figure}

G and IG of 2CI2 are constructed as explained above. G(2CI2) can be seen in Figure~\ref{CI2_network}. However, since the figure of IG(2CI2) was not clear due to high number of edges and nodes, it was not inserted here.

\begin{figure}[!htb] 
 \centering 
  \includegraphics[width=\textwidth]{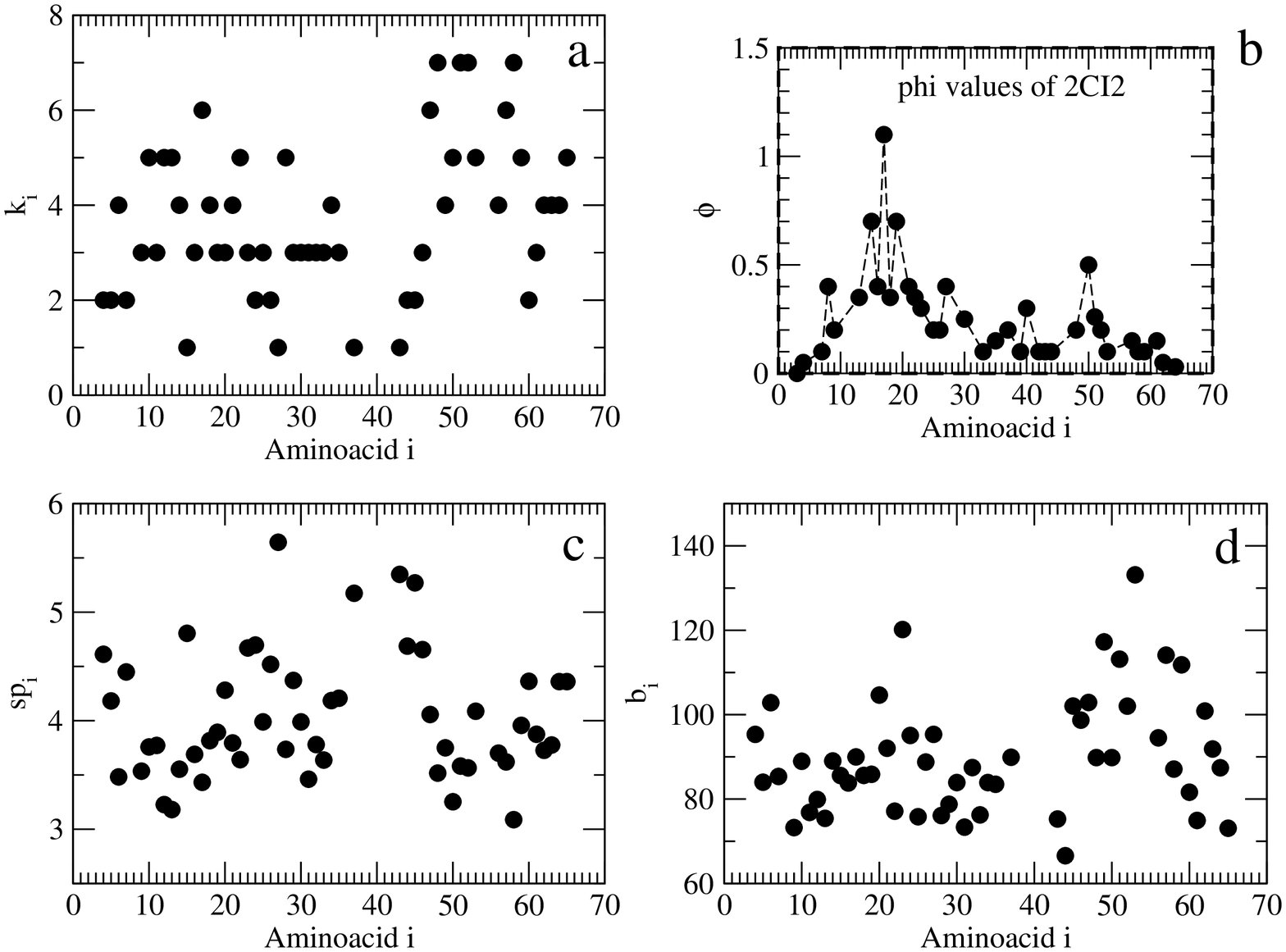}
 \caption[The comparison of $\phi$-values and topological features of normal network for 2CI2, G(2CI2)]{The comparison of $\phi$-values and topological features of normal network for 2CI2, G(2CI2); \textbf{a-)} degree of node i, \textbf{b-)} $\phi$-values of 2CI2, \textbf{c-)} node i's average shortest distance to other nodes, \textbf{d-)} betweenness of node i} 
 \label{2CI2_G_OtherTopologies} 
\end{figure}

\begin{figure}[!hbt] 
 \centering 
  \includegraphics[width=\textwidth]{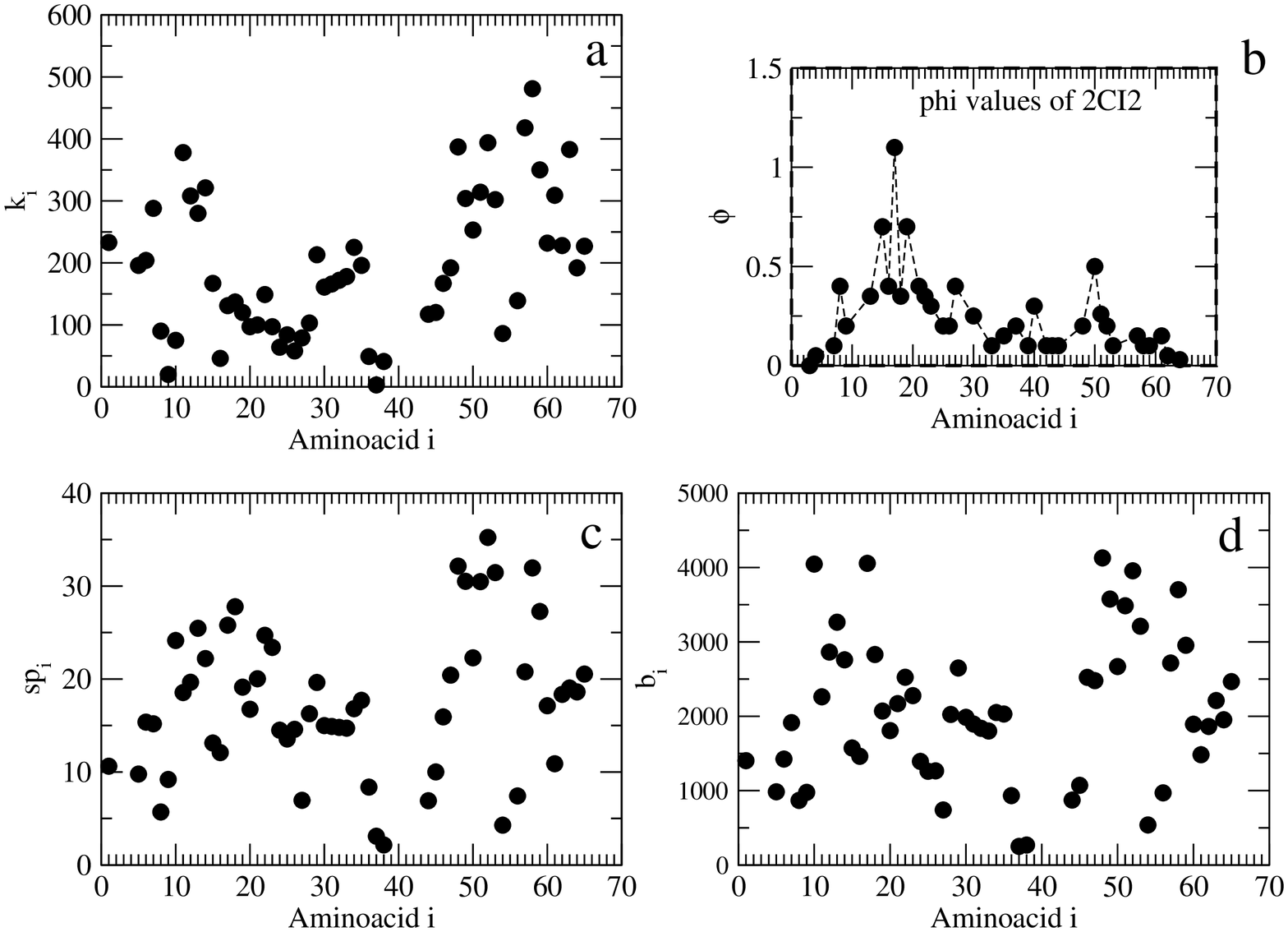}
 \caption[The comparison of $\phi$-values and topological features of incompatibility network for 2CI2, IG(2CI2)]{The comparison of $\phi$-values and topological features of incompatibility network for 2CI2, IG(2CI2); \textbf{a-)} degree of node i, \textbf{b-)} $\phi$-values of 2CI2, \textbf{c-)} node i's average shortest distance to other nodes, \textbf{d-)} betweenness of node i}
 \label{2CI2_IG_OtherTopologies} 
\end{figure}

The degree probability distributions of G(2CI2) and IG(2CI2) are given in Figure~\ref{2CI2_Degree_dist}. Other topological features related to nodes are explored for the sake of seeking a correlation to corresponding $\phi$-values (Figure~\ref{2CI2_G_OtherTopologies} and Figure~\ref{2CI2_IG_OtherTopologies}). When a similarity is seen by visual inspection, a statistical analysis is done in order to understand possible correlation. For the explanation of the statistical analysis, let me say $x_i$s are the candidates for being signals to $\phi_i$-values. I calculated the original root mean square deviation (RMSD) by $\sqrt{\sum_i{(\phi_i-x_i)^2}}$. Later, many times I shuffled the $\phi_i$ to have new $\phi$-values sets and each time I calculated another RMSD with new set of $\phi_i$ which gave me a distribution of RMSD at the end. If the original RMSD was in the first deviation part of this distribution, $x_i$s lost its candidateship for a signal. With this method, it was concluded that no correlation between topological features and $\phi$-values of 2CI2 exists.

\section{Other Proteins}\label{OtherProteins}

Apart from 2CI2, other proteins (2PTL, 1SHF, 1TEN and 1APS) are also studied, however, no correlation to $\phi$-values was seen for the limited data that is available. The $\phi$-value data of the proteins retrieved are given in Appendix~\ref{OtherPhiValues}.

\chapter{Conclusions \& Further Research}

Gene regulations are an important functional organization of the cells. Activation of the genes in eukaryotes has a dynamical structures. Because of the complexity of the systems, rough but powerful models: networks are used for both topological and dynamical investigations of the gene regulation. In particular, the boolean networks with synchronously updates are studied as generic models for the dynamical investigations. Although the individuals with interactions in the gene regulations are well established in literature, the functions that govern the activation of a gene are not fully understood. For this reason, different types of random functions are used in the studies, i.e. simple random, canalyzing, nested canalyzing and special subclasses of nested canalyzing functions.

Having fixed the network structures and the functions, the boolean dynamics possesses the state cycles called \textit{attractors} which is the main feature of the dynamics to quantify. In particular, the number of attractors, the length of the attractors, the transient length to attractors and basin of attractions are studied for understanding the boolean dynamics. It has been argued that the attractors correspond to some cell cycles in living organisms. Moreover, the robustness of a system is studied as an important property of biological system as ``Life at the edge of chaos'' hypothesis argues.

In this thesis, the model networks were introduced and investigated both in topology and dynamics. It has been shown that the fraction of dynamically relevant nodes to system size only depends on the average indegree $\langle k_{in}\rangle$ not on the topology type. The attractor features were explored for the network realization parameters $\langle k_{in}\rangle=2.0$ and $p=0.5$ which are discussed widely in literature. The distributions show power law decays for model networks. Average values of simple random functions are considerably bigger than canalyzing and nested canalyzing functions. Another important investigation was the scaling of the attractor features with the system size. I have shown that $\langle N_{attr} \rangle$, $\langle L_{attr} \rangle$ and $\langle \tau_{attr}\rangle$ of in-NK networks ($50<N<1000,\langle k_{in}=2.0\rangle$) and simple random functions($p=0.5$) scale with  $N^{0.53}$, $N^{0.87}$ and $N^{1.04}$, respectively. The result for the number of the attractor verifies $\sqrt{N}$ theorem while refutes for the length of the attractors. Also, $N^{0.53}$ scaling result refutes the recent study of Kauffman which argues that the scaling of the mean number of attractors is faster than linear. Apart from attractor studies, the robustness of the model networks was studied. It is shown that for all types of the topologies with simple random function, Derrida's expression $s=2p(1-p)\langle k_{in}\rangle$ is valid with a finite-size effect.

The yeast gene regulation network was investigated dynamically with a boolean approach. For the sake of comparison in attractor features for all function types, $p$ was set to $0.27$ which is the output $p$ value for the special subclasses of the nested canalyzing functions (SNCF) for the yeast GRN. The results show that SNCF are successful for the optimization of maximising the number of attractors and minimising the attractor lengths and transients which might be a desirable property for a biological system. Also, the distributions of the attractor features were observed with a nontypical distribution behavior for the number of attractors and the entropy. These distributions are not decreasing for all function types, especially for SNCF. As an important contribution, it was seen that SNCF type may be crucial to elucidate the actual dynamics of the yeast gene regulation. Moreover, the yeast GRN was compared with model networks whose indegree distributions decay similarly. The results show that yeast's attractor distributions and averages are not like model networks. But, it is observed that the robustness structures are similar. As a conclusion, it is seen that to know only the indegree distribution is not enough to produce attractor features of the yeast GRN while being very successful for the robustness behavior. As an another contribution, a recent model: \textit{Balcan et al.} model was discussed and it has been shown that it produces the similar topological networks of the yeast. However, dynamical analyses of this model networks established that the model is not successful at producing neither the attractor nor robustness structures. The main reason of this might be due to that \textit{Balcan et al.} model networks have low number of dynamically relevant nodes ($N=36 \mp 15$) comparing to actual yeast's ($N=82$).

At the last part of the thesis, a relation between the protein folding kinetics and a new network approach, incompatibility networks was asked. I have shown that no relation exists between some topological tools and known $\phi$-values for the limited data of some proteins.

\paragraph{Further Research:}

Mean attractor features were studied and the results contradict with Kauffman \textit{et al.}. An extended study which includes also SNCF and other $p$-values can be done a further work.

In the Appendix~\ref{FindingAttractorAlgorithms}, I discussed some finding attractor algorithms where I pointed a novel algorithm. I believe this algorithm would enhance the related research considerably and should be applied with a low-level programming language such as C++.

The SNCF type has been shown to give different results for the yeast GRN. It should be noted that this type of function is also very time-efficient since it uses a logical formalism (AND and OR functions). All these make me consider that only SNCF type can be used for a further the yeast attractor investigation.

I also saw the success of the Balcan \textit{et al.} model for producing the similar topologies. I believe that this model can be enhanced for the dynamical successes also with a consideration of the dynamically relevant subnetworks used in this thesis and \textit{motifs} in the literature.

Apart from the gene regulation part, I consider that the study I have done should be repeated when there is richer $\phi$-value data of the other proteins.

\appendix
\appendixpage

\chapter{Analytical Expression for $\langle k_{in}\rangle$}\label{ProducingNetworks}

For in-NK networks we have exactly $k_{in}$ edges going into each node, therefore, $\langle k_{in} \rangle=k_{in}$. However, for in-EXP and in-PL networks $k_{in}$ values vary for different nodes. The approximate analytical calculation of $\langle k_{in} \rangle$ follows from substituting the sum with an integral:

\underline{in-PL:}

\begin{equation}\label{PL_dist_Norm}
\begin{array}{ll}
1 &= \int_{k_{in}^{min}}^{k_{in}^{max}}A(\alpha)k^{-\alpha}dk \;\;(Normalization\; cond.) \\ 
\\ 
A(\alpha) &= \frac{1-\alpha}{(k_{in}^{max})^{1-\alpha}-(k_{in}^{min})^{1-\alpha}}
\end{array}
\end{equation} 
\begin{equation}\label{PL_dist_k_in}
\begin{array}{ll}
\langle k_{in}\rangle &= \int_{k_{in}^{min}}^{k_{in}^{max}}A(\alpha)k k^{-\alpha}dk \\
\\
 &= A(\alpha)\frac{({k_{in}^{max}}^{2-\alpha})-({k_{in}^{min}}^{2-\alpha})}{2-\alpha}
\end{array}
\end{equation}

\underline{in-EXP:}

\begin{equation}\label{EXP_dist_Norm}
\begin{array}{ll}
1 &= \int_{k_{in}^{min}}^{k_{in}^{max}}B(\lambda)e^{-\lambda k}dk \;\;(Normalization\;cond.)\\  
\\
B(\lambda)&= \frac{\lambda}{e^{\lambda {k_{in}^{max}}}-e^{\lambda {k_{in}^{min}}}}
\end{array}
\end{equation}
\begin{equation}\label{EXP_dist_k_in}
\begin{array}{ll}
\langle k_{in}\rangle &= \int_{k_{in}^{min}}^{k_{in}^{max}} B(\lambda)k e^{-\lambda k}dk \\
\\
 &= B(\lambda) \frac{({k_{in}^{max}}e^{\lambda {k_{in}^{max}}})-({k_{in}^{min}}e^{\lambda {k_{in}^{min}}})}{-\lambda} + \frac{1}{\lambda}
\end{array}
\end{equation} 

However, since in real case we have quantized k values, i.e. k=1,2,3,..,N. $\langle k_{in} \rangle$ deviates from the Eqs. \ref{PL_dist_k_in} and \ref{EXP_dist_k_in}. Correspondence of exponents of PL and EXP networks to $\langle k_{in} \rangle$ for both analytical and actual cases is presented in Figure \ref{PLandEXP_gammaVSavK}. This figure also discuss the main reason of this deviation.

\chapter{Finding Attractor Algorithms}\label{FindingAttractorAlgorithms}

Finding all the attractors of a large network is a challenging task. There are mainly two types algorithms for finding the attractors: \textbf{exhaustive} and \textbf{heuristic}.

An exhaustive kind algorithm is desired solution which finds all attractors. One can consider two types of exhaustive algorithm. First one is the straightforward method which starts from each initial network states and finds the attractors. However, since the numbers of network states ($=2^N$) grows exponentially with $N$, it is computationally infeasible with the available computers. After some trials I concluded this exact algorithm fails for $N>18$. Second type is a novel algorithm which I have seen during surveying the literature and it claims to find all attractors of N around $100$ \cite{Irons_FindingAttractor}. Its main idea is to go back from the partial network states (a state description includes only $2-$, $3-$, etc. node sates) and to try to clarify which partial states are impossible to be found in any attractors of that network realization. I have scripted this algorithm with using \textbf{python}\footnote{Available at http://www.python.org} \cite{Python} although I did not get the efficiency stated in the paper \cite{Irons_FindingAttractor}, mainly because of using a higher level script language rather than a low level programming language such as C++. Yet, it was more powerful than straightforward method approximately for $17<N<60$.

Second approach is the heuristic algorithms which sample from the random initial conditions and finds the algorithms. Since the biological networks in the dynamics of interest in this thesis are of size $N \cong 85$ or more I left the exhaustive algorithms and implemented a heuristic algorithm.
\clearpage

\chapter{$\phi$-values of Some Proteins}\label{OtherPhiValues}

$\phi$-values which are found by private communication with Prof.~Michele Vendruscolo are given in Table \ref{Phivalues_Vendruscolo_3}, Table \ref{Phivalues_Vendruscolo_2} and Table \ref{Phivalues_Vendruscolo_1}.

\begin{table}[!htb]
\begin{center}
\begin{tabular}{||c|c||c|c||c|c||}
\hline
1bf4	&		&	1bk2	&		&	1shf2	&		\\
res. i	& $\phi_i$ &	res. i	&	$\phi_i$ &res. i	&	$\phi_i$	\\
\hline
3	&	0.01	&	3	&	0.16	&	4	&	0.28	\\
14	&	0	&	6	&	0	&	6	&	0.18	\\
16	&	0	&	18	&	0.32	&	18	&	0.06	\\
26	&	1	&	19	&	0.29	&	20	&	0.22	\\
29	&	0.44	&	24	&	0.22	&	24	&	0.41	\\
30	&	0	&	31	&	0.25	&	26	&	0.15	\\
31	&	0.43	&	38	&	0.26	&	28	&	0.71	\\
34	&	0.3	&	39	&	0.48	&	39	&	0.86	\\
36	&	0.25	&	41	&	1	&	41	&	1	\\
40	&	0.22	&	47	&	0.58	&	44	&	0.74	\\
42	&	0.21	&	48	&	0.61	&	50	&	0.37	\\
44	&	0.59	&	50	&	0.53	&	55	&	0.01	\\
45	&	0.09	&	53	&	0.16	&		&		\\
50	&	0.22	&		&		&		&		\\
54	&	0.21	&		&		&		&		\\
55	&	0.27	&		&		&		&		\\
58	&	0.6	&		&		&		&		\\
\hline
\end{tabular}
\end{center}
\label{Phivalues_Vendruscolo_3}
\caption{$\phi$-values of 1bf4, 1bk2 and 1shf2.}
\end{table}


\begin{table}[!htb]
\begin{center}
\begin{tabular}{||c|c|c|c||c|c|c|c||c|c||}
\hline
2ci2	&		&	2ci2	&		&	2ptl	&		&	2ptl	&		&	1aps	&		\\
res. i	&	$\phi_i$	&	res. i	&	$\phi_i$	&	res. i	&	$\phi_i$	&	res. i	&	$\phi_i$	&	res. i	&	$\phi_i$	\\
\hline
3	&	0	&	42	&	0.1	&	4	&	0.58	&	31	&	0.4	&	11	&	0.93	\\
4	&	0.05	&	43	&	0.1	&	5	&	0.27	&	32	&	0.21	&	13	&	0.37	\\
7	&	0.1	&	44	&	0.1	&	6	&	0.41	&	33	&	0.3	&	17	&	0.1	\\
8	&	0.4	&	48	&	0.2	&	7	&	0.64	&	34	&	0.09	&	20	&	0.18	\\
9	&	0.2	&	50	&	0.5	&	8	&	0.56	&	35	&	0.29	&	22	&	0.09	\\
13	&	0.35	&	51	&	0.26	&	9	&	0.14	&	36	&	0.29	&	29	&	0.15	\\
15	&	0.7	&	52	&	0.2	&	10	&	0.41	&	37	&	0.12	&	30	&	0.42	\\
16	&	0.4	&	53	&	0.1	&	11	&	0.59	&	38	&	0	&	36	&	0.22	\\
17	&	1.1	&	57	&	0.15	&	12	&	0.18	&	40	&	0.16	&	39	&	0.14	\\
18	&	0.35	&	58	&	0.1	&	13	&	0.55	&	44	&	0	&	42	&	0.37	\\
19	&	0.7	&	59	&	0.1	&	14	&	0.79	&	45	&	0	&	45	&	0.58	\\
21	&	0.4	&	61	&	0.15	&	15	&	0.68	&	48	&	0.26	&	47	&	0.54	\\
22	&	0.35	&	62	&	0.05	&	17	&	0.4	&	49	&	0.33	&	51	&	0.39	\\
23	&	0.3	&	64	&	0.03	&	19	&	0.25	&	51	&	0.24	&	54	&	0.98	\\
25	&	0.2	&		&		&	20	&	0.59	&	52	&	0	&	61	&	0.21	\\
26	&	0.2	&		&		&	21	&	0.85	&	55	&	0.12	&	64	&	0.34	\\
27	&	0.4	&		&		&	22	&	0.5	&	56	&	0.25	&	65	&	0.27	\\
30	&	0.25	&		&		&	23	&	0.45	&	57	&	0.14	&	71	&	0.09	\\
33	&	0.1	&		&		&	24	&	0.3	&	58	&	0.28	&	75	&	0.02	\\
35	&	0.15	&		&		&	25	&	0.45	&	59	&	0.17	&	78	&	0.02	\\
37	&	0.2	&		&		&	26	&	0.8	&	60	&	0.19	&	83	&	0.04	\\
39	&	0.1	&		&		&	29	&	0.27	&	61	&	0.09	&	86	&	0	\\
40	&	0.3	&		&		&	30	&	0.08	&	62	&	0	&	89	&	0.07	\\
	&		&		&		&		&		&		&		&	94	&	0.76	\\
\hline
\end{tabular}
\end{center}
\label{Phivalues_Vendruscolo_1}
\caption{$\phi$-values of 2ci2, 2ptl and 1aps.}
\end{table} 

\begin{table}[!htb]
\begin{center}
\begin{tabular}{||c|c||c|c||c|c||}
\hline
1ten	&	&	1fmk	&		&	1imq		&		\\
res. i	&	$\phi_i$ &	res. i	& $\phi_i$	&	res. i	& $\phi_i$	\\
\hline
1	&	0.04	&	1	&	0	&	7	&	0.15	\\
4	&	0.04	&	2	&	0.1	&	13	&	0.98	\\
7	&	0.1	&	3	&	0.03	&	15	&	0.33	\\
9	&	0.23	&	4	&	0.05	&	16	&	0.52	\\
17	&	0.14	&	5	&	0	&	18	&	0.4	\\
19	&	0.39	&	8	&	0.03	&	19	&	0.32	\\
28	&	0.13	&	10	&	0.28	&	22	&	0.31	\\
31	&	0.19	&	15	&	0.13	&	27	&	0.12	\\
33	&	0.35	&	16	&	0.26	&	33	&	0.27	\\
35	&	0.53	&	17	&	0.03	&	36	&	0.25	\\
47	&	0.67	&	18	&	0.4	&	37	&	0.15	\\
49	&	0.42	&	22	&	0.62	&	40	&	0.01	\\
56	&	0.38	&	24	&	0.55	&	52	&	0.03	\\
58	&	0.6	&	27	&	0.77	&	53	&	0.07	\\
61	&	0.33	&	34	&	0.25	&	67	&	0.41	\\
63	&	0.47	&	35	&	0.15	&	68	&	0.23	\\
65	&	0.25	&	36	&	0.54	&	71	&	0.36	\\
67	&	0.42	&	37	&	1	&	76	&	0.37	\\
69	&	0.54	&	38	&	0.08	&	77	&	0.37	\\
71	&	0.29	&	39	&	0.95	&	83	&	0.31	\\
76	&	0.21	&	40	&	0.72	&		&		\\
80	&	0.03	&	42	&	0.86	&		&		\\
83	&	0.21	&	45	&	0.68	&		&		\\
85	&	0.08	&	47	&	0.56	&		&		\\
87	&	0.11	&	48	&	0.71	&		&		\\
89	&	0.11	&	49	&	0.24	&		&		\\
	&		&	53	&	0	&		&		\\
\hline
\end{tabular}
\end{center}
\label{Phivalues_Vendruscolo_2}
\caption{$\phi$-values of 1ten, 1fmk and 1imq.}
\end{table}



\bibliographystyle{unsrt}
\bibliography{myrefs}



\vita{Murat Tu\u{g}rul was born in Tirebolu in April 1, 1983. After living in Tirebolu, İstanbul, Batman, Ankara, Anamur, Afyon and Eskişehir; in September, 2000 he arrived at Middle East Technical University, Ankara where he got his BSc degree in physics and minor degree in philosophy in June, 2005. With an intention to do research in the biological sciences he started the ``Computational Sciences and Engineering'' master program of Koç University in İstanbul with full scholarship in September 2005. He worked there as a teaching assistant of mathematics and physics courses. He graduates in December 2007 with a thesis title \textit{The Structure and Dynamics of Gene Regulation Networks}. In February, 2008 he will join the Institute for Cross-Disciplinary Physics and Complex Systems in Palma de Mallorca for his PhD degree in physics with an intention to do research in ecology and evolution.}

\end{document}